\documentclass[12pt,preprint]{aastex}

\shorttitle{Spectral Energy Distribution of NGC 4631}
\shortauthors{Bendo et al.}

\begin{document}

\title{The Spectral Energy Distribution of Dust Emission in the
Edge-on Spiral Galaxy NGC~4631 as seen with {\it Spitzer} and JCMT}

\author{George J. Bendo\altaffilmark{1,2,3}, 
Daniel A. Dale\altaffilmark{4},
Bruce T. Draine\altaffilmark{5},
Charles W. Engelbracht\altaffilmark{2},
Robert C. Kennicutt, Jr.,\altaffilmark{2,6},
Daniela Calzetti\altaffilmark{7},
Karl D. Gordon\altaffilmark{2},
George Helou\altaffilmark{8},
David Hollenbach\altaffilmark{9},
Aigen Li\altaffilmark{10},
Eric J. Murphy\altaffilmark{11},
Moire K. M. Prescott\altaffilmark{2},
and John-David T. Smith\altaffilmark{2}
}
\altaffiltext{1}{Astrophysics Group, Imperial College, Blackett Laboratory,
    Prince Consort Road, London SW7 2AZ United Kingdom; g.bendo@imperial.ac.uk}
\altaffiltext{2}{Steward Observatory, University of Arizona, 933 North 
    Cherry Avenue, Tucson, AZ 85721}
\altaffiltext{3}{Guest User, Canadian Astronomy Data Centre, which is 
    operated by the Herzberg Institute of Astrophysics, National Research 
    Council of Canada.}
\altaffiltext{4}{Department of Physics and Astronomy, University of Wyoming, 
    Laramie, WY 82071.}
\altaffiltext{5}{Princeton University Observatory, Peyton Hall, Princeton, 
    NJ 08544-1001.}
\altaffiltext{6}{Institute of Astronomy, University of Cambridge, 
    Cambridge CB3 0HA, United Kingdom}
\altaffiltext{7}{Space Telescope Science Institute, 3700 San Martin Drive, 
    Baltimore, MD 21218.}
\altaffiltext{8}{California Institute of Technology, MC 314-6, Pasadena, CA 
    91101 USA.}
\altaffiltext{9}{NASA Ames Research Center, MS 245-3, Moffett Field, CA 
    94035-1000 USA.}
\altaffiltext{10}{Department of Physics and Astronomy, University of 
    Missouri, Columbia, MO 65211.}
\altaffiltext{11}{Department of Astronomy, Yale University, P.O. Box 208101, 
    New Haven, CT 06520 USA.}

\begin{abstract}
We explore the nature of variations in dust emission within an
individual galaxy using 3.6 - 160~$\mu$m {\it Spitzer} Space Telescope
observations and 450 and 850~$\mu$m James Clerk Maxwell Telescope
observations of the edge-on Sd spiral galaxy NGC~4631 with the goals
of understanding the relation between polycyclic aromatic hydrocarbons
(PAH) and dust emission, studying the variations in the colors of the
dust emission, and searching for possible excess submillimeter
emission compared to what is anticipated based on dust models applied
to the mid- and far-infrared data.  PAH emission at 8~$\mu$m is found
to correlate best with hot dust emission at 24~$\mu$m on kiloparsec
scales, although the relation breaks down on scales equal to hundreds
of parsecs, possibly because of differences in the mean free paths
between the photons that excite the PAHs and heat the dust and
possibly because the PAHs are destroyed by the hard radiation fields
in the centers of some star formation regions.  The ratio of 8~$\mu$m
PAH emission to 160~$\mu$m cool dust emission appears to be a function
of radius.  The 70/160 and 160/450~$\mu$m flux density ratios are
remarkably constant even though the surface brightness varies by a
factor of 25 in each wave band, which suggests that the emission is
from dust heated by a nearly-uniform radiation field.  Globally, we
find an excess of 850~$\mu$m emission relative to what would be
predicted by dust models.  The 850~$\mu$m excess is highest in regions
with low 160~$\mu$m surface brightness, although the strength and
statistical significance of this result depends on the model fit to
the data.  We rule out variable emissivity functions or $\sim4$~K dust
as the possible origins of this 850~$\mu$m emission, but we do discuss
the other possible mechanisms that could produce the emission.
\end{abstract}


\section{Introduction \label{s_intro}}

Before the launch of the {\it Spitzer} Space Telescope
\citep{wetal04}, research into the 1 - 1000~$\mu$m spectral energy
distributions (SEDs) of nearby galaxies was greatly restricted by the
angular resolution and signal-to-noise levels of far-infrared data.
Most of the {\it Infrared Astronomical Satellite} (IRAS) and {\it
Infrared Space Observatory} (ISO) data were only usable as global flux
density measurements.  In the few cases where galaxies were resolved
at far-infrared wavelengths, the spatial information was limited.
With {\it Spitzer} data, it is now possible to examine 1 - 1000~$\mu$m
SEDs for kiloparsec-sized regions in galaxies at distances up to
10~Mpc.  These higher-resolution, higher-sensitivity data allow for
new observational investigations into the properties of the polycyclic
aromatic hydrocarbons (PAHs) and other dust across this large
wavelength regime.  In this paper, we will use {\it Spitzer} data to
address two major issues related to dust emission that were not
completely resolved using IRAS and ISO data.

The first major question that can be asked is how best to describe the
SEDs of discrete regions within individual nearby spiral galaxies,
particularly in the context of describing the dust emission between 60
and 2000~$\mu$m. Studies combining IRAS and ISO results
with submillimeter or millimeter data suggested that a range of simple
one or two-component blackbodies modified with different emissivity
laws could describe the data.  The exact descriptions, however, were
confusing and contradictory.  The major point of debate was how to
describe the emission at submillimeter or millimeter wavelengths that
exceeded what is expected when extrapolating from the 60 - 200$\mu$m
regime using one or a series of $\sim20$ - 30~K blackbodies modified
by $\lambda^{-2}$ emissivity functions.  Some studies have suggested
that the power law describing the emissivity function varies among
nearby galaxies \citep[e.g][]{deeiac00, betal03}.  Other studies found
that the submillimeter or millimeter excess could be described by a
cold blackbody modified by a $\lambda^{-2}$ emissivity law.  This
cold dust temperature component could have a temperature as low as
$\lesssim10$~K \citep[e.g][]{kszc98, skc99, gmjwbl03, gmjwb05}.

\citet{rtbetal04} were among the first to examine this issue using
{\it Spitzer} data.  They examined the SEDs of the central starburst
ring and outer disk of NGC~7331 and found the 70 - 850~$\mu$m dust
emission from the ring was consistent with dust in the $\sim20$ - 25~K
range with an emissivity law of $\lambda^{-2}$.  However, a limitation
of the study by Regan et al. was the lack of significant
submillimeter emission at a high signal-to-noise level outside of the
ring.  They therefore could not determine whether the description of
the dust emission in the center of NGC~7331 applies throughout the
galaxy's disk.

This issue is particularly important in terms of modeling the dust
emission and estimating the dust mass within nearby galaxies.  If dust
as cold as $\sim6$~K is present, it will only be a significant source
of emission at submillimeter wavelengths, but it may contribute
substantially to the mass of the interstellar medium.  If the dust
emissivity is variable then dust models need to be adjusted to account
for this variability.  Without these results, the full picture of
dust emission in nearby galaxies is incomplete.

The other major question is to determine how PAH emission,
particularly the 7.7~$\mu$m feature, is correlated with dust emission
on scales of hundreds of parsecs.  Results from ISO gave differing
results.  \citet{frsc04}, among others, found a strong correlation
between 7 and 15~$\mu$m emission in galaxy disks, thus
demonstrating a correlation between PAH and hot ($\sim100$~K) dust
emission (which is commonly associated with transiently-heated very
small grain emission but which may also represent emission from grains
that are in thermal equilibrium at $\sim100$~K).  \citet{hkb02},
however, found that 8~$\mu$m PAH emission was correlated more closely
with 850~$\mu$m emission from cool ($\sim25$~K) grains than with the
15~$\mu$m emission from hot dust.  Using {\it Spitzer} observations of
NGC~300, \citet{hraetal04} found that 8~$\mu$m emission originated
from the rims of star formation regions within the galaxy whereas the
24~$\mu$m emission was more strongly peaked in the centers of the star
formation region, which suggested a correlation on kiloparsec scales
but not on smaller scales.  \citet{detal05}, however, found that
8~$\mu$m emission from kiloparsec sized regions within M~81, M~51, and
NGC~7331 appeared to be more closely correlated with the total
infrared flux than with the 24~$\mu$m flux density.  In a separate
analysis of the {\it Spitzer} data for M~81, P\'erez-Gonz\'alez et
al. (2006, in press) also found that the 8~$\mu$m luminosity was
closely correlated with the total infrared luminosity on kiloparsec
scales, although they did not examine the relation between 8 and
24~$\mu$m emission.

Identifying how PAH emission is correlated with dust emission will
lead to a better understanding of how it can be used as a star
formation indicator, as has been suggested by the results of
\citet{rsvb01} and \citet{frsc04} (although the contradictory results
of \citet{cetal05} should also be noted). If PAH emission is
correlated with 24~$\mu$m hot dust emission on scales of hundreds of
parsecs or kiloparsecs, then it should be as effective a star
formation indicator as 24~$\mu$m emission on those spatial scales.
However, if PAH emission is more closely correlated with 160~$\mu$m
cool dust emission, which may trace extended cirrus emission unrelated
to star formation \citep[e.g.][]{hrgetal04}, then the PAH emission
should be used with greater caution as a star formation tracer.

Among the data available in the {\it Spitzer} Infrared Nearby Galaxy
Survey \citep[SINGS;][]{ketal03}, the data for \object{NGC 4631} are
optimal for studying the complete SED of the dust emission from
mid-infrared to submillimeter wavelengths.  NGC~4631 is a nearly
edge-on \citep[inclination $85\deg$;][]{t88} Sd galaxy at a distance
of 9.0~Mpc with an optical disk of $15^\prime.5 \times 2^\prime.7$
\citep{ddcbpf91}.  The edge-on nature of the galaxy makes it a high
surface brightness source that is easier to detect at all wavelengths,
particularly submillimeter wavelengths.  However, the edge-on
orientation adds a geometry effect to the observed surface brightness
variations; regions may appear bright because they are intrinsicly
luminous or because they represent regions with high column densities.
Nonetheless, in terms of this study, the advantages of the edge-on
orientation outweigh the disadvantages.

In addition to the Infrared Array Camera \citep[IRAC;][]{fetal04} and
Multiband Imaging Photometer for {\it Spitzer}
\citep[MIPS;][]{ryeetal04} data taken as part of the SINGS legacy
program, deep archival SCUBA \citep{hetal99} 450 and 850~$\mu$m data
covering most of the optical disk \citep[previously published
in][]{adb99, sag05} are available for this galaxy.  These data almost
completely cover the extended dust emission in this galaxy and allow
for extraction of 3.6-850~$\mu$m SEDs from regions throughout the
disk.  Mid-infrared-to-submillimeter SEDs of this galaxy have been
studied before \citep{betal03, dkw04, sag05}.  However, these previous
studies have relied on IRAS and ISO data that have lacked the
resolution, the sensitivity, or the spatial coverage to study anything
more than the global SED or the SED of the center of the object.  With
these new data, we can study infrared color variations across the
entire optical disk of the galaxy and examine the SEDs within discrete
regions throughout the disk.  These results will lead to answers as to
how to characterize far-infrared to submillimeter dust emission
throughout a spiral galaxy and how PAH and dust emission at different
wavelengths are interrelated.

In Section~\ref{s_obs}, we present information on the observations and
data reduction for this galaxy.  We briefly discuss the images in
Section~\ref{s_images}.  In Section~\ref{s_color}, we present simple
color information from flux densities integrated over set apertures
distributed throughout the plane of the galaxy.  Then, in
Section~\ref{s_sed}, we present the 3.6 - 850~$\mu$m spectral energy
distributions of the galaxy as a whole and of discrete regions within
the galaxy.  We then summarize the results in
Section~\ref{s_conclusions}.

\section{Observations and Data Reduction \label{s_obs}}

\subsection{IRAC (3.6, 4.5, 5.7, and 8.0~$\mu$m) Data}

The 3.6 - 8.0~$\mu$m data were taken with IRAC on the {\it Spitzer}
Space Telescope on 2004 May 23 and 26 in IRAC campaign 5.  The
observations consist of a series of dithered $5^\prime \times 5^\prime$
individual frames that are offset $2.^\prime5$ from each other.  The
two separate sets of observations allow asteroids to be recognized and
provide observations at orientations up to a few degrees apart to ease
removal of detector artifacts.  Points in the center are imaged eight
times in 30 s exposures.  The full-width half-maxima (FWHM) of the
point spread functions (PSFs) are
$1.^{\prime\prime}6-2.^{\prime\prime}0$ for the four wavebands.

The data are processed using a special SINGS IRAC pipeline.  First, a
geometric distortion correction is applied to the individual frames.
Data from the second set of observations are rotated to the same
orientation as the first set of observations.  Bias structure is
subtracted from the data by subtracting a bias frame from each frame.
This bias frame is made by median combining all data frames for the
observations; the median filtering effectively removes any target
signal from the data, leaving only the bias signal.  Next, the image
offsets are determined through image cross-correlation.  Following
this, bias drift is removed by determining the relative offset between
overlapping regions of individual frames and subtracting the offset.
Finally, cosmic ray masks are created using standard drizzle methods,
and final image mosaics are created using a drizzle technique.  The
final pixel scales are set at $\sim0^{\prime\prime}.75$.  A final
residual background is measured in small regions outside the target
that are free of bright foreground/background sources, and then this
residual is subtracted from the data.  The uncertainty in the
calibration factor applied to the final mosaics is 10\%.

\subsection{MIPS (24, 70, and 160~$\mu$m) Data}

The 24, 70, and 160~$\mu$m data were taken with MIPS on the {\it
Spitzer} Space Telescope on 2005 January 24 and 26 in MIPS campaign
18. The observations were obtained using the scan-mapping mode in two
separate visits to the galaxy, which has the same benefits as for the
IRAC data. As a result of redundancy inherent in the scan-mapping
mode, each pixel in the core map area was effectively observed 40, 20,
and 4 times at 24, 70, and 160~$\mu$m, respectively, resulting in
integration times per pixel of 160 s, 80 s, and 16 s, respectively.
The FWHM of the PSFs are $6{\prime\prime}$ at 24~$\mu$m,
$18{\prime\prime}$ at 70~$\mu$m, and $40{\prime\prime}$ at 160~$\mu$m.

 The MIPS data were processed using the MIPS Data Analysis Tools
\citep{getal05} version 2.96.  Since the processing for the 24~$\mu$m
data differed from the 70 and 160~$\mu$m data, the 24~$\mu$m data
processing will be discussed separately from the 70 and 160~$\mu$m
wavebands.

First, the 24~$\mu$m images were processed through a droop correction
(that removes an excess signal in each pixel that is proportional to
the signal in the entire array) and were corrected for non-linearity in the
ramps.  The dark current was then subtracted.  Next, scan-mirror-position
dependent flats (created from all SINGS MIPS off-target data from MIPS
campaign 18) were applied to the data.  Following this,
scan-mirror-position independent flats (made from off-target data
frames within each visit's scan map) were applied.  A correction to
the offsets between the four channels reading out the 24~$\mu$m
detector data was applied to each data frame. Latent images from
bright sources, erroneously high or low pixel values, and unusually
noisy frames were also masked out.  For background subtraction, the
median background value as a function of time was determined for each
scan leg and then subtracted.  The data from both visits were then
mosaicked together, and a final residual background offset was
measured outside the optical disk of the galaxy and subtracted.  We
applied the S10 calibration factors in the MIPS Data
Handbook\footnote[11]{http://ssc.spitzer.caltech.edu/mips/dh/}
\citep{s06} to the final mosaics.  The calibration factor has an
uncertainty of $\sim$10\%.  Color corrections, which are smaller than
the calibration uncertainties, are not applied to these data except
for making comparisons to IRAS 25~$\mu$m data.

For the 70 and 160~$\mu$m data, ramps were fit to the 70 and
160~$\mu$m reads to derive slopes.  In this step, readout jumps and
cosmic ray hits were also removed, and an electronic nonlinearity
correction was applied.  Next, the stim flash frames taken by the
instrument were used as responsivity corrections.  The dark current
was subtracted from the data, an illumination correction was applied,
and then short term variations in drift were subtracted from the data
(which also subtracted the background).  Following this, erroneously
high or low pixel values were removed from the data.  The data from
both visits were then mosaicked together, and a residual offset
measured in regions staggering the target was subtracted from the
final maps.  We applied the S10 calibration factors in the MIPS Data
Handbook, which have uncertainties of $\sim$20\%, to the final
mosaics.  These uncertainties can also be treated as encompassing the
uncertainties from nonlinearity effects in the 70~$\mu$m wave band.
The color corrections are smaller than the calibration uncertainties
and are not applied to these data.  Note that the 70~$\mu$m data were
affected by a latent image effect that created negative streaks to the
south of the central $\sim2^{\prime}$ of the data. However, we will
not be using this region in the analysis.

To check the accuracy of the flux density measurement at 24~$\mu$m, we
can compare the integrated flux density with the 25~$\mu$m integrated
flux density measured from IRAS data.  The 24~$\mu$m global flux
density measured here is $8.0 \pm 0.8$~Jy.  The MIPS Data Handbook
suggests a color correction term of $\sim0.95$ for blackbody emission
from a 100-200~K source\citep{s06}, which is what is anticipated for
the dust emission that dominates the 24~$\mu$m wave band.  Dividing
the observed flux density by this factor gives a color-corrected
24~$\mu$m flux density of 8.4~Jy.  \citet{rlsnkldh88} published a
measured 25~$\mu$m flux density of $9.65 \pm 1.4$~Jy and a
color-corrected flux density of 8.35~Jy.  Given the uncertainties in
the MIPS and IRAS measurements, the color-corrected flux densities are
in good agreement with each other.

\subsection{SCUBA (450 and 850~$\mu$m) Data}

The 450 and 850~$\mu$m data were taken from the JCMT SCUBA
archives maintained by the Canadian Astronomy Data Centre.  The data
used were taken in jiggle map mode.  Each jiggle map consists of data
taken in a $\sim2.3^\prime$ region.  During the observations, the
secondary shifts through a 64-point pattern so that the bolometers may
fully sample the image plane.  Additionally, the secondary chops
off-source so as to measure the background during the observations.
In the data used here, the chop was $120^\prime$-$180^\prime$ in the
north-south direction.  The FWHM of the PSFs are $8{\prime\prime}$ and
$14{\prime\prime}$ at 450 and 850~$\mu$m, respectively.

Data from projects with identification numbers U43 (5-6 Oct 1997),
M98BU19 (8-9 Mar 1998), M99AI16 (14 May 1999), and M00BU51 (14 Jan
2001) were used to build the submillimeter images.  These data consist
of simultaneous observations at 450 and 850~$\mu$m, although the data
from 14 May 1999 and one frame of data from 14 Jan 2001 were not
usable for constructing the 450~$\mu$m map.  Note that the SCUBA
narrow-band filters were used for the 1997-1999 observations and the
wide-band filters were used for 2001 data; except for some
improvements in the signal-to-noise in the 450~$\mu$m wide-band filter
data, the two filter sets are otherwise functionally similar and
treated that way in the data processing.  All together, the data
consist of 182 integrations (194 min) of on-target observations (of
which, 149 integrations (159 min) were usable 450~$\mu$m data).

The data were processed with the SCUBA User Reduction Facility
\citep{jl98}.  The data were first flatfielded and corrected for
atmospheric extinction.  Noisy bolometers were removed next, followed
by spike removal.  Then the background signal was subtracted using the
signal from several bolometers at the north and south edges of each
frame that were selected so as to avoid noisy pixels and any possible
negative images (which are an artifact related to the chopping
performed during the observations).  The data were calibrated using
observations of the planets Mars or Uranus or the submillimeter
standard CRL~618 and then regridded onto the sky plane.  An additional
background subtraction was then performed to remove any residual
offset.  A $6^{\prime\prime}$ border at the edge of the observed field
was masked out so as to remove anomalously ``hot'' or ``cold'' pixels
that are artifacts of the sky subtraction.  These anomalous pixels are
identifiable by the amplitude of the signal, which is greater than the
signal from the target, and by the shapes of their radial profiles,
which are narrower than the PSFs of true sources.  Additional
anomalous pixels were identified by eye and masked out.  Negative
pixels near the south edge of the map just east of the nucleus were
identified as a negative image of the plane of the galaxy and masked
out in the final map, and the region around a noisy bolometer at the
west end of the 850~$\mu$m map was masked out as well.  Typical
uncertainties in the calibration are $\sim25$\% at 450~$\mu$m and
$\sim10$\% at 850~$\mu$m.

To test the data reduction, the flux densities can be compared to the
450 and 850~$\mu$m flux densities measured in the same data by
\citet{sag05} and the 870~$\mu$m flux density measured by
\citet{dkw04} with the 19-channel bolometer array at the
Heinrich-Hertz Telescope (HHT).  The flux densities measured with the
archival SCUBA data in this paper (without adjustments for CO, free-free,
or synchrotron emission) are $24 \pm 6$~Jy at 450~$\mu$m and $4.9 \pm
0.5$~Jy at 850~$\mu$m.  \citet{sag05} measured statistically identical
450 and 850~$\mu$m global flux densities (without any adjustments)
from the same data, thus demonstrating that the data reduction
techniques used here can accurately reproduce the data.  \citet{dkw04}
measured the 870~$\mu$m flux density (without any adjustments) to be
$3.8 \pm 0.6$~Jy.  This is only a $\sim2\sigma$ discrepancy between
the SCUBA and HHT measurements.  The SCUBA data, however, have a higher
signal-to-noise ratio throughout the disk than the HHT data.  The
plane of the disk from $\sim200^{\prime\prime}$ east of the nucleus to
$\sim300^{\prime\prime}$ west of the nucleus is detected at the
$3\sigma$ level in the unsmoothed 850~$\mu$m map presented here.  In
the smoothed 870~$\mu$m data of Dumke et al., however, some regions
inside the disk of the galaxy with signal-to-noise levels of less than
$2\sigma$ are visible, particularly to the east of the center.  In the
central $135^{\prime\prime}$, where the signal-to-noise levels of both the
SCUBA data and the Dumke et al. data are greater than $5\sigma$, the
two maps give results that agree with each other.  We measure the flux
density to be $1.8 \pm 0.2$~Jy within this region, whereas Dumke et
al. obtained 1.8~Jy.  Therefore, we conclude that the disagreement is
for the flux density measurements in the outer disk.  Because the
signal-to-noise levels of the SCUBA data are superior to the Dumke et
al. data, we will rely only on the SCUBA data for 850~$\mu$m flux
densities.  (The signal-to-noise ratio of the 1230~$\mu$m data in Dumke et
al. are higher than $3\sigma$ throughout the disk, so these data
should be more reliable.  We will use these 1230~$\mu$m data in
constructing global SEDs.)

\subsection{Convolution of the Data \label{s_obs_kernel}}

To compare data between wavelengths of different resolutions on scales
smaller than an arcminute, it is best to convolve the data with
kernels that will match not only the FWHM of the PSFs but also the
shapes of the PSFs.  We are using convolution kernels that will
convert an input PSF into a lower resolution output PSF using the
ratio of Fourier transforms of the output to input PSFs.  High
frequency noise in the input PSF is suppressed when these kernels are
created. See K. D. Gordon et al.  (2006, in preparation) for details.

For comparing flux densities between two wave bands, we match the
resolutions of the data to either the $18{\prime\prime}$ resolution of
the 70~$\mu$m data or the $40{\prime\prime}$ resolution of the
160~$\mu$m data.  In a comparison of PAH 8~$\mu$m and 24~$\mu$m maps
in Section~\ref{s_color_smallgrain}, we degraded the IRAC data to the
$6{\prime\prime}$ resolution of the 24~$\mu$m data.  Note that the
kernel for degrading the resolution of the 450~$\mu$m maps to
$18{\prime\prime}$ was not available at the time of this writing.

\section{Images \label{s_images}}

The global 3.6-850~$\mu$m images of these data are presented in
Figure~\ref{f_globalimages}.  The PAH and dust emission structures
from 5.7~$\mu$m to 850~$\mu$m all appear very similar, and they do
trace similar structures seen in the 3.6 and 4.5~$\mu$m bands, which
contains the Rayleigh-Jeans side of the blackbody starlight emission.
Most of the emission lies on the major axis of the optical disk of the
galaxy.  No bulge is visible, even in the 3.6 and 4.5~$\mu$m bands.

The brightest 3.6-850~$\mu$m source is a central structure with a
diameter of $\sim2^\prime$ that looks like an edge-on ring.  Magnified
images of this structure are shown in Figure~\ref{f_centerimages}.  
Both the east and west lobes of this central structure as well as the
nucleus are the three single brightest sources in the central region,
although the relative brightness of the nucleus compared to the lobes
varies.  The nucleus is the brightest 3.6-24~$\mu$m source,
but the east lobe in the central structure is brighter at
70-850~$\mu$m.  Aside from this major variation in the colors, some
other more subtle variations can be seen.  Two knots of infrared
emission to the west of the nucleus are stronger 24 and 70~$\mu$m
sources than the filamentary structure just east of the nucleus, but
this filament to the east of the nucleus is a brighter 450 and
850~$\mu$m source than the knots to the west of the nucleus.  Despite
these minor color variations, however, all wavebands effectively trace
the same structure.

Outside of the nucleus, all wavebands from 5.7 to 850~$\mu$m still
trace similar structures.  A series of knots loosely connected by some
filamentary dust structures can be seen throughout the disk of this
galaxy.  The appearance of the disk is notably asymmetric.  The
infrared emission clearly decreases more sharply to the east of the
nucleus than it does to the west.  The knots to the west of the
central structure all lie in a single line.  The dust structures to
the east, however, are more jumbled in appearance.  Some of the
emission above and below the major axis of the optical disk of the
galaxy may origniate from outside the plane of the galaxy or near the
edge of the stellar disk, but it is difficult to distinguish which
interpretation is correct from these data alone.

The interacting elliptical galaxy NGC~4627 can be seen to the
northwest of the center outside the plane of the galaxy.  The
infrared emission from this galaxy is mostly stellar; in these data,
it is brightest at 3.6~$\mu$m, and it generally decreases in
brightness towards longer wavelengths.  Faint emission is still
vaguely visible at 24~$\mu$m, but none is visible at
160~$\mu$m.  No significant dust structures connecting NGC~4627 and
NGC~4631 are visible.

\section{Color Variations across the Dust Disk \label{s_color}}

\subsection{Measured Flux Densities within Discrete Regions 
\label{ss_measuredfd}}

To study the infrared color variations within NGC~4631, we have
extracted flux densities from a series of circular regions within the
optical disk of the galaxy.  These regions were selected by eye
to include all the bright knots in the 24-70~$\mu$m images as well as
a few other prominent infrared structures.  Two different diameters
for the regions are used: $15^{\prime\prime}$ (to approximately match
the FWHM of the 70 and 850~$\mu$m PSFs) and $40^{\prime\prime}$ (to
match the FWHM of the 160~$\mu$m PSF).  At a distance of 9.0~Mpc
\citep{ketal03}, the $15^{\prime\prime}$ and $40^{\prime\prime}$
angular sizes of these regions correspond to diameters of 650 and
1700~pc.  In these comparisons, the data are first convolved with the
kernels described in Section~\ref{s_obs_kernel} so that all the image
resolutions match.

The regions in which the flux densities are measured are shown in
Figure~\ref{f_regionimages}.  Right ascensions, declinations, and flux
densities are listed in Tables~\ref{t_fd15arcsec} and
\ref{t_fd40arcsec}.  Uncertainties in the calibrations are listed as
percentages, and uncertainties from background noise (measured in
off-source regions of the sky) are listed in Jy.  Note that the
uncertainties from the background noise is negligible compared to the
calibration uncertainty except in the submillimeter bands.  Also note
that the uncertainties of the mean background levels (which are
subtracted from each waveband) correspond to uncertainties in the flux
densities that is negligible compared to the uncertainties from
background noise.  For comparing the ratios of flux density
measurements from two individual wavebands, only the background noise
is relevant.  In plots of flux density versus the ratio of flux
densities, changes in the calibration factors will have systematic
effects on individual wavebands that will only change the offsets in
the plots.  Since the goal is simply to identify whether the ratio
varies as a function of flux density, any offset introduced by a
calibration uncertainty does not change the scientific results.  For
composing SEDs in Section~\ref{s_sed}, however, the calibration
uncertainties will also be used, since changes in calibration factors
could change the shapes of the SEDs as well as the models fit to the
data and the physical interpretation of the results.

The IRAC measurements require the application of aperture corrections
to account for a the scattering of light through the detector material
\citep{retal05}, a problem which, among the data in this paper, is
unique to the IRAC data.  Note that this IRAC aperture correction is
not the same as an aperture correction that accounts for the fraction
of the PSF created by the optics of the telescope that falls outside a
measurement region (which does not need to be applied if the data
being compared are convolved to matching resolutions). Aperture
corrections are not applied to the flux densities measured in the
$15^{\prime\prime}$ regions because the flux densities measured in
these regions are used for only studying color variations (so aperture
corrections would only offset the data points in the plots), because
no definitive aperture corrections have been published at the time of
this writing, and because preliminary work on IRAC aperture
corrections implies that the aperture corrections will be
approximately equal to or less than the calibration uncertainties
(T. Jarrett, private communication).  However, since the flux
densities measured in the $40^{\prime\prime}$ regions are used for
modeling the SEDs (where precise calibration is more important), the
``infinite'' aperture corrections of \citet{retal05} have been applied
to the flux densities measured in the $40^{\prime\prime}$ regions.
The flux densities listed in Table~\ref{t_fd40arcsec} include these
corrections.

To use the 8~$\mu$m band as a tracer of PAH emission, we need to
subtract the stellar continuum emission.  To do this, we will apply
the formula
\begin{equation}
f_\nu(PAH 8\mu m) = f_\nu(8\mu m (total)) - 0.232 f_\nu(3.6\mu m)
\label{e_pahsub}
\end{equation}
from \citet{hraetal04} to the measurements in
Tables~\ref{t_fd15arcsec} and \ref{t_fd40arcsec}. We use the term
``PAH 8~$\mu$m'' to indicate where the continuum has been subtracted.
Note that this continuum subtraction only removes the contribution of
starlight.  In regions with low PAH emission, a significant fraction
of the non-stellar emission may originate from hot dust \citep[see][,
for example]{cetal06}.  Therefore, the actual variations PAH emission
may be stronger than what is represented by the continuum-subtracted
``PAH 8~$\mu$m''.

\subsection{Comparison between PAH and Hot Dust Emission 
    \label{s_color_smallgrain}}

First we will investigate how 8~$\mu$m PAH emission and 24~$\mu$m hot
dust emission are interrelated.

Figure~\ref{f_comp_8_24} shows how PAH 8~$\mu$m and 24~$\mu$m flux
densities measured within the $15^{\prime\prime}$ (650~pc) regions in
Table~\ref{t_fd40arcsec} and the $40^{\prime\prime}$ (1700~pc) regions
in Table~\ref{t_fd40arcsec} are related to each other.  The best
fitting functions are
\begin{equation}
  log \left( \frac{f_{PAH 8\mu m}}{f_{24\mu m}} \right) = (-0.03\pm0.04) +
  (0.032\pm0.019)log(f_{24\mu m}/Jy) ~ ~ ~ (15^{\prime\prime} \mbox{regions})
  \end{equation}
and
  \begin{equation}
  log \left( \frac{f_{PAH 8\mu m}}{f_{24\mu m}} \right) = (-0.13\pm0.03) +
  (0.08\pm0.02)log(f_{24\mu m}/Jy) ~ ~ ~ (40^{\prime\prime} \mbox{regions}).
  \end{equation}
Note that the correspondence between the two wavebands is very good
when the flux densities are measured within $40^{\prime\prime}$, but
the relation begins to break down in $15^{\prime\prime}$.  In both
aperture sizes, the (PAH 8)/24~$\mu$m flux density ratio does not vary
significantly with the 24~$\mu$m flux density.  However, the standard
deviation of the (PAH 8)/24~$\mu$m flux density ratio is a factor of 2
higher in the $15^{\prime\prime}$ data compared to the
$40^{\prime\prime}$ data, partly because more regions with higher (PAH
8)/24~$\mu$m flux density ratios are probed by the $15^{\prime\prime}$
regions.  The map of the ratio of PAH 8~$\mu$m to 24~$\mu$m surface
brightness in Figure~\ref{f_ratio_8_24} shows exactly why the data
exhibit this scatter.  The map reveals a significant amount of
variation in the ratio within the disk of the galaxy on sub-kiloparsec
scales.  24~$\mu$m emission appears to peak more strongly than PAH
emission in the centers of the dusty knots of star formation in this
galaxy.  The PAHs may be destroyed in the centers of these knots,
whereas the hot dust emission rises because the dust is heated more
strongly by the intense radiation fields in these knots.  This
phenomenon had also been reported for NGC~300 in \citet{hraetal04},
and further {\it Spitzer} observations may reveal that this phenomenon
is common among nearby galaxies.

To further examine the difference between the distribution of PAH and
hot dust emission, we extracted the PAH 8~$\mu$m and 24~$\mu$m radial
profiles of multiple individual regions: the center of the galaxy, the
lobes at either end of the central structure, and the seven brightest
sources outside the center.  These radial profiles are shown in
Figure~\ref{f_radialprofiles}.  As can be seen in these figures, the
PAH 8~$\mu$m emission is clearly broader than the 24~$\mu$m emission.
The PAH 8~$\mu$m data used to create these radial profiles have been
convolved to match the PSFs of the 24~$\mu$m data, but in several of
the regions considered here, the PAH 8~$\mu$m radial profiles
extracted from unconvolved data are still broader than the 24~$\mu$m
radial profiles.

In summary, these data demonstrate that, on kiloparsec scales, PAH and
hot dust emission are correlated, but on smaller scales, the two can
be spatially separated from each other.  The breakdown of the
correlation on small scales can be physically interpreted in two ways.
One possibility is that the primary source of heating for the dust
that produces the 24~$\mu$m emission may be ultraviolet light that is
absorbed close to the centers of the star formation regions, but the
PAHs can be excited by visible light \citep{ld02} that has a greater
mean free path.  The PAH emission would therefore appear to be more
extended than the 24~$\mu$m emission.  The other possibility is that
PAHs may be destroyed by the hard radiation in the centers of star
formation regions \citep[e.g.][]{metal06}, which is why the PAH 8~$\mu$m
emission does not appear to be as peaked as the 24~$\mu$m emission.
Either of these mechanisms would be an adequate explanation for the
effects observed here, and both mechanisms may be involved in creating
the observed color variations.

\subsection{Comparison of PAH and Hot Dust Emission to Cool 
Dust Emission \label{s_color_smalllargegrain}}

Next, we will investigate how the PAH 8~$\mu$m emission and 24~$\mu$m
hot dust emission are related to cooler dust emission at 70 and
160~$\mu$m.

Figure~\ref{f_comp_24_70_160} shows comparisons of hot dust emission
at 24~$\mu$m to the Wien side of the cool dust emission at 70~$\mu$m
and the Rayleigh-Jeans side of the peak of the cool dust emission at
160~$\mu$m.  Both plots show a general trend where the 24/70~$\mu$m
and 24/160~$\mu$m flux density ratios increase with surface
brightness.  The functions that best fit these data are
  \begin{equation}
  log \left( \frac{f_{24\mu m}}{f_{70\mu m}} \right) = (-1.007\pm0.012) +
  (0.237\pm0.014)log(f_{70\mu m}/Jy)
  \end{equation}
and
  \begin{equation}
  log \left( \frac{f_{24\mu m}}{f_{160\mu m}} \right) = (-1.70\pm0.02) +
  (0.24\pm0.03)log(f_{160\mu m}/Jy).
  \end{equation}
Because of the edge-on nature of this galaxy, it is unclear as to
whether this color variation is linked to variations in the dust
heating with the intrinsic source luminosity (as anticipated, for
example, by the models of \citep{dhcsk01} and \citep{ld01}) or if the
variation simply demonstrates that locations with significant hot dust
emission are associated with large column densities of cool dust.
However, if the color variation is linked to the cool dust column
densities, then the amount of dust integrated over the line of sight
is not physically linked to the hot dust, so some significant scatter
should be visibile in the relations between surface brightness and
color.  The absence of such scatter in Figure~\ref{f_comp_24_70_160}
implies that the color variations are more closely related to dust
heating, although column effects may still play a significant role.

Note the two data points in the plot of the 24/70~$\mu$m flux density
ratio versus 70~$\mu$m flux density that deviate from the general
trend.  The high surface brightness deviant data point represents the
nucleus, and the other deviant data point represents the point-like
source at right ascension 12:42:21.4 and declination +32:33:05 located
above the major axis of the optical disk of the galaxy to the east of
the center.  These regions may deviate from the trend because the star
formation in these regions is strong enough yet compact enough that
the infrared colors deviate significantly from the rest of the disk,
that something other than star formation (possibly an active galactic
nucleus in the center, for example) is heating the dust, or that the
surface brightness of stars is high enough that the stellar emission
in the 24~$\mu$m band is non-negligible.  Also note that the
extraplanar region may be a foreground/background source not
associated with NGC~4631.

Figure~\ref{f_comp_8_160_850} shows comparisons of PAH 8~$\mu$m
emission to the large grain emission at 160~$\mu$m.  For comparison
with \citet{hkb02}, we also include a comparison of PAH 8~$\mu$m
emission to 850~$\mu$m emission, although the 850~$\mu$m wave band may
not trace the same dust as the 160~$\mu$m wave band (see
Section~\ref{s_color_largegrain}).  These data show that the (PAH
8)/160~$\mu$m and (PAH 8)/850~$\mu$m flux density ratios increase with
surface brightness, which means that the PAH emission is higher in
systems with higher surface brightnesses.  The functions that best fits
these data are
  \begin{equation}
  log \left( \frac{f_{PAH 8\mu m}}{f_{160\mu m}} \right) = (-1.97\pm0.02) +
  (0.34\pm0.04)log(f_{160\mu m}/Jy).
  \end{equation}
  \begin{equation}
  log \left( \frac{f_{PAH 8\mu m}}{f_{850\mu m}} \right) = (0.82\pm0.14) +
  (0.54\pm0.08)log(f_{850\mu m}/Jy).
  \end{equation}
These trends clearly show that the PAH emission does not have a
one-to-one correspondence with the 160 or 850~$\mu$m emission in this
galaxy, which contradicts the results of \citet{hkb02}.  Note that
Haas et al. observed variations in the PAH 8/850~$\mu$m ratio of less
than a factor of 2, whereas the ratios presented here vary by more
than a factor of 10.  At least on kiloparsec scales within this galaxy,
PAH emission is more closely related to hot dust emission than to cool
dust emission.

The variations in the ratio of PAH emission to large grain emission
could arise from either luminosity effects (where infrared-luminous
regions produce relatively more PAH emission) or radial effects (where
fewer PAHs are found at larger radii).  To investigate this further,
we compared plots of the total infrared luminosity \citep[calculated
using the MIPS 24, 70 and 160~$\mu$m data and equation 4 in][]{dh02},
the 24/160~$\mu$m flux density ratio, and the (PAH 8)/160~$\mu$m flux
density ratio as a function of radius for all regions listed in
Table~\ref{t_fd40arcsec} that lie along the major axis of optical disk
of the galaxy.  These radial plots are shown in
Figure~\ref{f_radialcolor}.  The plots show that the 24/160~$\mu$m and
(PAH 8)/160~$\mu$m flux density ratios both peak in the center, which
is the most infrared-luminous region in the galaxy.  Outside the
center, however, a correspondence can be found between local peaks in
the 24/160~$\mu$m flux density ratio and local peaks in the total
infrared emission, which indicates that variations in the
24/160~$\mu$m flux density ratio are linked to infrared surface
brightness.  In contrast, the (PAH 8)/160~$\mu$m flux density ratio
monotonically decreases with radius.  No local peaks in the (PAH
8)/160~$\mu$m flux density ratio corresponding to local peaks in the
total infrared emission are found.  The variation in the (PAH
8)/160~$\mu$m flux density ratio is therefore a radial effect.

At first, it appeared that the variations in the (PAH 8)/160~$\mu$m
flux density ratio may be related to metallicity since metallicity
also generally varies with radius.  Variations in the strength of PAH
emission with metallicity have been found previously in observations
of the integrated SEDs of galaxies \citep{tsm99, detal05, eetal05,
gmjwb05}.  However, \citet{ogr02} find variations in the nitrogen
abundance in this galaxy that are smaller than the variations in the
(PAH 8)/160~$\mu$m flux density ratio found here, though the examined
region does not span the entire optical disk of the galaxy.  Moreover,
based on the data of \citet{eetal05}, the variation between PAH and
dust emission is expected to look like a step function, not a smooth
gradient, although the edge-on orientation could be responsible for
smoothing out the emission so that this step function disappears.
Therefore, we are hesitant to indicate that the radial variations in
the (PAH 8)/160~$\mu$m flux density ratio are connected to
metallicity.  The cause of the radial variations needs to be examined
further.

\subsection{Variations in Large Grain Emission \label{s_color_largegrain}}

Color variations in the large grain emission from 70-850~$\mu$m show
how the color temperatures of the large grains vary with surface brightness
and may also reveal hints of where the coolest dust emission
is located within the galaxy.

Figure~\ref{f_comp_70_160_450} shows how the 70~$\mu$m, 160~$\mu$m,
and 450~$\mu$m emission are interrelated.  The functions that best fit
these data are
  \begin{equation}
  log \left( \frac{f_{70\mu m}}{f_{160\mu m}} \right) = (-0.444\pm0.013) +
  (0.010\pm0.016)log(f_{160\mu m}/Jy)
  \end{equation}
and
  \begin{equation}
  log \left( \frac{f_{160\mu m}}{f_{450\mu m}} \right) = (0.96\pm0.03) +
  (-0.06\pm0.05)log(f_{450\mu m}/Jy).
  \end{equation}
Functionally, this is an indicator of the heating of the large dust
grains.  Virtually no variation in the 70/160~$\mu$m or 160/450~$\mu$m
colors is apparent, even though the flux densities vary by more than
an order of magnitude.  At first, this appears to contradict the
results of dust models such as the semi-empirical dust models of
\citet{dhcsk01} and the physical dust models of \citet{ld01}, which
anticipate that these colors should vary dramatically with changes in
the radiation fields or surface brightnesses of the regions.  However,
the absence of variations in the 70/160~$\mu$m and 160/450~$\mu$m flux
density ratios could indicate that the interstellar radiation field
that heats the large grains does not significantly vary throughout the
disk of this galaxy.  The cause of the variations in surface
brightness would therefore be related to variations in the mass of the
dust producing the infrared radiation.  However, since this galaxy is
viewed edge-on, the color variations in the disk may be averaged out
along the line of sight, although we may still expect some differences
between infrared-bright and infrared-faint regions.

For comparison, ISO observations of targets such as M~51, M~101
\citep{hetal96}, NGC~6946 \citep{tetal96}, and the disk of M~31
\citep{hetal98} show that far infrared colors do not vary with surface
brightness.  However, some {\it Spitzer} observations of nearby
galaxies such as M~81 (P\'erez-Gonz\'alez et al. 2006, in
preparation), M~51, NGC~7331 \citep{detal05}, and NGC~55
\citep{eetal04} do show variations in the 70/160~$\mu$m color with
surface brightness.  In light of these results, the invariance of the
70/160~$\mu$m colors found here might be unusual.  More observations
of far-infrared color variations within other galaxies are needed to
interpret the galaxy-to-galaxy differences.

Figure~\ref{f_comp_70_160_450_850} shows how 850~$\mu$m emission
varies with the large grain emission at 70~$\mu$m, 160~$\mu$m, and
450~$\mu$m.  The functions that best fit these data in logarithm space
are given by
  \begin{equation}
  log \left( \frac{f_{70\mu m}}{f_{850\mu m}} \right) = (1.49\pm0.11) +
  (0.21\pm0.07)log(f_{850\mu m}/Jy),
  \end{equation}
  \begin{equation}
  log \left( \frac{f_{160\mu m}}{f_{850\mu m}} \right) = (2.14\pm0.14) +
  (0.47\pm0.14)log(f_{850\mu m}/Jy),
  \end{equation}
  and
  \begin{equation}
  log \left( \frac{f_{450\mu m}}{f_{850\mu m}} \right) = (1.06\pm0.08) +
  (0.37\pm0.07)log(f_{850\mu m}/Jy).
  \end{equation}
If the 160~$\mu$m, 450~$\mu$m, and 850~$\mu$m emission sample the
Rayleigh-Jeans side of $\sim25$~K dust emission and if the emissivity
law does not vary, then the 160/850~$\mu$m and
450/850~$\mu$m flux density ratios should remain constant.
Furthermore, since the 70~$\mu$m, 160~$\mu$m, and 450~$\mu$m surface
brightnesses are well correlated, the dust heating should be
uniform, and the ratio of 850~$\mu$m emission to emission at shorter
wavelengths should not change significantly.  Despite these
expectations, strong variations in 850~$\mu$m relative to
70-450~$\mu$m emission is visible in these data.  As the surface
brightness of the regions decreases, the 850~$\mu$m emission increases
relative to the 70-450~$\mu$m emission.  These trends suggest that the
850~$\mu$m emission cannot also be described as originating entirely
from dust with a temperature of $\sim25$~K and an emissivity of
$\lambda^{-2}$.

\subsection{Conclusions on Color Variations}

In summary, this comparison of flux densities measured in different
wavebands has yielded interesting results.  The PAH 8~$\mu$m and
24~$\mu$m hot dust emission are found to be correlated on kiloparsec
scales, but the correlation breaks down on the sub-kiloparsec level.
The 24~$\mu$m hot dust emission is found to increase relative to the
70 and 160~$\mu$m cool dust emission as the far-infrared surface
brightness increases, as had been anticipated based on previous dust
models.  The PAH 8~$\mu$m emission is also found to increase relative
to the 160 and 850~$\mu$m wave bands as the far-infrared surface
brightness increases, partly because of variations with radius.  The
70/160~$\mu$m and 160/450~$\mu$m flux density ratios are invariant
relative to infrared surface brightness, which suggests that the
radiation field that heats the cool grains is relatively invariant.
However, the 850~$\mu$m emission decreases relative to 70-450~$\mu$m
emission as the far-infrared surface brightness increases, which
suggests that the 850~$\mu$m emission may originate partially from a
source other than $\sim25$~K dust with a $\lambda^{-2}$ emissivity.

\section{Spectral Energy Distributions \label{s_sed}}

\subsection{Global Spectral Energy Distribution}

To examine the global SED, we will use the global flux density
measurements published by \citet{detal05} with some adjustments.  The
3.6 - 8.0~$\mu$m IRAC measurements did not include aperture
corrections to account for the scattering of light through the
detector material; we have applied the ``infinite'' aperture
correction of \citet{retal05} to those data.  The 450 and 850~$\mu$m
flux density measurements were repeated for this paper.  The SCUBA
observations cover almost all of the infrared-bright regions in the
plane of the galaxy as is revealed when comparing the SCUBA images to
the 24~$\mu$m images, so a direct measurement of the flux densities
from the images within the optical disk (a $15.5^\prime \times
2.7^\prime$ ellipse) defined in the Third Reference Catalogue of
Bright Galaxies \citep{ddcbpf91} should be accurate.  The
uncertainties related to missing extended emission will probably be
less than the the submillimeter calibration
uncertainties.\footnote[12]{Note that \citet{detal05} applies
correction factors of 1.27 and 1.17 to the 450 and 850~$\mu$m
observations, respectively.  The correction factor is calculated from
the fraction of the integrated flux density in the 70 and 160~$\mu$m
bands that falls outside the region covered by the 450 and 850~$\mu$m
images.  In the case of NGC~4631, the 70 and 160~$\mu$m emission
falling outside the region observed in the 450 and 850~$\mu$m regions
is dominated by extended emission above and below the plane of the
galaxy that may represent the outer regions of the PSF from the
central structure and other bright sources in the plane of the galaxy.
The regions not covered in the 450 and 850~$\mu$m maps above and below
the plane probably do not contain any true sources that contributes
significantly to the integrated flux density of the galaxy.  The 450
and 850~$\mu$m PSF is effectively truncated in the north-south
direction because the observations include chopping to the sky outside
the plane of the galaxy, so emission from the wings of the PSF would
not be present.  Therefore, the 450 and 850~$\mu$m flux densities have
probably been overcorrected in \citet{detal05}. We apply no correction
to the data for the analysis here.}

Additionally, we will combine these results with JHK data from the
2MASS Large Galaxy Atlas \citep{jccsh03}; 12, 60, and 100~$\mu$m
integrated flux densities from IRAS \citep{rlsnkldh88}; and the
1.23~mm measurements from \citet{dkw04} to fill out the spectral
energy distribution from 1 to 2000~$\mu$m.  (The 25~$\mu$m flux
density from \citet{rlsnkldh88}, the 870~$\mu$m flux density from
\citet{dkw04}, and the 450 and 850~$\mu$m measurements from
\citet{sag05} are discussed in the context of checks on the 24, 450,
and 850~$\mu$m measurements presented here rather than included as
redundant flux density measurements; see Section~\ref{s_obs}.)
Additionally, CO, free-free, and synchrotron emission (calculated in
\citet{dkw04}) have been subtracted from the 850 and 1230~$\mu$m flux
densities (although the correction is less than 10\%).

The results are presented in Table~\ref{t_globalsed} and
Figure~\ref{f_globalsed}.  This global SED is typical for nearby
spiral galaxies.  Stellar blackbody radiation is seen shortward of
4.5~$\mu$m, PAH and hot dust emission are seen from 5.7 to
24~$\mu$m, and large cool grain emission is seen longward of
70~$\mu$m.  The large cool grain emission has the highest flux density
of any source within the 1-1000~$\mu$m range, as is typical for normal
nearby galaxies.

The slope of the 160-1230~$\mu$m emission is shallower than what is
expected from blackbody emission modified by a $\lambda^{-2}$
emissivity law for a single temperature component of dust (discussed
in more detail in Section~\ref{s_sed_global_bb}).  This is consistent
with the results found by \citet{betal03} and \citet{dkw04} although
the latter only found excess emission at 1230~$\mu$m.  CO, free-free,
and synchrotron emission have already been subtracted from the 850 and
1230~$\mu$m flux densities, so none of these other emission mechanisms
can account for the observed excess emission.  Either a dust component
in the 5-15~K range must be present or the dust that dominates the 850
- 1230~$\mu$m regime must have an emissivity that does not vary as
$\lambda^{-2}$.  We examine the possibilities further in the following
section.  

As a test of the models, we are searching for models that give
gas-to-dust ratios similar to the ratio of $\sim165$ for the Milky Way,
which was calculated by examining the depletion of elements in the gas
phase of the ISM \citep{l05}.  Models with significantly higher or
lower dust mass predictions are probably implausible.  This test
assumes that the ISM in NGC~4631 is similar to the Milky Way's.
However, the metallicity of NGC~4631 is $\sim0.5$ solar metallicity
\citep{ogr02}.  If the gas-to-dust mass ratio varies with metallicity,
then the expected ratio for NGC~4631 may be higher than the Milky Way's
ratio.

\subsubsection{Simple Modified Blackbody Fitting \label{s_sed_global_bb}}

The simplest way to describe the large grain emission longward of
70~$\mu$m is to fit the data with one or two blackbodies modified with
an emissivity function that scales as a function of $\lambda^{-\beta}$,
where $\beta$ is the emissivity index.  This approach is useful just
for approximating the temperature and emissivity of the bulk of the
far-infrared dust emission within the galaxy, which can then be used
to estimate the mass of the dust.

The first fit is with a blackbody modified by a $\lambda^{-2}$
emissivity law.  This emissivity law is what is suggested by simple
physical models for far-infrared absorption by small particles
\citep[e.g.][]{d04}.  Figure~\ref{f_globalsed_bb}a shows this fit
applied to only the 70 - 450~$\mu$m data, which are the data best fit
by this single thermal compoment.  The component has a temperature of
$23 \pm 2$~K.  Excess emission at 850 and 1230~$\mu$m is clearly
visible in the figure.  Note that if a single thermal component is fit
to the 70 - 1230~$\mu$m data, the function significantly overestimates the
160 and 450~$\mu$m flux densities while still underestimating the
1230~$\mu$m flux densities, which is what is expected if the emission
at 850 and 1230~$\mu$m is in excess above the thermal emission expected when
extrapolating from shorter wavelengths.

The mass of this dust can be calculated using
\begin{equation}
M_{dust} = \frac{D^2 f_{\nu}}{\kappa_{\nu} B_\nu(T)}
\label{e_dustmass}
\end{equation}
where $D$ is the distance to the galaxy, $f_{\nu}$ is the flux
density, $\kappa_{\nu}$ represents the absorption opacity of the dust
(given by \citet{ld01}), and $B_\nu(T)$ is the blackbody function,
with $T$ the best-fit dust temperature.  Using the 450~$\mu$m data, we
estimate the dust mass to be $4.9 \times 10^7$~M$_\odot$.  For
comparison, the total gas mass in this galaxy is $1.4 \times
10^{10}$~M$_\odot$ \citep{ketal03}.  The gas-to-dust ratio implied by
this simple modified blackbody fit is 290, which is within a factor of
2 of the Milky Way's ratio. The mass of the hot dust is not included
in this calculation, but the additional mass from hot dust should
bring the gas-to-dust ratio closer to the Milky Way's ratio.

The results show that, to account for the emission at 850 and
1230~$\mu$m, either a second thermal component with a cooler
temperature is needed or the emissivity law must be shallower than
$\lambda^{-2}$.  If the source of the excess 850 and 1230~$\mu$m
emission is a new dust component that produces negligible emission at
shorter wavelengths, then this new component should not be
significantly more massive than the 23~K dust (i.e. the new dust
component should not increase the dust mass by a factor of 10 or
more).  The 23~K dust alone can account for at least half of the dust
mass expected for this galaxy.

The next fit is with a single blackbody modified by an emissivity law
in which the emissivity index is a variable fit to the data.  This
fit, applied to the 70 - 1230~$\mu$m data, is shown in
Figure~\ref{f_globalsed_bb}b.  The fit has a temperature of $28 \pm
2$~K and an emissivity index of $1.2 \pm 0.1$.  Note that this
accounts for the excess seen at 850 and 1230~$\mu$m.  However, the
slope between the 160 and 450~$\mu$m data is only marginally
consistent with such an emissivity function.  Since the dust
emissivity index is treated as variable, the absorption opacities may
deviate away from the \citet{ld01} values by an uncertain amount.
However, if the Li \& Draine absorption opacities are still applicable
in this fit, the corresponding dust mass is $3.3 \times
10^7$~M$_\odot$ and the gas-to-dust ratio is 420.  This is close to
the value derived when the emissivity index was fixed to -2, although
the result with the fixed emissivity are slightly closer to the ratio
for the Milky Way.

Another possibility is that the dust should be represented with two
blackbodies, each of which is modified by $\lambda^{-2}$ emissivity
law.  To perform this fit properly on dust at temperatures near the
temperature of the cosmic microwave background (CMB), we
cannot use $f_\nu \propto \lambda^{-2} B_\nu(T_{dust})$ because it may
give temperatures colder than T$_{CMB}$.  Instead, we start with an
equation that represents the thermal emission from both the CMB and
dust in the galaxy
\begin{equation}
I_\nu=B_\nu(T_{CMB})e^{-\tau}+B_\nu(T_{dust})(1-e^{-\tau}),
\end{equation}
where $\tau$ is the optical depth of the dust.  When the background is
subtracted, the equation is rewritten as
\begin{equation}
I_\nu(sub)=B_\nu(T_{CMB})e^{-\tau}+B_\nu(T_{dust})(1-e^{-\tau})-B_\nu(T_{CMB})
\end{equation}
\begin{equation}
\approx \tau(B_\nu(T_{dust})-B_\nu(T_{CMB})) ~ ~ ~ \tau \ll 1
\end{equation}
Since $\tau$ is proportional to $\kappa_{\nu}$ (which we are assuming to
be proportional to $\lambda^{-2}$), we can fit
\begin{equation}
f_\nu \propto \lambda^{-2}(B_\nu(T_{dust})-B_\nu(T_{CMB}))
\end{equation}
to the data.  Figure~\ref{f_globalsed_bb}c shows the sum of the two
thermal components with this modification applied to the 70 -
1230~$\mu$m data.  The two thermal components have temperatures of $23
\pm 2$ and $3.5^{+1.4}_{-0.8}$~K.  Even though the two components are
fit simultaneously to the 70 - 1230~$\mu$m data, the results show that
the warmer component represents the majority of the 70 - 450~$\mu$m
emission and that the colder component only contributes a significant
amount of emission at 850 and 1230~$\mu$m.  This very cold dust
emission is physically implausible because of the implied dust masses.
If the dust emissivities in \citet{ld01} apply to this cold dust
component, then the dust mass of the 3.5~K dust is enormous.  Based on
the fit, approximately half of the 850~$\mu$m emission comes from this
cold dust.  Inserting this result into equation~\ref{e_dustmass}, we
calculate the mass of the 3.5~K dust to be $3.1 \times
10^9$~M$_\odot$.  This implies a gas-to-dust ratio of $\sim5$, which,
when compared to the ratio of $\sim165$ for the Milky Way, is too low to be
believable.  Therefore, we conclude that 3.5~K dust with an emissivity
index of -2 cannot give rise to the excess emission observed at 850 -
1230~$\mu$m.

All together, these results show that, using simple blackbodies
modified by emissivities that scale with wavelength, the excess
emission at 850 - 1230~$\mu$m can be represented in two ways.  The
scenario with 3.5~K dust, however, must be rejected because of the
implausibly high dust masses needed to produce the emission.  Further
analysis on the SEDs of smaller regions within the plane of the galaxy
will allow us to further examine whether the variable emissivity index
scenario is an appropriate descriptions for the 70 - 850~$\mu$m
emission.

\subsubsection{Semi-Empirical Model Fitting}

Another option for modeling the dust in this galaxy is to apply the
semi-empirical model of \citet{dhcsk01} to the data.  This model is
built using a series of PAH, very small grain, and large grain SEDs.
These SEDs are illuminated by a range of radiation fields with
relative contributions described by the equation
\begin{equation}
dM_{dust} \propto U^{-\alpha} dU
\label{e_dustscale}
\end{equation}
where U is the radiation field.  The two variables fit to the data are
$\alpha$ and a scaling term.  A low $\alpha$ indicates relatively
strong dust heating, whereas a high $\alpha$ indicates the
predominance of cooler dust.  This is presented as a more complex way
to model the dust emission than the simple modified blackbody fit.

The top part of Figure~\ref{f_globalsed_2modelfit} shows the
semi-empirical model from \citet{dhcsk01} that best fits the 5.7 -
850~$\mu$m data.  This model has an $\alpha$ of $2.31\pm 0.10$.  This
fit does not change significantly if the model is fit to only the 3.6
- 850~$\mu$m data or the 70 - 850~$\mu$m data.  Of interest here is
whether this model describes the dust emission longward of 20~$\mu$m.
The model accurately fits the 70 - 450~$\mu$m data, but it
underestimates the 850 - 1230~$\mu$m emission.  The large grain
emission in the \citet{dhcsk01} model is represented by a series of
blackbodies modified with an emissivity function that scales as
$\lambda^{-\beta}$, where $\beta$ is a function of the radiation field
\citep[see equation 2 in][]{dh02}.  The inability of this model to
reproduce these observed results demonstrates that the summation of a
range of such modified blackbodies cannot reproduce the excess 850 -
1230~$\mu$m emission seen here.  Either additional dust components or
modifications to the emissivity function are needed to reproduce the
excess submillimeter emission.

This model can be used to determine the dust mass of the galaxy using
the analysis in Section 5.2 of \citet{dh02}.  First, we calculate the
dust mass predicted for a blackbody modified by a emissivity law with
a variable index fit to the 70 - 450~$\mu$m data.  (We will not use
the 850 or 1230~$\mu$m data because the model does not adequately
describe the emission at those wavelengths.)  The best fitting
modified blackbody has a temperature of $23 \pm 2$~K and a $\beta$ of
$1.8 \pm 0.1$.  Note that this fit statistically differs little from
the blackbody modified by the $\lambda^{-2}$ emissivity function fit
to the 70 - 450~$\mu$m in Section~\ref{s_sed_global_bb}.  Next, we use
the 160~$\mu$m flux density and the absorption opacities from
\citet{ld01} in equation~\ref{e_dustmass} to calculate a dust mass,
which is found to be $5.8 \times 10^7$~M$_\odot$.  We then apply the
appropriate correction factor from Figure 6 in \citet{dh02}.  For a
model with $\alpha \cong 2.0 - 2.5$, the correction factor is $\sim9$.
This gives the total dust mass as $5.2 \times 10^8$~M$_\odot$, and the
gas-to-dust mass ratio as $\sim30$.  This is still within a factor of
five of the ratio of $\sim165$ for the Milky Way, but it is higher than
expected, especially in light of NGC~4631's metallicity. Nevertheless,
the dust emission shortward of 450~$\mu$m is able to account for all
of the expected dust mass.  This further implies that the
modifications that should be applied to this model to explain the
excess 850 and 1230~$\mu$m emission should not substaintially increase
the total dust mass.

\subsubsection{Physical Modeling}

As a third option, we have fit the physical model of \citet{ld01} as
updated by Draine \& Li (2006, in preparation) to the global SED.
This model is designed to reproduce the wavelength-dependent dust
extinction and emission within the Milky Way and the Magellanic
Clouds.  The model treats the dust as a mixture of PAHs, amorphous
silicate grains and carbonaceous grains with a size distribution
ranging from tens of atoms to greater than 1~$\mu$m.  A fraction
$\gamma$ of the dust mass is illuminated by a range of radiation
fields described by the power law in equation~\ref{e_dustscale} but
with the exponent fixed to -2 and the maximum fixed to $10^6$
multiplied by the local interstellar radiation field given by
\citep{mmp83}.  This can be thought of as the mass fraction of dust in
photodissociation regions.  The remainder of the dust mass is modeled
as a cool cirrus component heated by a single radiation field equal to
the minimum of the range of radiation fields described by the
power-law.  Five parameters are fit to the data: the minimum value for
the illuminating radiation fields, the fraction of the total mass in
PAH molecules that are smaller than $10^3$ carbon atoms, the starlight
continuum intensity in the IRAC bands, $\gamma$, and the overall dust
mass.  This model is designed to simultaneously represent the
starlight that dominates at 4.5~$\mu$m and shorter wavelengths, the
PAH emission in the the 5.7 and 8~$\mu$m wave bands, the 12 and
24~$\mu$m hot dust emission, and the 60-1230~$\mu$m cool cirrus
emission.

The bottom part of Figure~\ref{f_globalsed_2modelfit} shows the
physical model that best fits the 1.0-1230~$\mu$m data.  This physical
model is able to reproduce the observed 850~$\mu$m flux densities to
within the calibration uncertainty of 10\%.  Furthermore, it is able
to account for $\sim75$\% of the total 1230~$\mu$m emission.  In
contrast, the single blackbody modified with the $\lambda^{-2}$
emissivity and the semi-empirical model both could account for only
$\sim30$-60\% of the 850~$\mu$m and 1230~$\mu$m emission.
Unfortunately, the physical model does not accurately reproduce the
steep slope of the dust emission between 160 and 450~$\mu$m.  The
model overestimates the 450~$\mu$m flux density by $\sim40$\%,
although note that this is only $\sim1.5$ times larger than the 25\%
uncertainty in the 450~$\mu$m measurement.  In summary, it appears
that this physical model can fit the 160-1230~$\mu$m data points to
within $2.5\sigma$.  This could be interpreted as marginally
successful, although the nature of the mismatch between the model and
observations implies that the model underpredicts the 1230~$\mu$m emission or
overpredicts the 450~$\mu$m emission.

Unlike the single thermal component or the semi-empirical models, the
physical model is able to accurately describe the emission at 850 and
1230~$\mu$m because the colors and scale of the 160-1230~$\mu$m
emission can be adjusted independently of the hot dust emission at
shorter wavelengths.  In contrast, the single thermal component model
treats the emission between 70-450~$\mu$m as originating from a single
temperature when some of the emission at 70~$\mu$m may be from warmer
dust, and the semi-empirical models assume that the colors of all dust
emission between 5.7 and 1000~$\mu$m are interdependent.

The dust mass of the best-fitting model is $9.7 \times
10^7$~M$_\odot$.  This suggests a gas-to-dust mass ratio of $\sim140$,
which is very close to the ratio of $\sim165$ for the Milky Way.  The dust
mass is only a factor of $\sim2$ higher than the dust mass predicted
by the 23~K blackbody modified by the $\lambda^{-2}$ emissivity given
in Section~\ref{s_sed_global_bb}.  This is because the temperature of
most of the dust in this model is $\sim15$-18~K, which implies,
according to equation~\ref{e_dustmass}, a mass that is a factor of
$\sim2$ greater than that calculated using 23~K.  Note that this
factor is significantly lower than the correction factors of
$\sim10-15$ given for the far-infrared or submillimeter bands by
Figure 6 in \citet{dh02}.  Also note that the minimum temperature
predicted by this model is $\sim15$~K.  A very cold ($<$ 10~K) dust
component that constitutes approximately 50\% or more of the global
dust mass may not necessarily be needed to explain the 850 or
1230~$\mu$m emission.

Among the other parameters in the fit, the $\gamma$ term (the mass
fraction of dust in the photodissociation regions) yields the most
interesting information.  The $\gamma$ determined in the fit is 0.015
(or 1.5\%), which indicates that most of the mass of the dust in the
galaxy is heated by a weak baseline radiation field in diffuse cirrus
emission.  The model also predicts that the small fraction of dust in
photodissociation regions produces $\sim10$\% of the total dust
emission.  

The other parameters from the physical model fit are the PAH mass
fraction, which is 0.034, and the minimum value of the illuminating
radiation field, which is 2.0 times the local interstellar radiation
field.

\subsection{Spectral Energy Distributions in Discrete Regions}

The SEDs of discrete regions within this galaxy will reveal where the
excess 850~$\mu$m emission is present in the galaxy.  This will provide
additional clues on where the submillimeter excess arises from, which
can then be used to better understand the nature of the excess.  We
composed SEDs for individual regions throughout the disk using the data
for the regions in Table~\ref{t_fd40arcsec} but excluding locations where
either 450 or 850~$\mu$m data were not available.  Below, we will attempt
to describe the data with both simple and complex models of the
emission.

\subsubsection{Simple Modified Blackbody Fitting \label{s_sed_discrete_bb}}

First, we will fit the 70-450~$\mu$m data with a single blackbody
modified by a $\lambda^{-2}$ emissivity law.  Example plots of the
fits for the central structure, bright knots in the disk, and faint
regions in the disk are shown in Figure~\ref{f_discretesed_bbfit}.
The parameters for these fits and the difference between the measured
850~$\mu$m flux density and the fits (henceforth referred to as the
excess 850~$\mu$m emission) is presented in
Table~\ref{t_discretesed_bbfit}.  Additionally,
Figures~\ref{f_comp_tir_temp_bbfit} and
\ref{f_comp_tir_excess850_bbfit} show how the temperature and excess
850~$\mu$m emission from these fits vary with the total infrared
luminosity \citep[calculated using the MIPS 24, 70 and 160~$\mu$m data
and equation 4 in][]{dh02}.

These results show that the function fit to the 70 - 450~$\mu$m data
represents those data very well.  As seen in
Figure~\ref{f_comp_tir_temp_bbfit}, very little variation in the
temperature is present, as was expected from the invariance of the
70/160~$\mu$m and 160/450~$\mu$m colors in
Section~\ref{s_color_largegrain}.  Additionally, these results provide
us with two pieces of evidence that suggest that the excess 850~$\mu$m
emission cannot be explained simply by an emissivity function that
deviates from the $\lambda^{-2}$ law given by \citet{ld01}.  First are
the results from the fits to the 70 - 450~$\mu$m data that show that
dust with an emissivity index of -2 is adequate for describing the
dust emission.  Second is the elbow-like bend in the SEDs from 160 to
850~$\mu$m found in many of the SEDs.  This bend cannot be produced by
varying the emissivity index for the 70-1000~$\mu$m dust emission.  If
the emissivity index is variable, it only varies longward of
450~$\mu$m.  It is possible, however, that some constituent of the
interstellar medium other than large solid grains contributes to the
emission at 850~$\mu$m.

The variations in the excess 850~$\mu$m emission in
Figure~\ref{f_comp_tir_excess850_bbfit} show that the excess becomes
more prominent in the 850~$\mu$m emission from infrared-faint regions.
In the infrared-bright regime, the contribution of the excess to the
total emission is only $\sim20$\%.  This is close to the calibration
uncertainties from SCUBA as well as other similar submillimeter data.
The excess therefore becomes statistically indistinguishable from the
$\sim25$~K dust emission found within the waveband.  In infrared-faint
regions, however, the excess 850~$\mu$m emission can contribute
$\sim80$\% of the total 850~$\mu$m flux density.  

Note that the excess 850~$\mu$m emission observed here (and in
Sections~\ref{s_sed_discrete_semimodel} and
\ref{s_sed_discrete_physicalmodel}) as well as the variations in the
70/850, 160/850, and 450/850~$\mu$m flux density ratios is unlikely to
be produced by either a background offset or a calibration error.  If
the background were incompletely subtracted, the excess 850~$\mu$m
emission listed in Table~\ref{t_discretesed_bbfit} would be constant.
However, the excess 850~$\mu$m emission instead varies significantly,
with low excess values generally more likely to be found in
infrared-faint regions (although the trend between the 850~$\mu$m
excess emission and the infrared luminosity is statistically weak).  If
the calibration of the 850~$\mu$m emission was incorrect, then the
trend in Figure~\ref{f_comp_tir_excess850_bbfit} would be flat.
Moreover, since the 450 and 850~$\mu$m data were processed using the
same software, calibration standards, and background regions, any
offset in the 850~$\mu$m data should also manifest itself in the
450~$\mu$m data.  Since the 450~$\mu$m data are consistent with
expectations based on shorter wavelengths but the 850~$\mu$m data are
not, we conclude that the excess 850~$\mu$m emission is real and not
an artifact of the data processing.

\subsubsection{Semi-Empirical Model Fitting \label{s_sed_discrete_semimodel}}

Next, we have fit the 5.7 - 450~$\mu$m data with the semi-empirical
model of \citet{dhcsk01}.  Example plots of these fits are shown in
Figure~\ref{f_discretesed_semimodelfit}, and parameters from the fits
are shown in Table~\ref{t_discretesed_semimodelfit}.  We did attempt
fits that included the 850~$\mu$m data points, but the match to the
850~$\mu$m data did not improve significantly despite its inclusion,
and the fits failed to converge on appropriate $\alpha$ values in
some situations.  We therefore excluded the 850~$\mu$m data from the
fits.

As with the single blackbodies modified with the $\lambda^{-2}$
emissivity laws, these fits still reveal the presence of excess
850~$\mu$m emission.  Plots of how the excess 850~$\mu$m emission
derived from these model fits vary with total infrared luminosity are
shown in Figure~\ref{f_comp_tir_excess850_semimodelfit}.  Again, the
fraction of 850~$\mu$m flux density from the excess increases as the
total infrared luminosity decreases.  In terms of determining the
presence and magnitude of the excess 850~$\mu$m emission, few
differences exist between using these semi-empirical model and using
single blackbodies modified with $\lambda^{-2}$ emissivity laws.

The $\alpha$ parameter does show a correlation with total infrared
luminosity, as shown in Figure~\ref{f_comp_tir_alpha_semimodelfit}.
Given that the 70/160~$\mu$m flux density ratio does not vary
for these regions but the 24/160~$\mu$m does vary, the
increase in heating must be reflected primarily in changes in the
amount of hot dust emission.  Again, this is somewhat
consistent with the results of \citet{dhcsk01}, which showed that the
24~$\mu$m emission would vary most strongly with dust heating.  The
large grain emission responsible for the 70 and 160~$\mu$m emission,
however, must be relatively unaffected.

\subsubsection{Physical Modeling \label{s_sed_discrete_physicalmodel}}

Finally, we present the physical models that have been fit to the 3.6
- 850~$\mu$m data.  Example plots of these fits are shown in
Figure~\ref{f_discretesed_drainemodelfit}.  The parameters from the
fits are given in Table~\ref{t_discretesed_drainemodelfit_param} and
the 450 and 850~$\mu$m excess flux densities are given in
Table~\ref{t_discretesed_drainemodelfit}.  Unlike the single blackbody
fits and the semi-empirical model fits, these physical models perform
much better at describing the 850~$\mu$m data.  However, a significant
fraction of the 850~$\mu$m flux density (up to 30\%) can still be
identified as in excess above the model predictions.  Moreover, as
seen in Figure~\ref{f_comp_tir_excess850_drainemodelfit}, this ratio
of the excess to total 850~$\mu$m flux density still varies with the
infrared luminosity.  Table~\ref{t_discretesed_drainemodelfit} also
shows that the model overestimates the 450~$\mu$m flux density in the
same infrared-faint regions where it underestimates the 850~$\mu$m
emission.  (The overestimates are expressed as negative 450~$\mu$m
excess emission.)  The overestimates of the 450~$\mu$m emission may
reach up to $\sim80$\% in some locations.  Nonetheless, these
mismatches only reach a maximum of $\sim3\sigma$ significance within
individual regions.  So, we conclude that these physical models
perform marginally well in explaining the 450 and 850~$\mu$m emission
but that excess 850~$\mu$m emission may still be present.

The PAH mass fraction determined for the $40^{\prime\prime}$
subregions do show a variation with radius.  This is shown in
Figure~\ref{f_comp_dist_pahdraine}.  The monotonic decrease in the PAH
mass fraction with radius is similar to the variation with radius seen
in the (PAH 8)/160~$\mu$m flux density ratio in
Figure~\ref{f_radialcolor}.  Again, no local peaks in the PAH mass
fraction corresponding to local peaks in the total infrared surface
brightness are seen outside the center of the galaxy, at least on the
kiloparsec scales of the $40^{\prime\prime}$ regions.  This additional
analysis demonstrates that variations in PAH emission in this galaxy
are a function of distance from the center, not infrared surface
brightness.

Also of interest are the variations in $\gamma$ (the mass fraction of
dust in photodissociation regions) with radius.  This is shown in
Figure~\ref{f_comp_dist_gammadraine}.  Note that this parameter stays
at the level of 0.01 - 0.02 except in the central structure, where it
reaches up to $\sim0.03$.  This shows that the relative mass of dust
in photodissociation regions is twice as high as in the central
structure, although it is still a relatively small amount of the total
dust mass.

\subsection{Discussion}

In summary, the SED results show that the far-infrared~/~submillimeter
dust emission is somewhat unusual in two respects.  First is the
invariance in the color temperature of dust emission in the 70 -
450~$\mu$m range.  This result is unusual in comparison to other
recent {\it Spitzer} observations of galaxies.  However, this
invariance could be explained if the bulk of the dust emission
originates from dust outside of regions with strong heating by young
stars, which may be consistent with the relatively low mass fraction
of dust in photodissociation regions found when the physical model was
fit to the global SED.  The increase in infrared luminosity would then
primarily be tracing an increase in dust mass integrated along the
line of sight rather than an increase in dust heating.  Note that,
because this is an edge-on galaxy, increases in the integrated dust
mass can be achieved if the line of sight passes through either a
greater fraction of diffuse dust distributed evenly throughout the galaxy's
disk or massive individual dust clouds within the galaxy.

The other unusual result from these observations is the slope of the
dust emission between 160 and 1230~$\mu$m.  The single modified
blackbody fits and the semi-empirical model fits reveal the presence
of 850 and 1230~$\mu$m emission in excess of what is anticipated from
models of thermal dust emission.  The physical models describe the
data better, but the results from the fits imply that such an excess
at 850 and 1230~$\mu$m may be present.

The characteristics of the source of the excess emission - if it is
present - are complex.  First, if it originates from dust, then the
modifications that need to be applied to the models to explain the
excess emission should not substantially increase the total estimated
dust mass.  This is because both simple modified blackbodies and more
sophisticated models fit to the 5.7 - 450~$\mu$m emission give dust
masses that suggest dust-to-gas mass ratios similar to what is
measured in the Milky Way.  Second, if the excess emission represents
variability in the emissivity index of the dust emission, the
variation only occurs longward of 450~$\mu$m, as explained in
Section~\ref{s_sed_discrete_bb}.  Third, the 850~$\mu$m excess appears
to be found throughout the disk.  However, if the excess is present in
the brightest infrared regions, it is relatively weak compared to
20-25~K large grain dust emission at 850~$\mu$m.  Finally, the
850~$\mu$m excess also diminishes in strength in infrared-faint
regions, although it may still contribute 30-80\% of the 850~$\mu$m
emission.

\citet{ketal01}, \citet{de01}, and \citet{betal03} had reported a
degeneracy in trying to represent the far-infrared to submillimeter
dust emission either as one blackbody modified by an emissivity power
law with a variable index or as multiple blackbodies modified with
fixed $\lambda^{-2}$ emissivity laws.  The results here have shown
that the variable emissivity index scenario cannot accurately describe
the 70-850~$\mu$m emission of discrete regions within a galaxy, so the
approach should not be applied globally. The 850~$\mu$m excess is best
described as a component of the SED that is separate from the thermal
dust emission seen at 70-450~$\mu$m.  However, the 850~$\mu$m excess
should not be described as a very cold thermal dust component because
the implied dust masses are too high.

Some previous observations, such as \citet{rtbetal04}, may have missed
similar excess 850~$\mu$m emission from nearby galaxies because of the
nature of the observations.  Most global SEDs will be dominated by a
few bright regions.  If SEDs for subregions are extracted, the
subregions are usually of the brightest sources within the targets.
The results for this galaxy demonstrate that excess submillimeter and
millimeter emission is small compared to the $\sim20$~K large grain
emission and cannot be easily detected.  However, the excess becomes
easier to detect in fainter regions in the disk, as it may constitute
80\% of the emission in some regions.  Therefore, we suggest that the
search for excess submillimeter or millimeter emission in other
galaxies should not be performed with global or nuclear SEDs but
instead should be carried out in fainter regions with galaxies'
optical disks.

Through a process of elimination, only a couple of options are
available for describing the 850 and 1230~$\mu$m excess if it is
indeed present.  One possibility is that the excess emission
originates from dust grains with high submillimeter emissivities
relative to their absorption cross sections.  Such grains, such as the
fractal grains proposed by \citet{retal95} for the Milky Way and
\citet{dkw04} for NGC~4631, may be able to maintain relatively low
temperatures, so they would be only detectable at submillimeter
wavelengths.  This scenario, however, may be physically implausible
\citep{l05}.  Another possibility, also considered by \citet{retal95}
for the Milky Way, is the possibility that the emissivity of the
$\sim20$-25~K large grains is enhanced above what is expected by
\citet{ld01} by resonances in the dust grains caused by impurities.
Further physical modeling as well as more observations of
submillimeter excesses will be needed to better understand this
phenomenon.

\section{Conclusions \label{s_conclusions}}

The primary results from this research can be summarized as follows:

1. PAH emission at 8~$\mu$m is closely related to hot dust
   emission at 24~$\mu$m.  The relation holds on 1.7~kpc scales but
   begins to break down at 650~pc scales.  Furthermore, the PAH
   emission is not as centrally peaked as the hot dust
   emission on scales of hundreds of pc.

2. Variations in the strength of PAH emission relative to cool dust
   emission and in the fraction of dust mass in PAHs depend primarily
   on radius, not infrared surface brightness.

3. The 70 - 450~$\mu$m color temperature does not appear to vary with
   surface brightness in this galaxy, even though the
   24/70~$\mu$m and 24/160~$\mu$m flux density
   ratios do vary with surface brightness.  This implies that a substantial
   part of the dust emission in this wavelength regime may originate from
   a cool, diffuse cirrus component.

4. The 850 and 1230~$\mu$m emission in this galaxy is found to exceed
   what is anticipated from either the 23~K blackbody modified by a
   $\lambda^{-2}$ emissivity law or the semi-empirical model that
   describes the emission shortward of 850~$\mu$m.  The
   physical dust models marginally describe the data, but they
   leave open the possibility that excess emission at 850~$\mu$m is
   present.  If present, the 850~$\mu$m excess is highest in regions
   of moderate infrared brightness and constitutes the greatest
   fraction of the 850~$\mu$m emission in infrared-faint regions.

Two major implications for using PAH emission as a star formation
indicator arise from this research.  First, these results demonstrate
that the spatial correlation between PAH emission and 24~$\mu$m dust
emission breaks down on scales of hundreds of parsecs.  This implies
that PAH emission cannot be used as an accurate star formation tracer
on such scales (if 24~$\mu$m emission primarily traces star formation
regions).  Second, these results suggest that the ratio of PAH to dust
mass varies radially, although the reason for the radial variation
needs further study.  If PAH emission is to be used as a tracer of
dust or star formation, then these radial effects need to be taken
into account.  Further comparisons using 8, 24, and 160~$\mu$m data of
nearby galaxies are needed to understand how PAH emission relates to
dust emission in other galaxies and to understand the radial
variations in the PAH emission.

The color variations and SEDs found within NGC~4631 have some major
implications for dust modeling.  The absence of variations in the
70/160~$\mu$m and 160/450~$\mu$m flux density ratios in this galaxy
but the presence of variations in the 70/160~$\mu$m ratio in other
galaxiesneeds to be studied further so as to understand the conditions
that cause the ratios to vary or remain constant.  Second, dust models
may need to be adjusted to better account for the excess emission at
850~$\mu$m if it is present.  Further analysis with {\it Spitzer} 70
and 160~$\mu$m data as well as 850~$\mu$m data with comparable
resolution are needed not only to confirm the presence of the
850~$\mu$m excess emission in other galaxies but to provide better
descriptions of where the excess can be found.  Furthermore,
additional observations of nearby galaxies at multiple wavelengths
between 160 and 850~$\mu$m are needed so as to better constrain the
shape of the SED in this wavelength regime.

\acknowledgments 
Support for this work, part of the {\it Spitzer} Space Telescope
Legacy Science Program, was provided by NASA through contract 1224769
issued by the Jet Propulsion Laboratory, California Institute of
Technology under NASA contract 1407.  BTD was supported in part by NSF
grant AST-0406833.  AL is supported by the University of Missouri
Summer Research Fellowship, the University of Missouri Research Board,
and NASA/{\it Spitzer} Theory Programs.

Facilities: \facility{{\it Spitzer}(IRAC,MIPS)}, \facility{JCMT(SCUBA)}

\clearpage

\begin{figure}
\epsscale{0.65}
\plotone{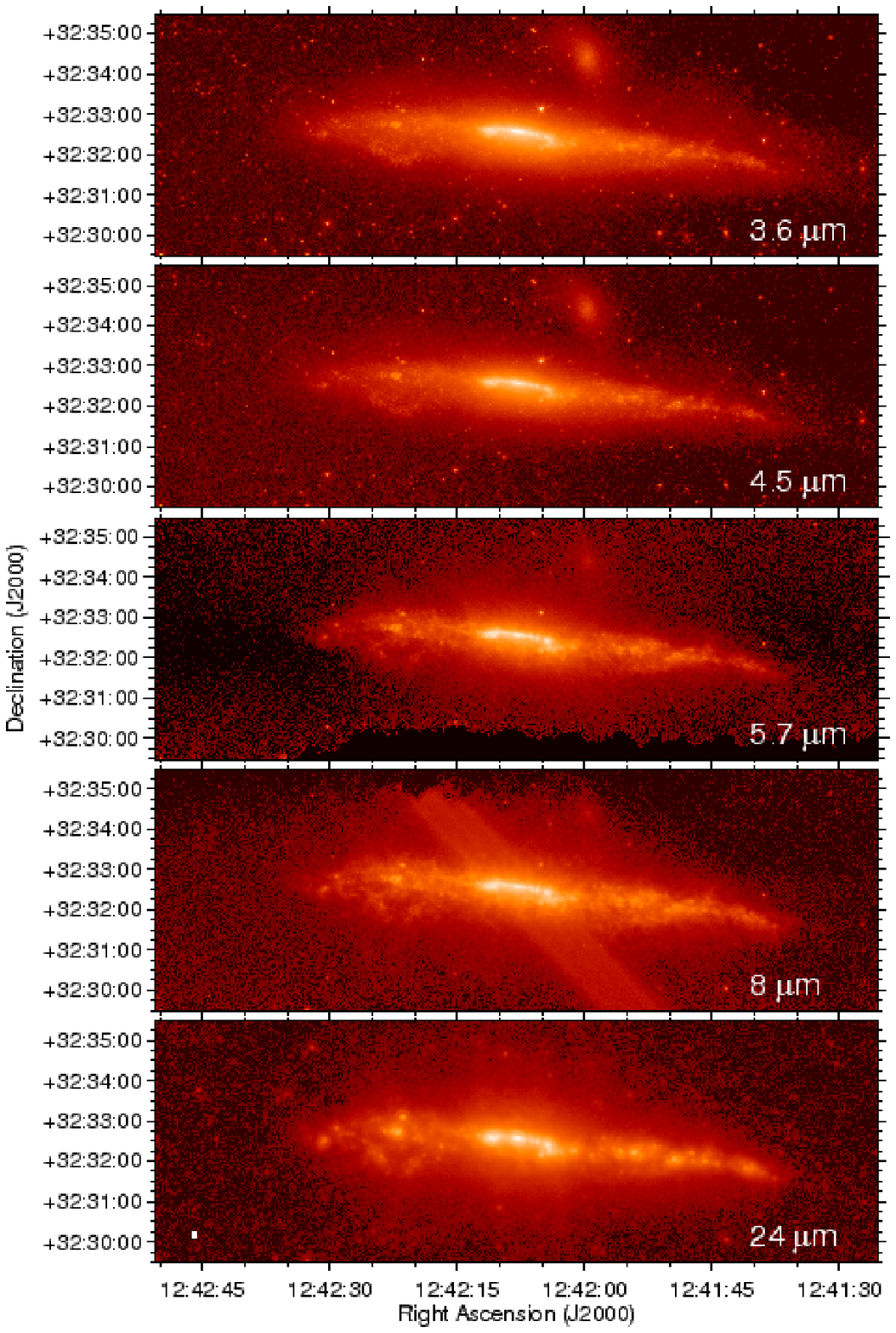}
\caption{Global images of NGC 4631 from 3.6 to 850~$\mu$m.  Each image is
$18^\prime \times 6^\prime$, with north up and east to the left.
Circles representing the FWHM of the PSF have been plotted in the
lower left corner of the 24 - 850~$\mu$m images.  In the 8~$\mu$m
image, a muxbleed effect is visible as streaking to the northeast and
southwest of the center of the galaxy.  Also note that if the stellar
continuum is subtracted from the 8~$\mu$m image using
equation~\ref{e_pahsub}, the resulting PAH emission map is
superficially similar to the above 8~$\mu$m image.  In the 70~$\mu$m
image, a latent image effect has created negative streaking to the
south of the center.  This latent image effect also results in an
artifact in the background subtraction that creates the positive
streaking to the north of the center.  The edges of the usable parts
of the 450 and 850~$\mu$m images is outlined in green.  For display
purposes only, the 450~$\mu$m image has been smoothed with an
$8^{\prime\prime}$ Gaussian, and the 850~$\mu$m image has been
smoothed with a $15^{\prime\prime}$ Gaussian.}
\label{f_globalimages}
\end{figure}

\clearpage

\addtocounter{figure}{-1}
\begin{figure}
\epsscale{0.65}
\plotone{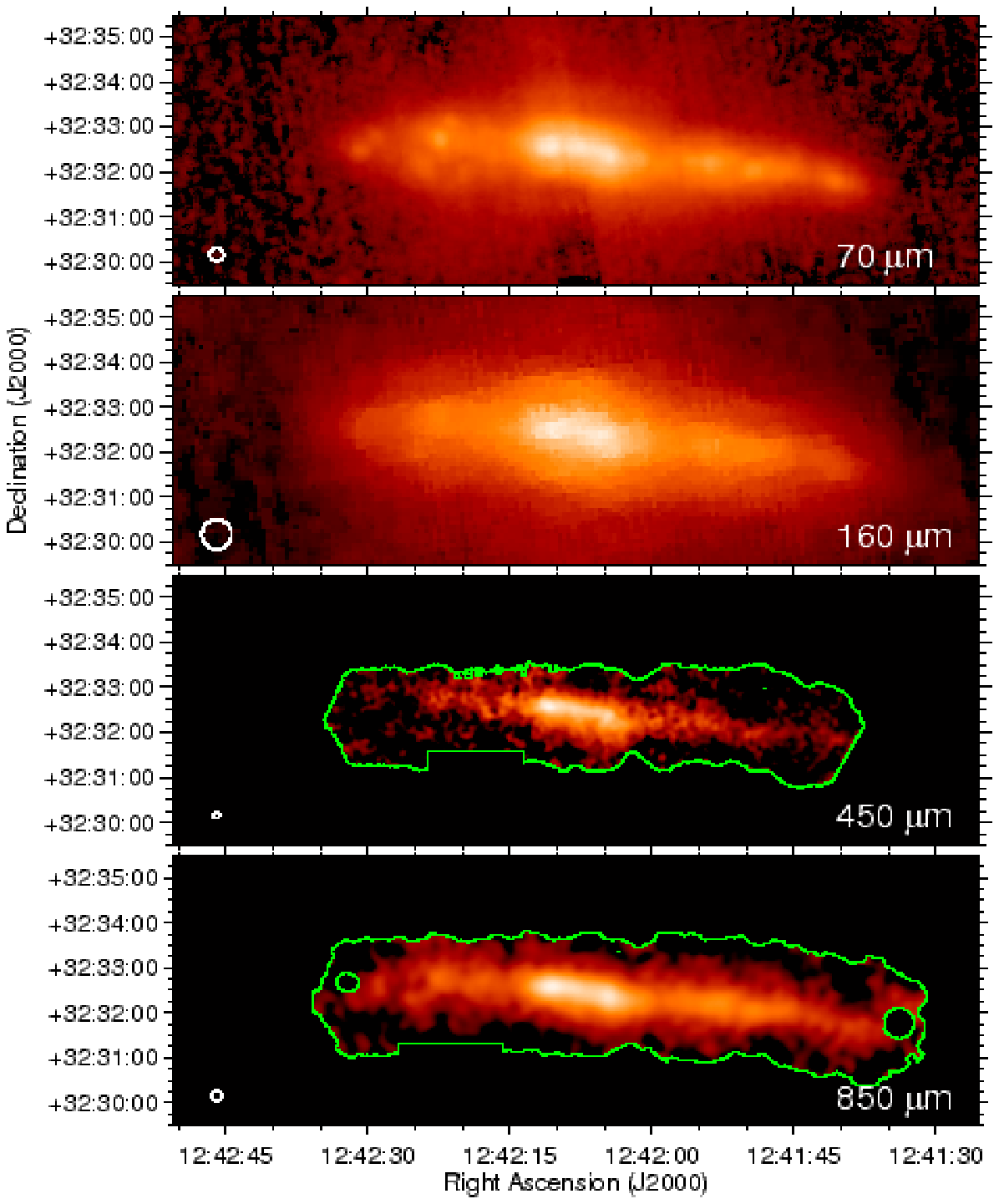}
\caption{Continued.}
\end{figure}

\clearpage

\begin{figure}
\epsscale{0.4}
\plotone{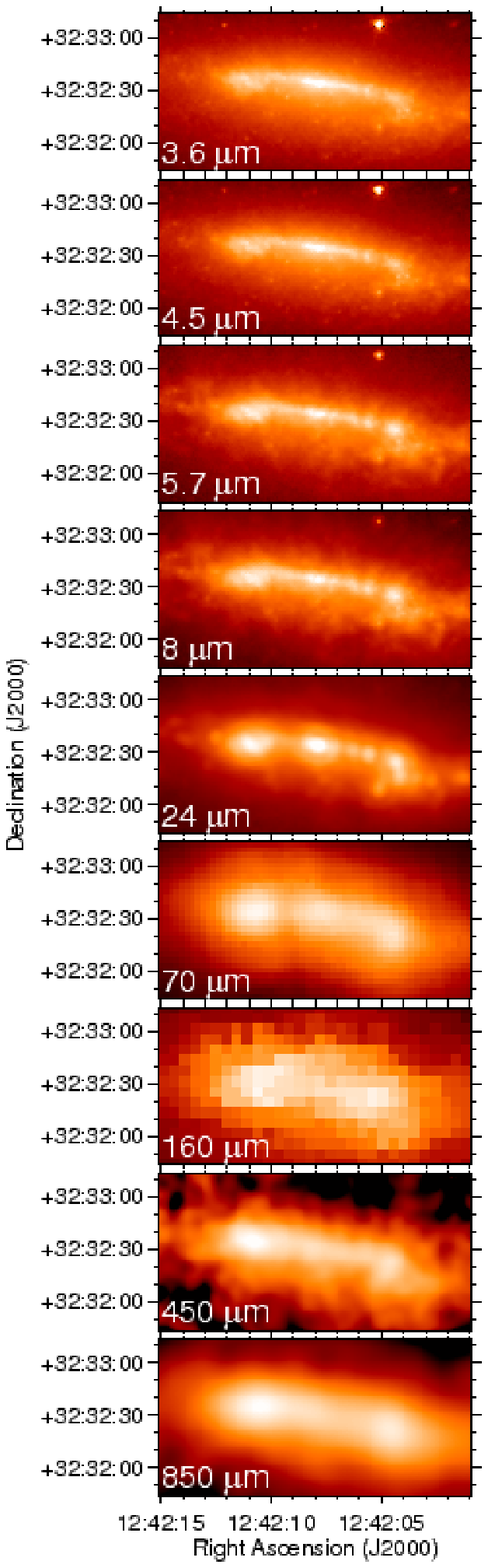}
\caption{Images of the center of NGC 4631 from 3.6 to 850~$\mu$m.  Each image
is $3^\prime \times 1^\prime.5$, with north up and east to the left.
For display purposes only, the 450~$\mu$m image has been smoothed with
an $8^{\prime\prime}$ Gaussian, and the 850~$\mu$m image has been
smoothed with a $15^{\prime\prime}$ Gaussian.  The edge of the usable
part of the 450~$mu$m observations is outlined in green.}
\label{f_centerimages}
\end{figure}

\clearpage

\begin{figure}
\epsscale{1.0}
\plotone{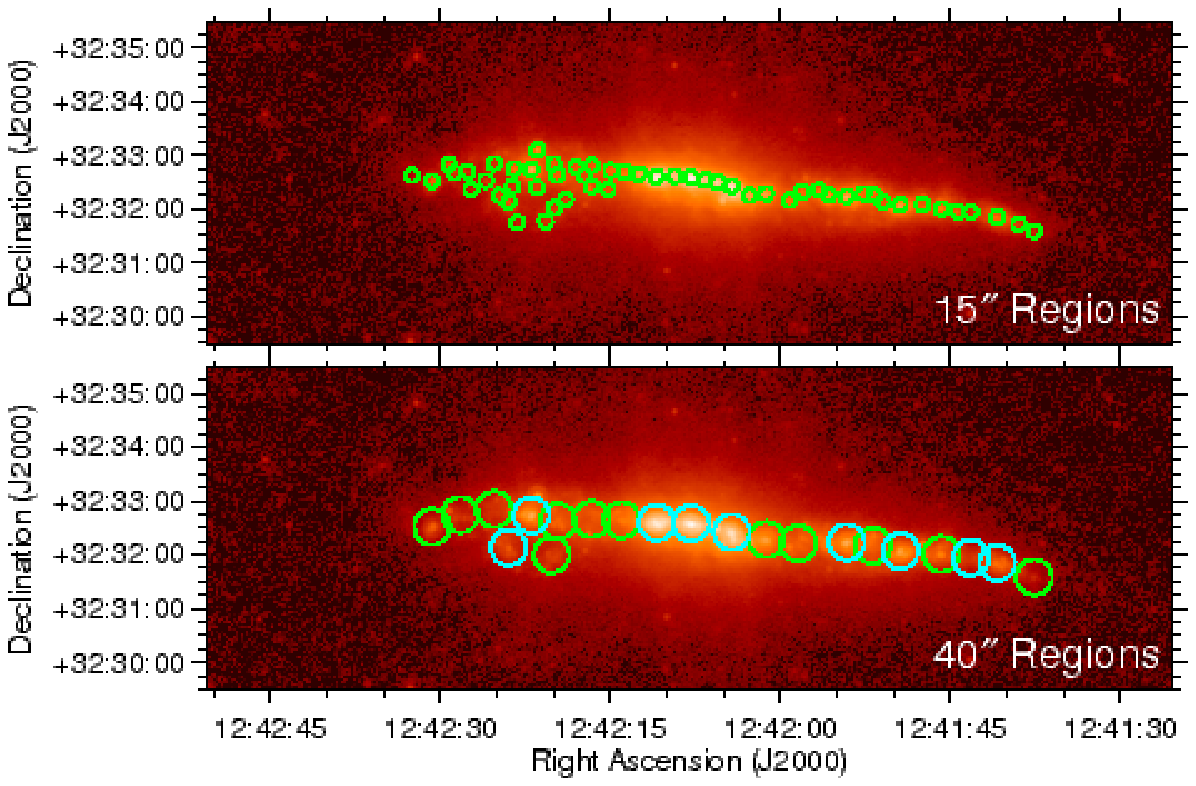}
\caption{Global images of NGC 4631 at 24~$\mu$m with the $15^{\prime\prime}$
and $40^{\prime\prime}$ extraction regions overlaid as circles on the
image.  The cyan circles mark the regions plotted in
Figures~\ref{f_discretesed_bbfit}, \ref{f_discretesed_semimodelfit},
and \ref{f_discretesed_drainemodelfit}. Each image is $18^\prime
\times 6^\prime$, with north up and east to the left.}
\label{f_regionimages}
\end{figure}

\clearpage

\begin{figure}
\plotone{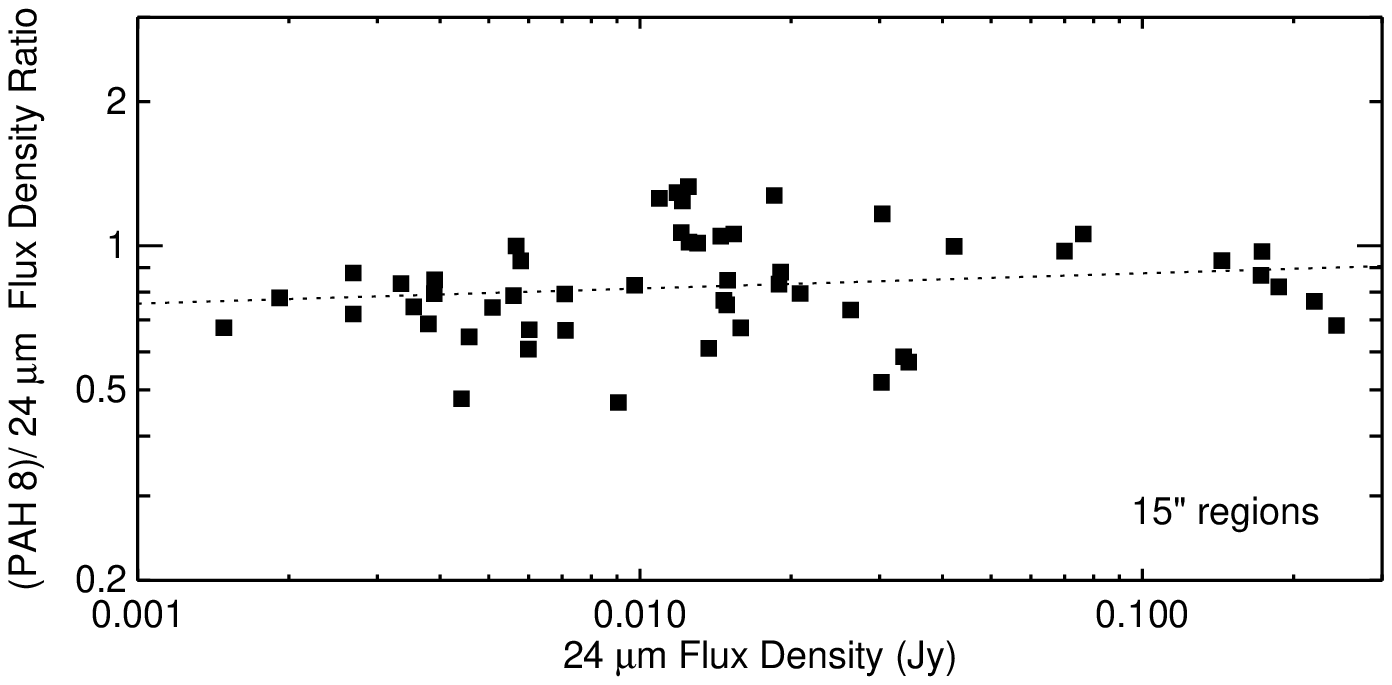}
\plotone{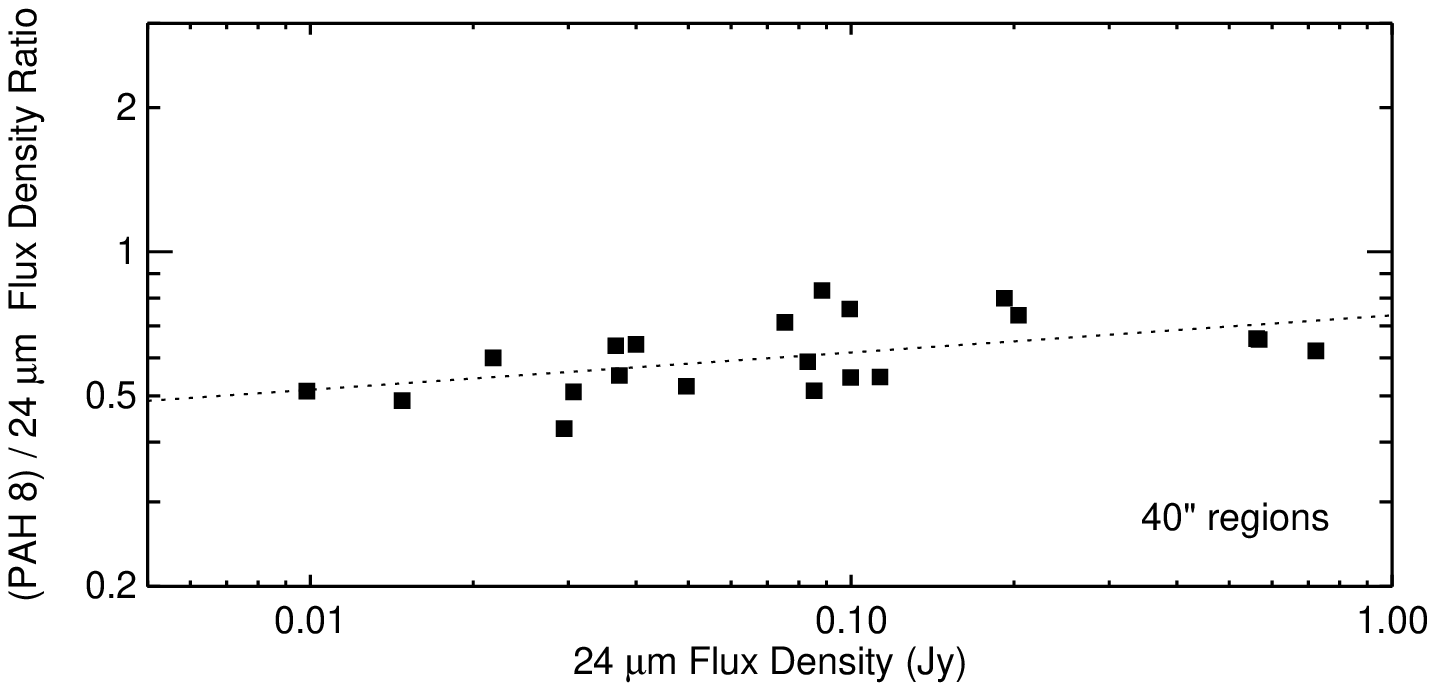}
\caption{A comparison of the PAH 8~$\mu$m and 24~$\mu$m flux densities
measured in the $15^{\prime\prime}$ (650~pc) regions listed in
Table~\ref{t_fd15arcsec} and the $40^{\prime\prime}$ (1.7~kpc) regions
listed in Table~\ref{t_fd40arcsec}.  Uncertainties from background
noise are smaller than the symbols.  The lines show the best fitting
relations between the (PAH 8)/24~$\mu$m flux density ratio and the
24~$\mu$m flux density.}
\label{f_comp_8_24}
\end{figure}

\clearpage

\begin{figure}
\plotone{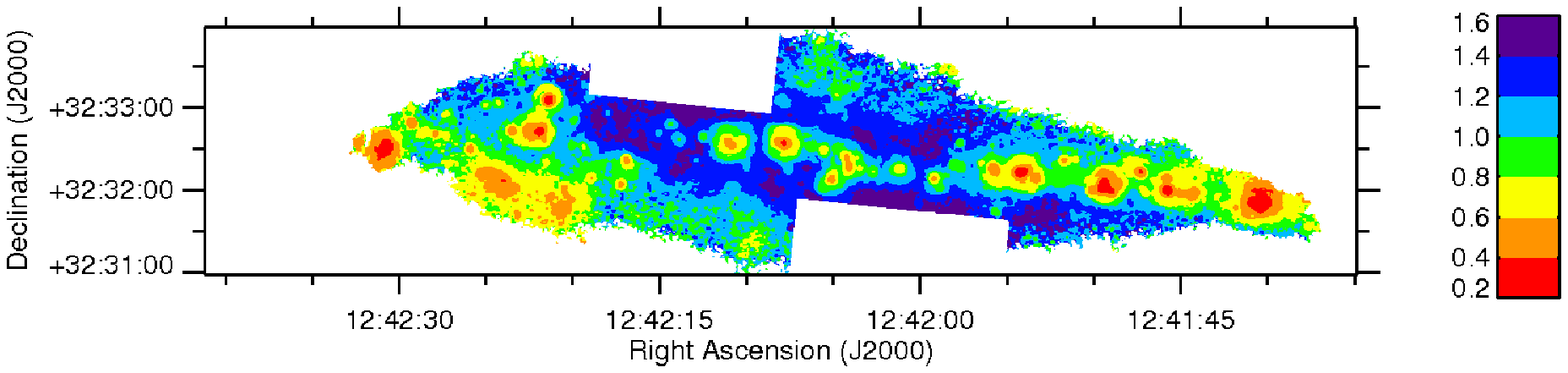}
\caption{Global image of the (PAH 8)/24~$\mu$m flux density ratio in
NGC~4631.  This image is $14^\prime \times 3^\prime$, with north up
and east to the left.  The PAH 8~$\mu$m data were convolved with a
kernel described in Section~\ref{s_obs_kernel} to match the resolution
of the 24~$\mu$m data.  Blue colors correspond to regions with
relatively strong PAH 8~$\mu$m emission; red colors correspond to
regions with relatively strong 24~$\mu$m emission.  Pixels with low
signal-to-noise ratios have been left blank, and the regions to the
northeast and southwest of the center that are heavily affected by
muxbleed in the 8~$\mu$m image have been masked out.  The color bar on
the right shows the correspondence between the shading and the
numerical values for the (PAH 8)/24~$\mu$m flux density ratio.}
\label{f_ratio_8_24}
\end{figure}

\clearpage

\begin{figure}
\epsscale{0.85}
\plotone{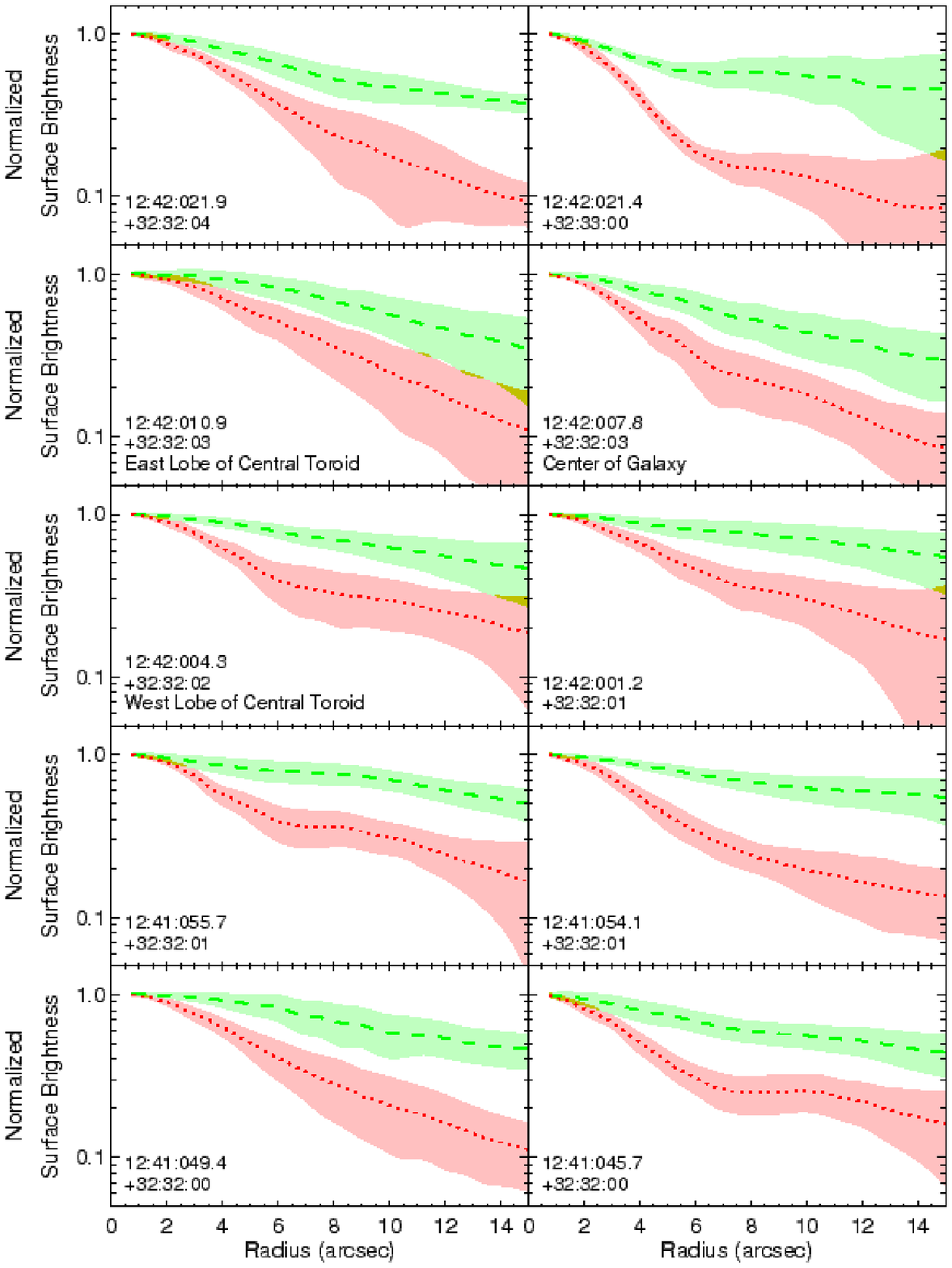}
\caption{PAH 8~$\mu$m surface brightness radial profiles (shown as green dashed
lines with uncertainties shown in light green) and 24~$\mu$m surface
brightness radial profiles (shown as red dotted lines with
uncertainties shown in pink) of ten regions in the disk of NGC~4631.
All PAH 8~$\mu$m data are convolved with the kernels described in
Section~\ref{s_obs_kernel} before radial profiles are extracted.  The
radial profiles are all normalized to 1 at the innermost radii.  The
region centers are identified by J2000 right ascension and declination
coordinates in the lower left corner of each plot.  Regions where
uncertainties overlap are shown in dark yellow.}
\label{f_radialprofiles}
\end{figure}

\clearpage

\begin{figure}
\plotone{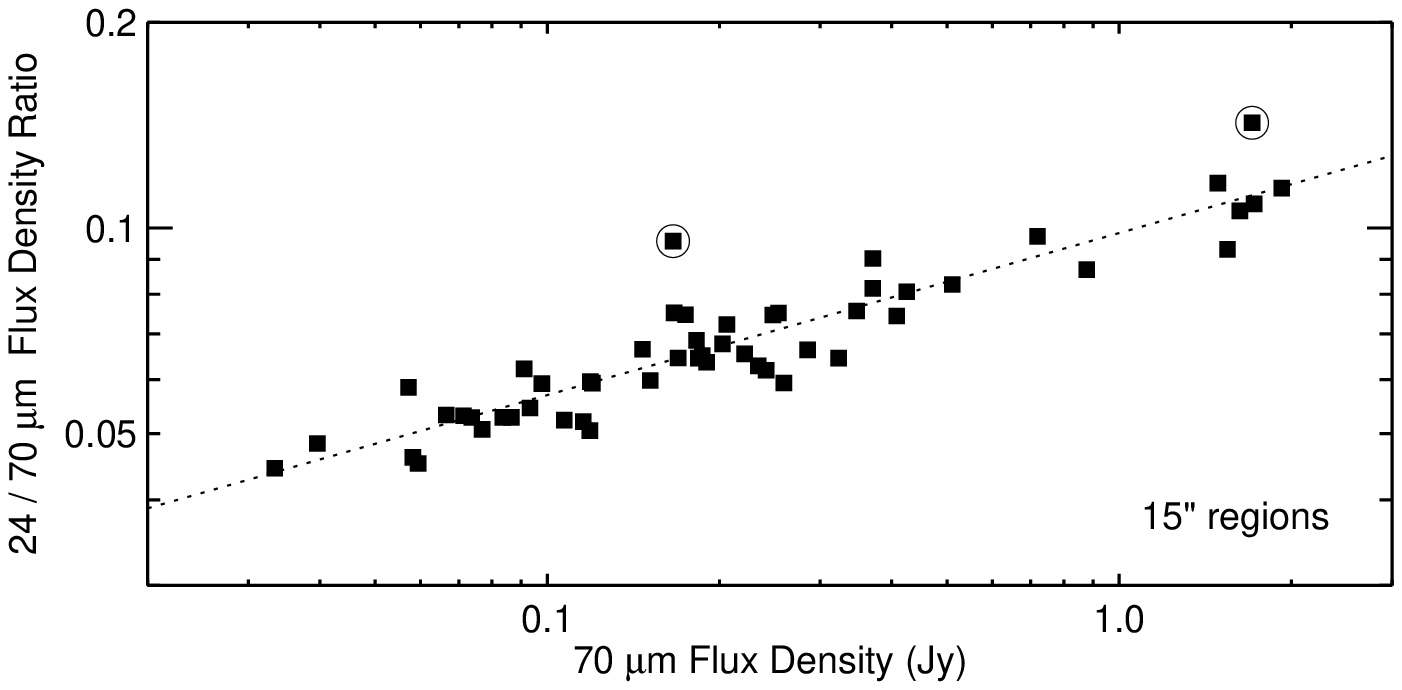}
\plotone{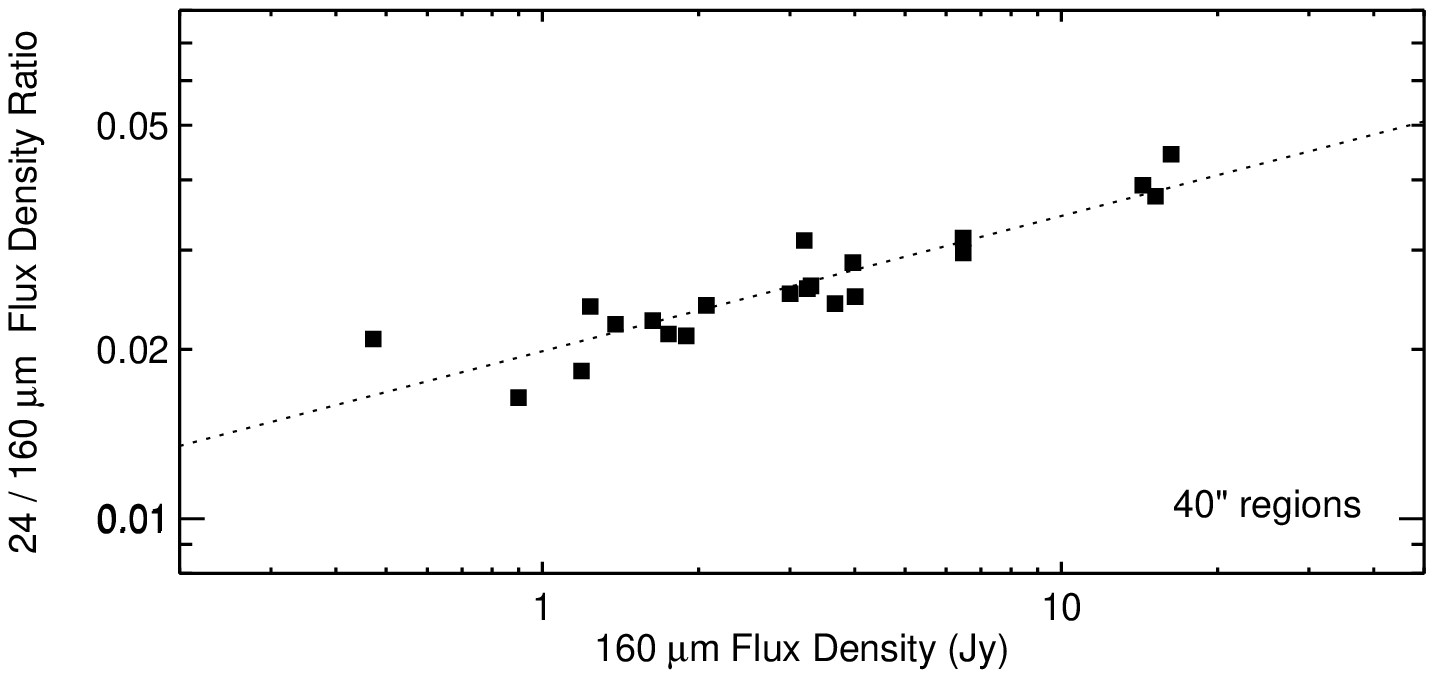}
\caption{A comparison of the 24 and 70~$\mu$m flux densities measured
in the $15^{\prime\prime}$ regions listed in Table~\ref{t_fd15arcsec},
and a comparison of the 24 and 160~$\mu$m flux densities measured in
the $40^{\prime\prime}$ regions listed in Table~\ref{t_fd40arcsec}.
Uncertainties from background noise are smaller than the symbols.  The
lines show the best fitting relations between the flux density ratios
on the y-axes and the flux densities on the x-axes.  The two data
points that deviate from the relation in the top plot are marked with
circles.  These data points are discussed in the text.}
\label{f_comp_24_70_160}
\end{figure}

\clearpage

\begin{figure}
\plotone{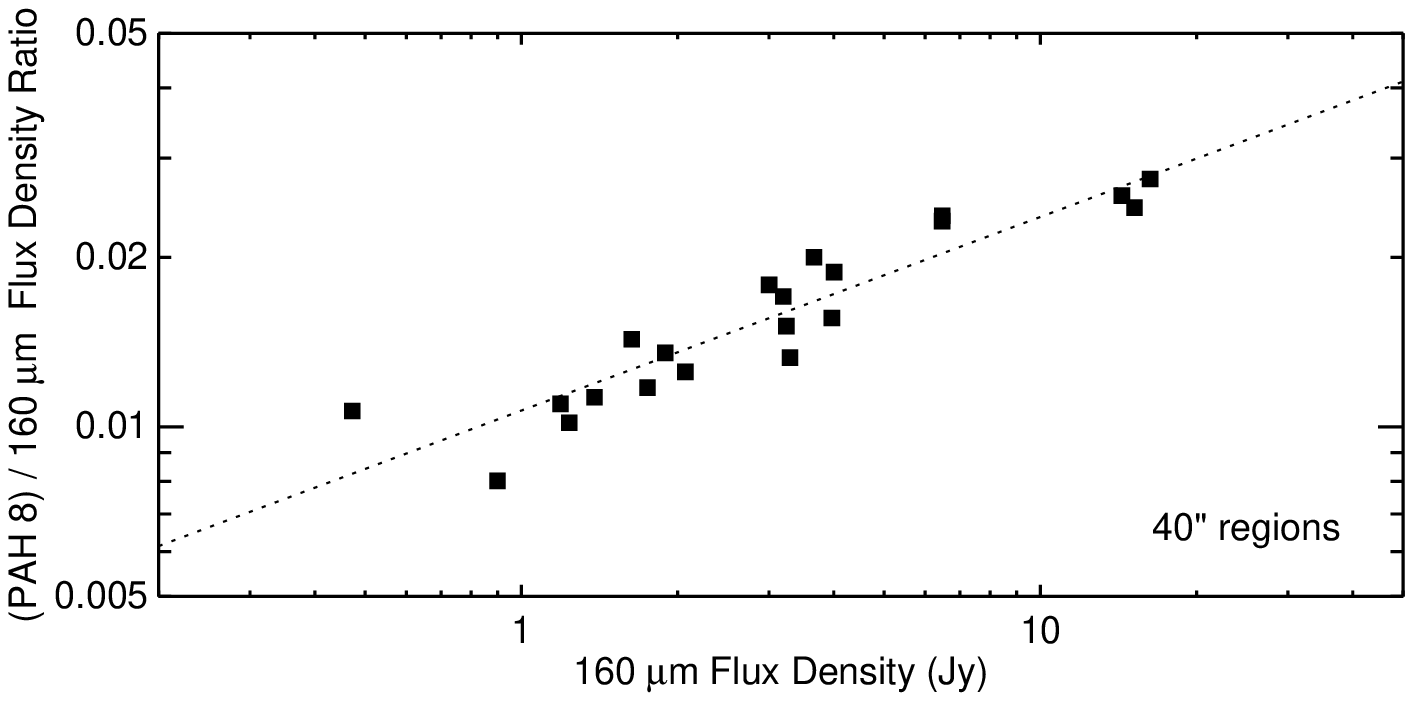}
\plotone{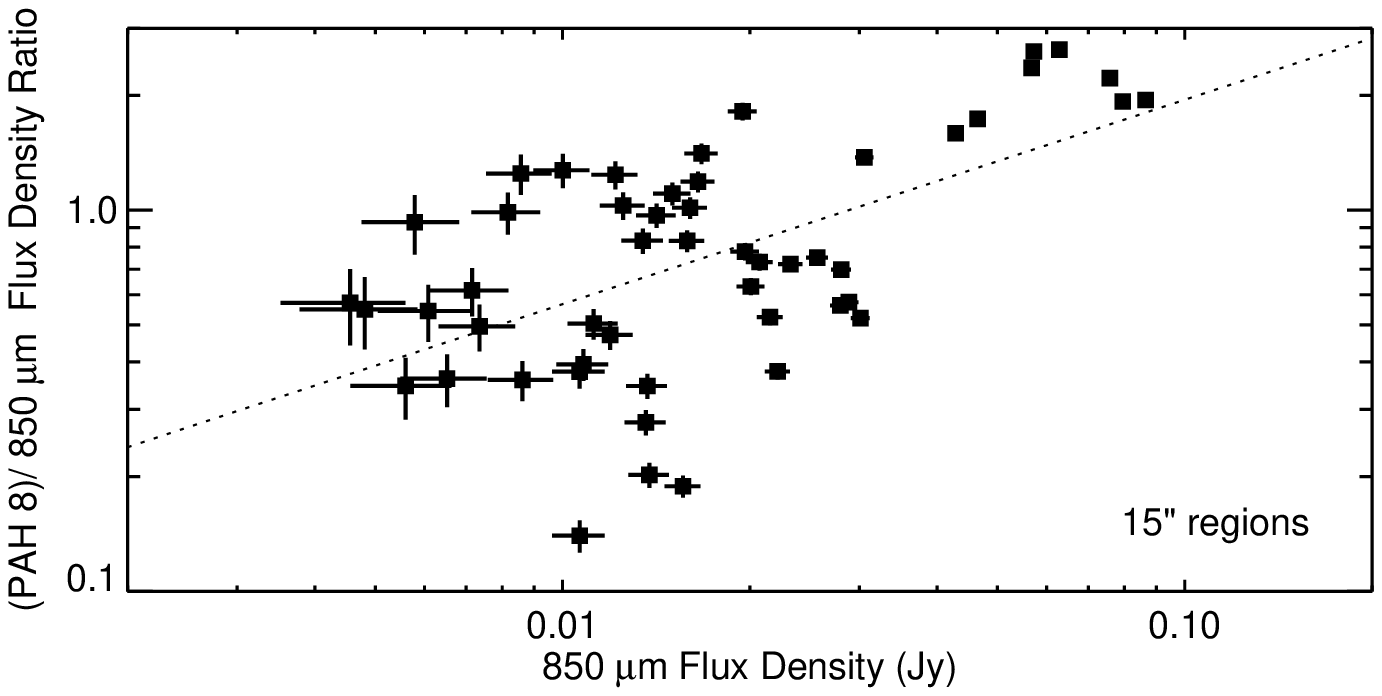}
\caption{A comparison of the PAH 8 and 160~$\mu$m flux densities
measured in the $40^{\prime\prime}$ regions listed in
Table~\ref{t_fd40arcsec}, and a comparison of the PAH 8 and 850~$\mu$m
flux densities measured in the $15^{\prime\prime}$ regions listed in
Table~\ref{t_fd15arcsec}.  Uncertainties from background noise are
smaller than the symbols except where error pars are shown.  The
lines show the best fitting relations between the flux density ratios
on the y-axes and the flux densities on the x-axes.}
\label{f_comp_8_160_850}
\end{figure}

\clearpage

\begin{figure}
\plotone{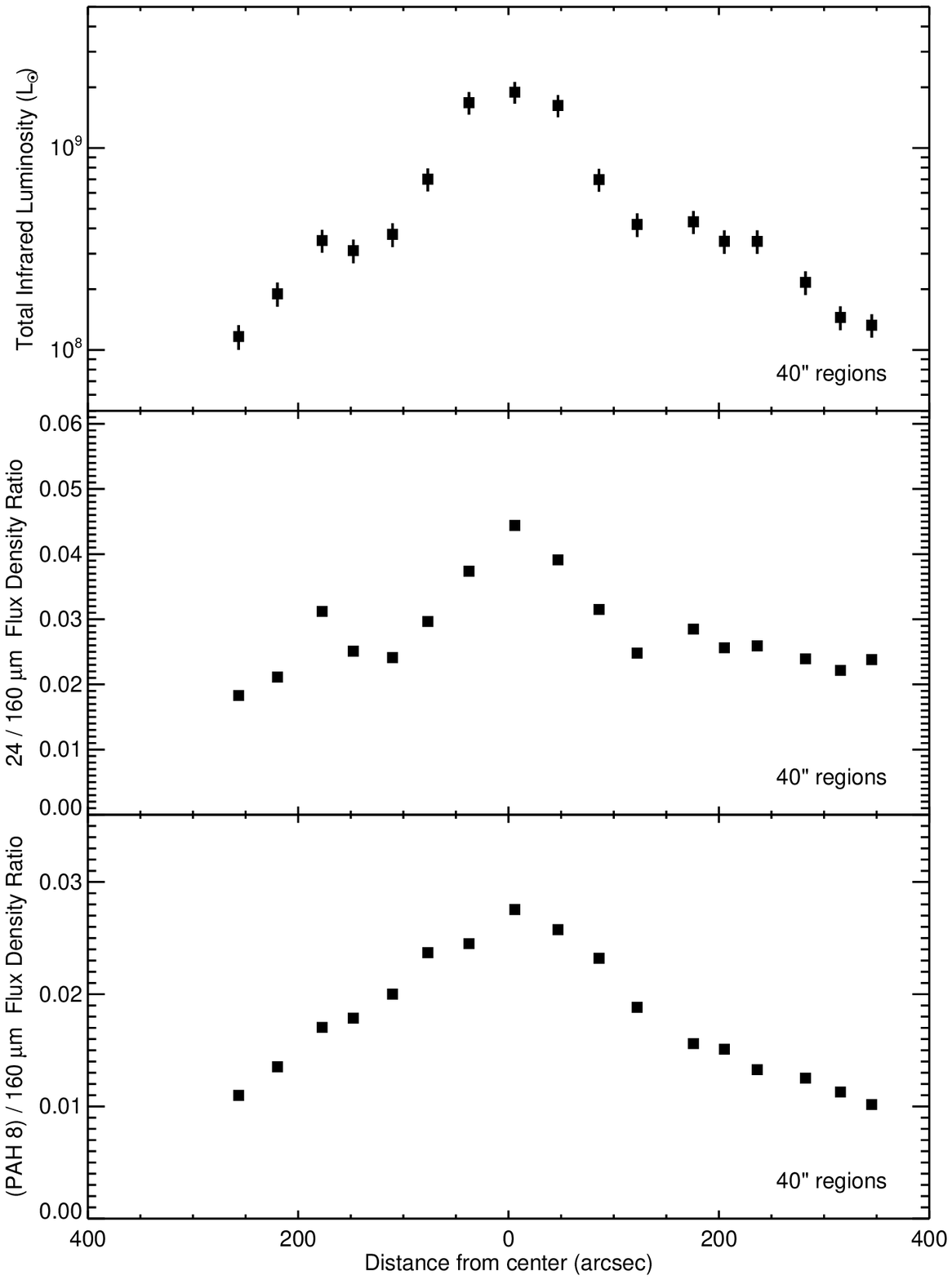}
\caption{A plot of the total infrared luminosity (top), the
24/160~$\mu$m ratio (middle), and the (PAH 8)/160~$\mu$m ratio
(bottom) plotted as a function of distance from the center.  These
data are calculated using the flux densities measured in the
$40^{\prime\prime}$ regions listed in Table~\ref{t_fd40arcsec}.  The
two regions listed in Table~\ref{t_fd40arcsec} that are located off
the major axis of optical disk of the galaxy are excluded from this
plot.  Uncertainties from background noise in the middle and bottom
plots are smaller than the symbols.  Data from the east side of the
galaxy are plotted to the left, and data from the west side are
plotted to the right.}
\label{f_radialcolor}
\end{figure}

\clearpage

\begin{figure}
\plotone{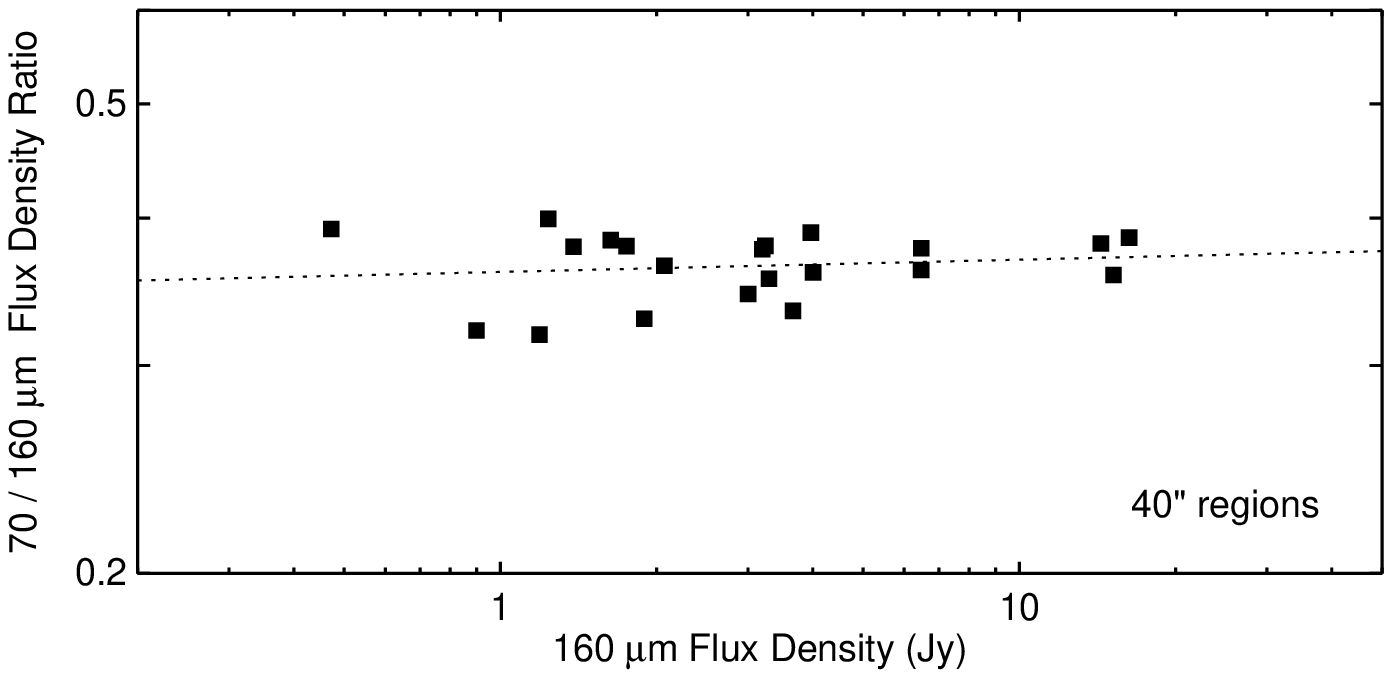}
\plotone{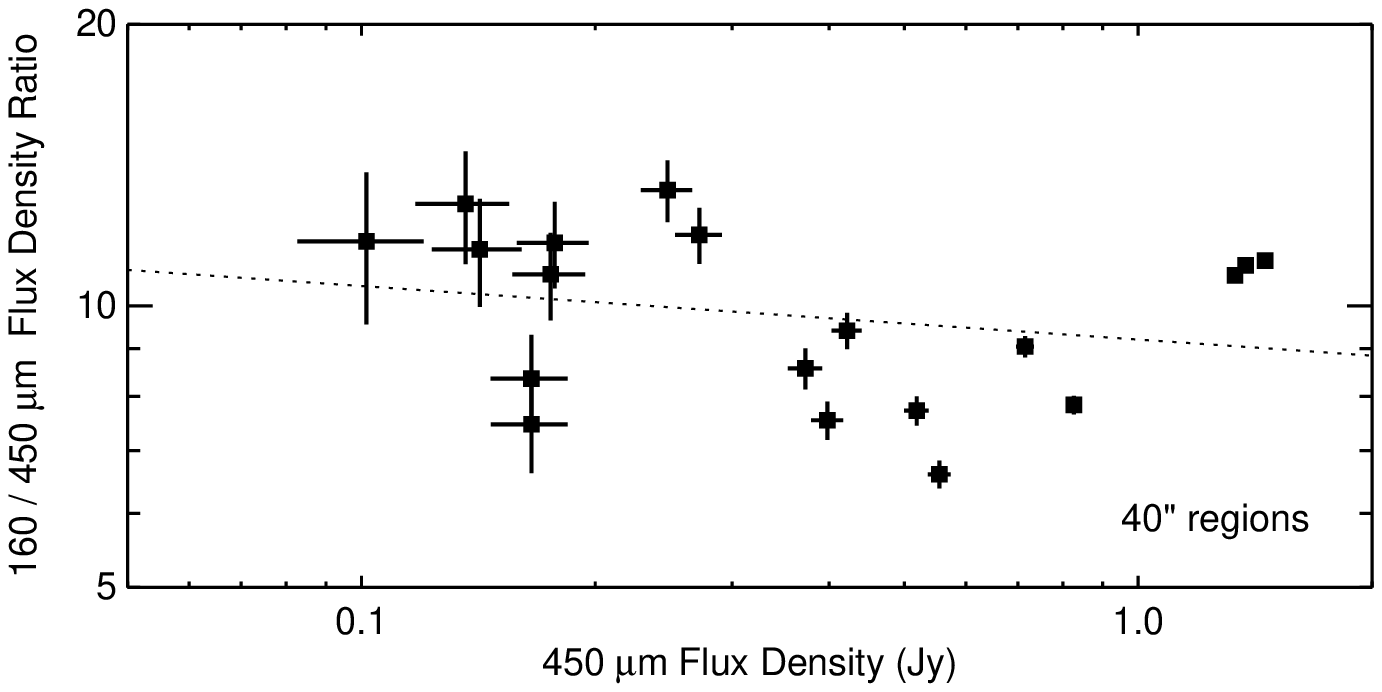}
\caption{A comparison of the 70, 160, and 450~$\mu$m flux densities
measured in the $40^{\prime\prime}$ regions listed in
Table~\ref{t_fd40arcsec}.  Except where uncertainty bars are shown,
uncertainties from background noise are smaller than the symbols.  The
lines show the best fitting relations between the flux density ratios
on the y-axes and the flux densities on the x-axes.}
\label{f_comp_70_160_450}
\end{figure}

\clearpage

\begin{figure}
\epsscale{0.8}
\plotone{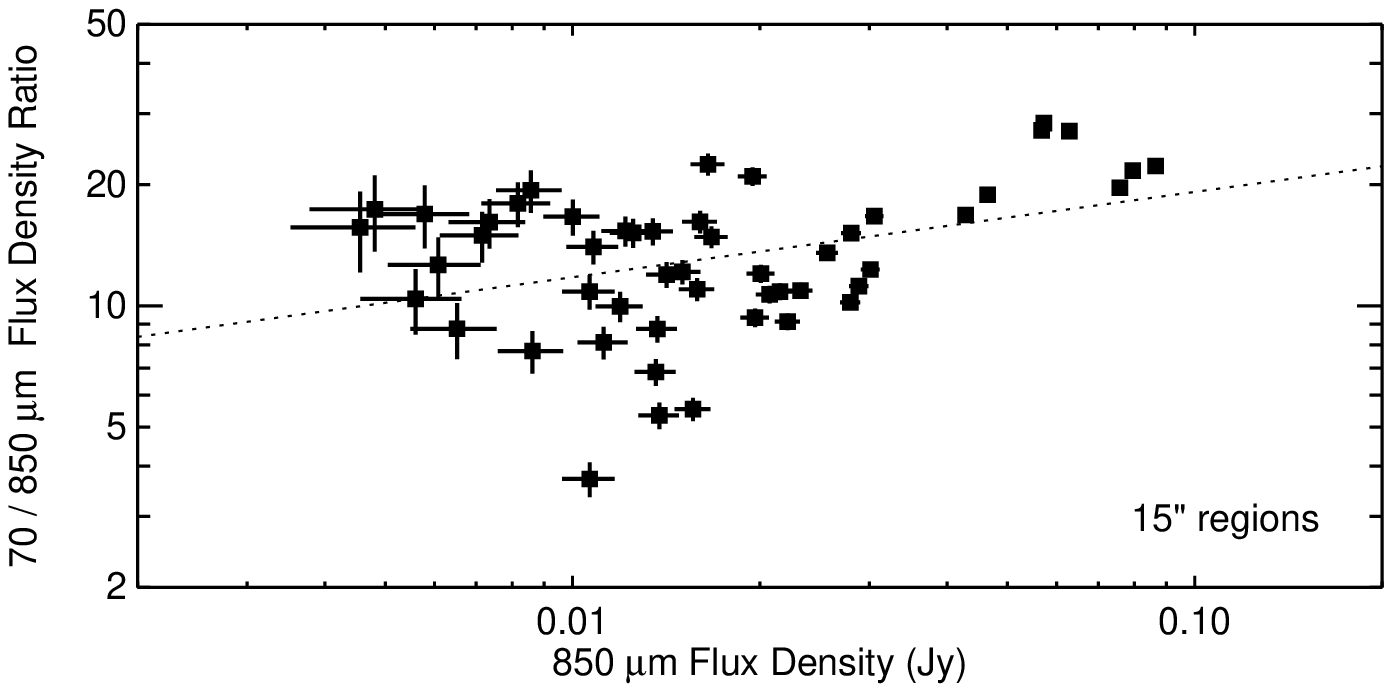}
\plotone{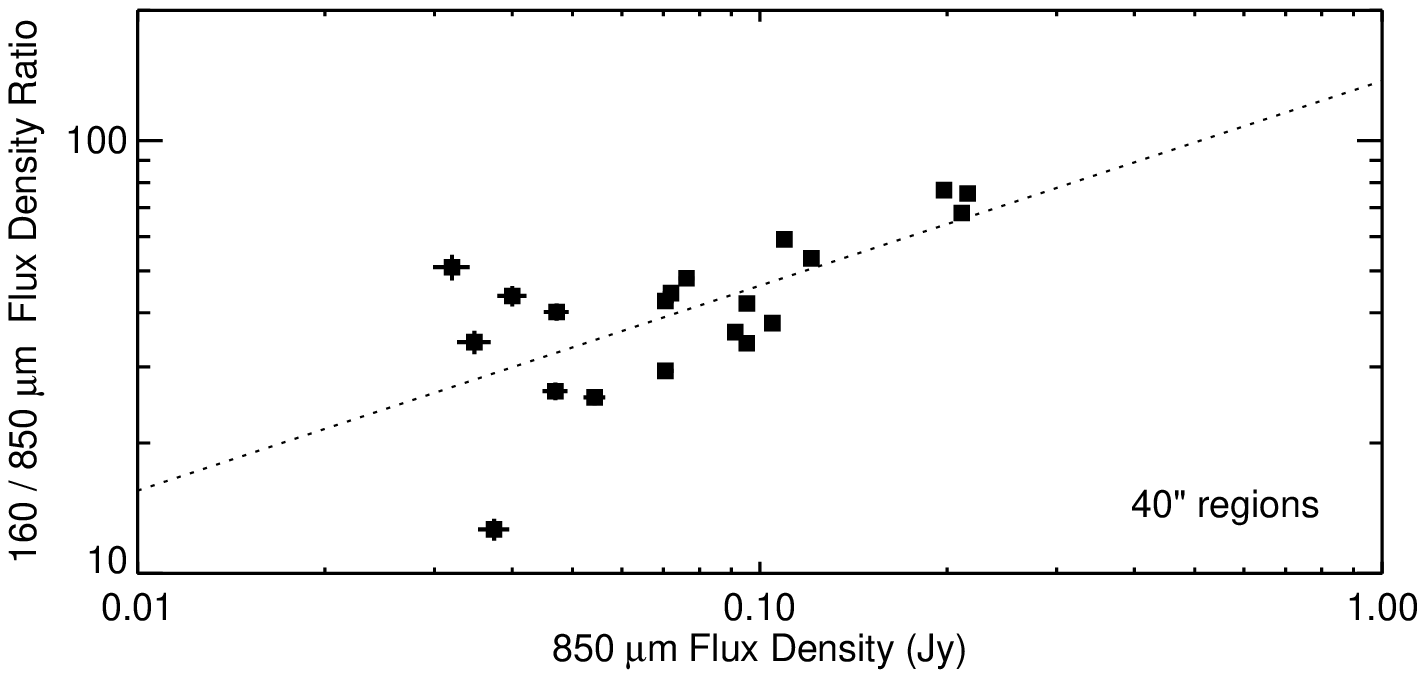}
\plotone{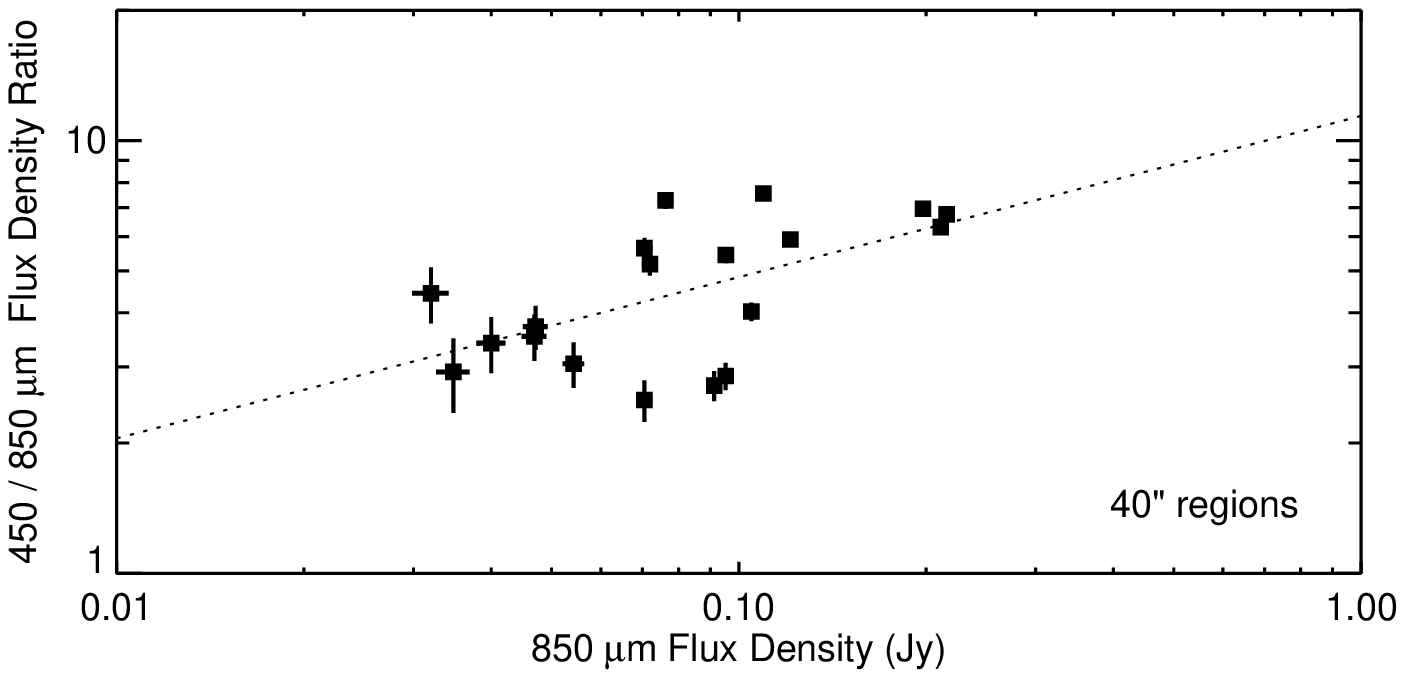}
\caption{A comparison of the 70 and 850~$\mu$m flux densities measured
in the $15^{\prime\prime}$ regions listed in Table~\ref{t_fd15arcsec},
and comparisons of the 160, 450~$\mu$m, and 850~$\mu$m flux densities
measured in the $40^{\prime\prime}$ regions listed in
Table~\ref{t_fd15arcsec}.  Except where uncertainty bars are shown,
uncertainties from background noise are smaller than the symbols.  The
lines show the best fitting relations between the flux density ratios
on the y-axes and the flux densities on the x-axes.}
\label{f_comp_70_160_450_850}
\end{figure}

\clearpage

\begin{figure}
\epsscale{1.0}
\plotone{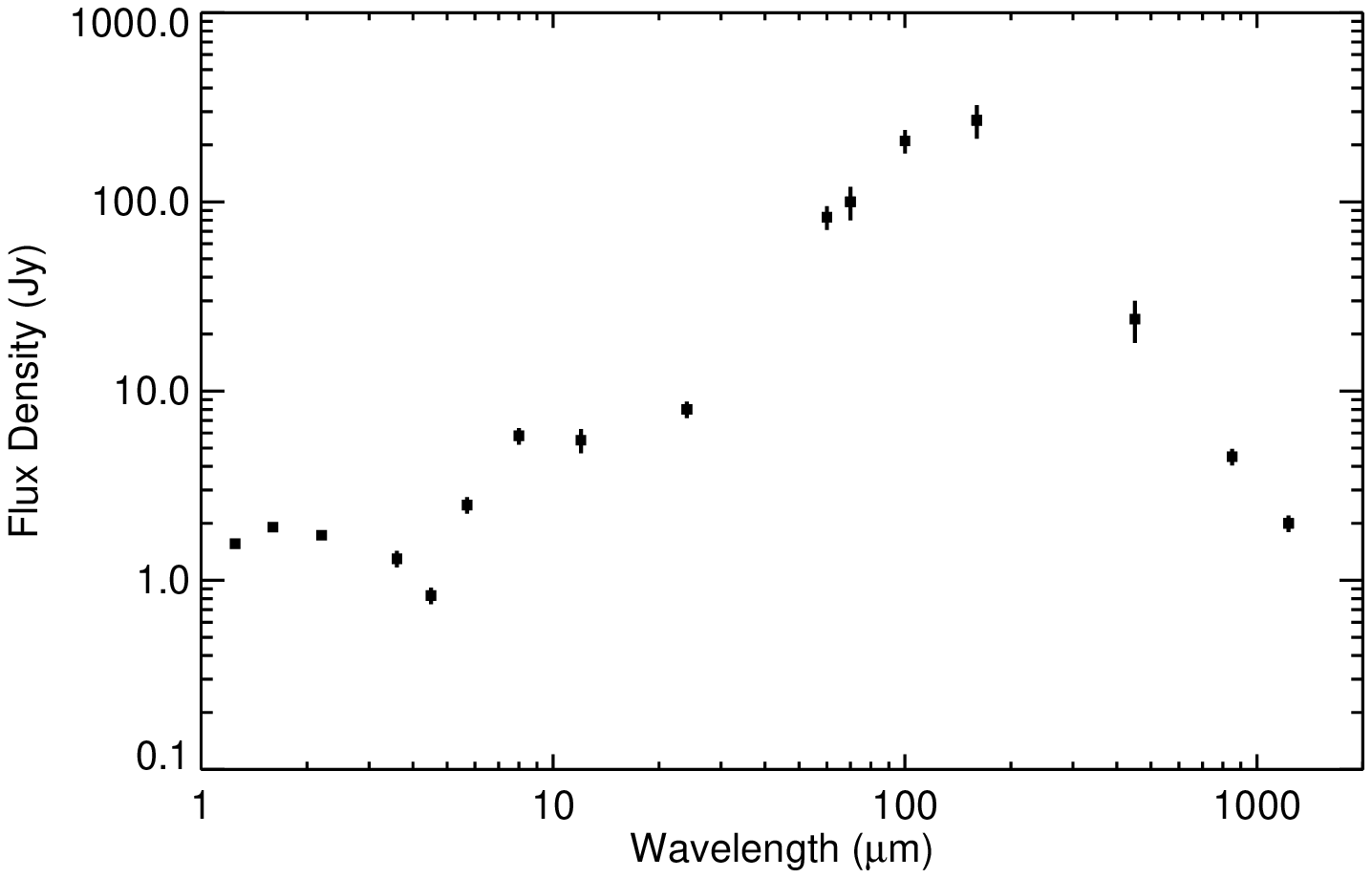}
\caption{The 1.3 - 1230~$\mu$m global SED for NGC 4631.  Except where
uncertainty bars are shown, the uncertainties are smaller than the
symbols in this plot.  Note that the 25~$\mu$m measurement from
\citet{rlsnkldh88}, the 870~$\mu$m measurement from \citet{dkw04}, and
the 450 and 850~$\mu$m measurements from \citet{sag05} are not
included in this plot, although they are discussed in the text.}
\label{f_globalsed}
\end{figure}

\clearpage

\begin{figure}
\plotone{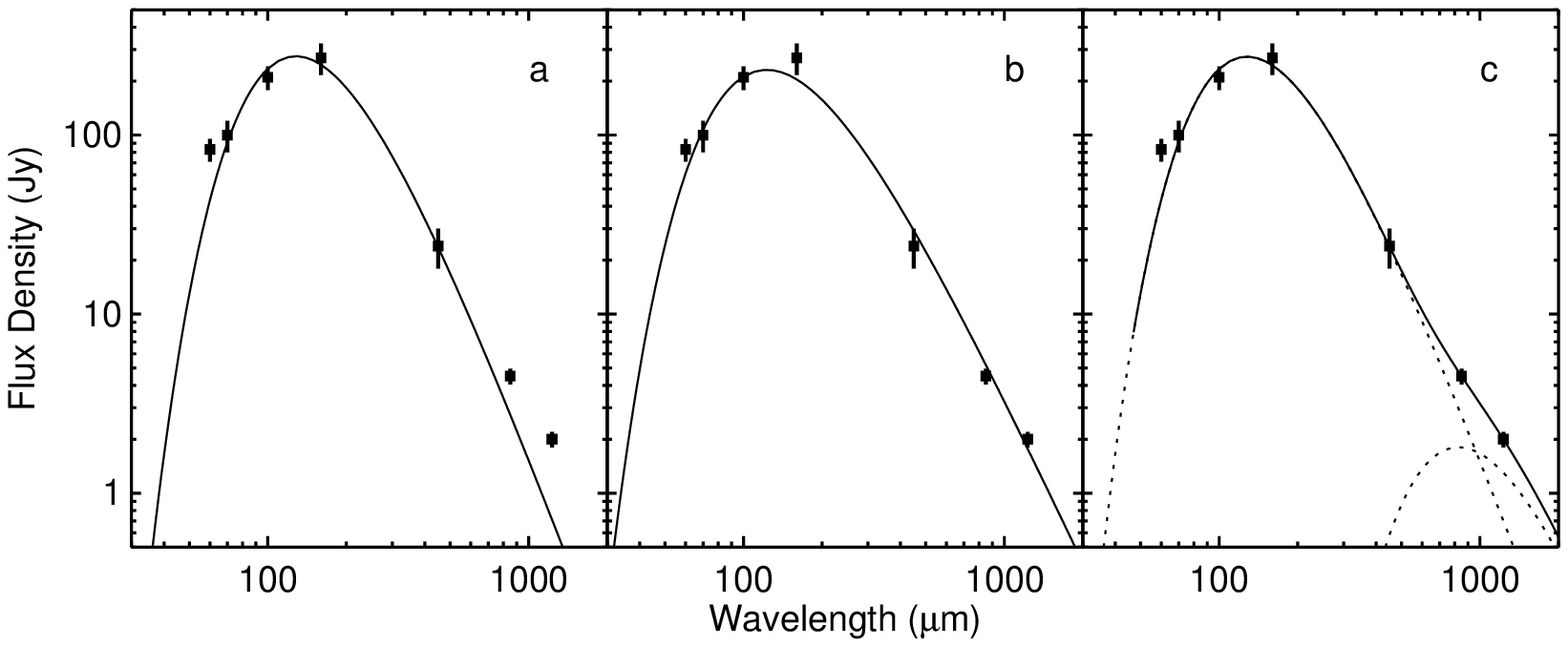}
\caption{The 30 - 1230~$\mu$m global SED for NGC 4631 with various
simple modified blackbodies fit to the data.  Except where the
uncertainty bars are shown, the uncertainties are smaller than the
symbols in this plot.  The functions are fit to the 70 - 1230~$\mu$m
data except where noted.  Part~a features one blackbody modified with
a $\lambda^{-2}$ emissivity law (fit only to the 70 - 450~$\mu$m
data), with the best fit having a temperature of $23 \pm 2$~K.  Part~b
features one blackbody modified with a emissivity law with an index
that was fit to the data.  The best fit has a temperature of $28 \pm
2$~K and a $\beta$ of $1.2 \pm 0.1$.  Part~c features two blackbodies
modified with $\lambda^{-2}$ emissivity laws (with an adjustment to
account for the CMB; see the text for details).  The individual
thermal components are shown as dotted lines.  The best fitting
functions had temperatures of $23 \pm 2$ and $3.5^{+1.4}_{-0.8}$~K.}
\label{f_globalsed_bb}
\end{figure}

\clearpage

\begin{figure}
\plotone{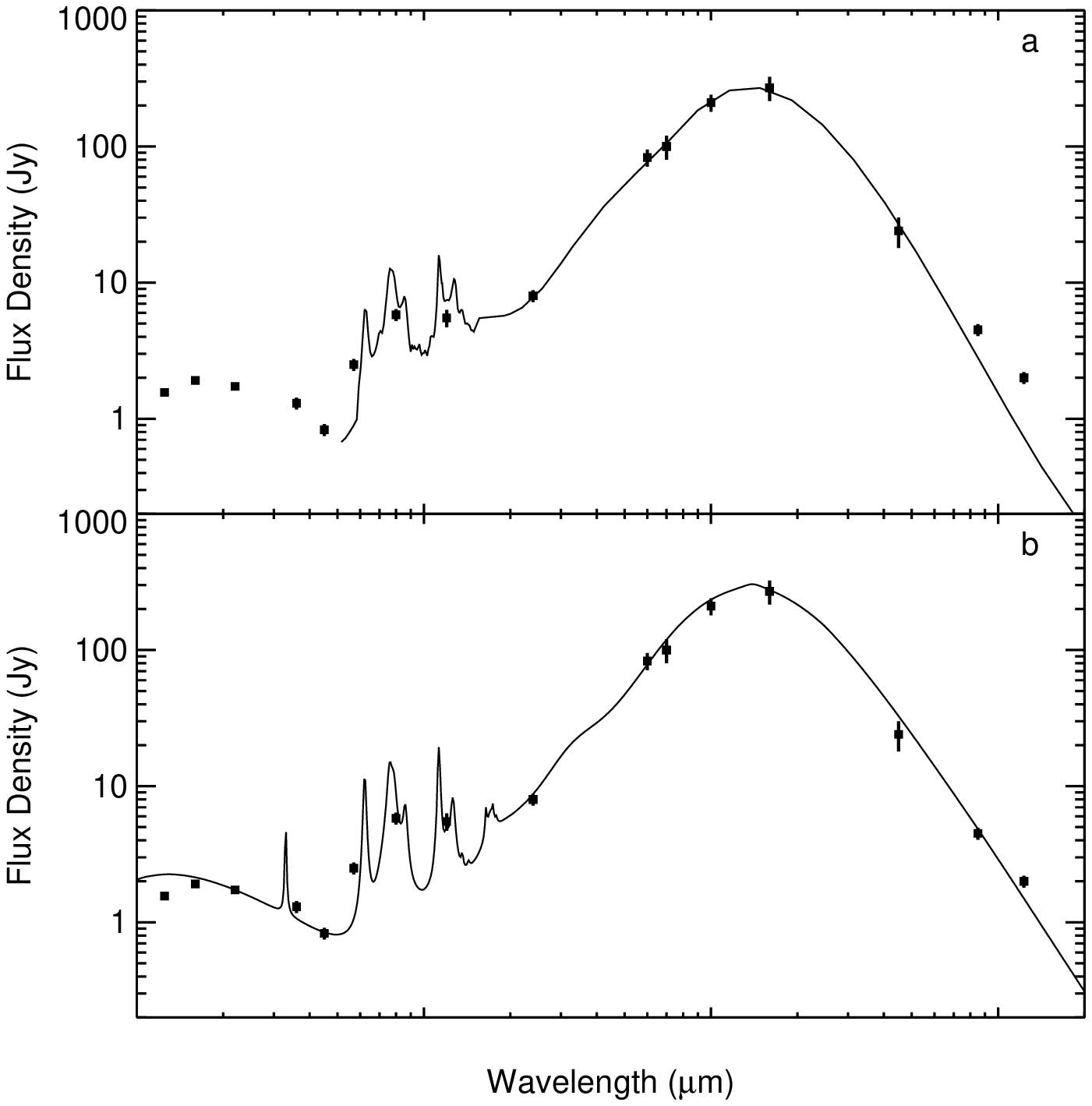}
\caption{The 1.3 - 1230~$\mu$m global SED for NGC 4631 with the best fitting
semi-empirical dust emission model of \citet{dhcsk01} (with $\alpha$
of $2.32 \pm 0.10$) overlaid on the data in panel a and the best
fitting physical dust emission model overlaid on the data in panel b.  
See the text for the parameters for the model in part b.  Except where
uncertainty bars are shown, the uncertainties are smaller than the
symbols in this plot.}
\label{f_globalsed_2modelfit}
\end{figure}

\clearpage

\begin{figure}
\plotone{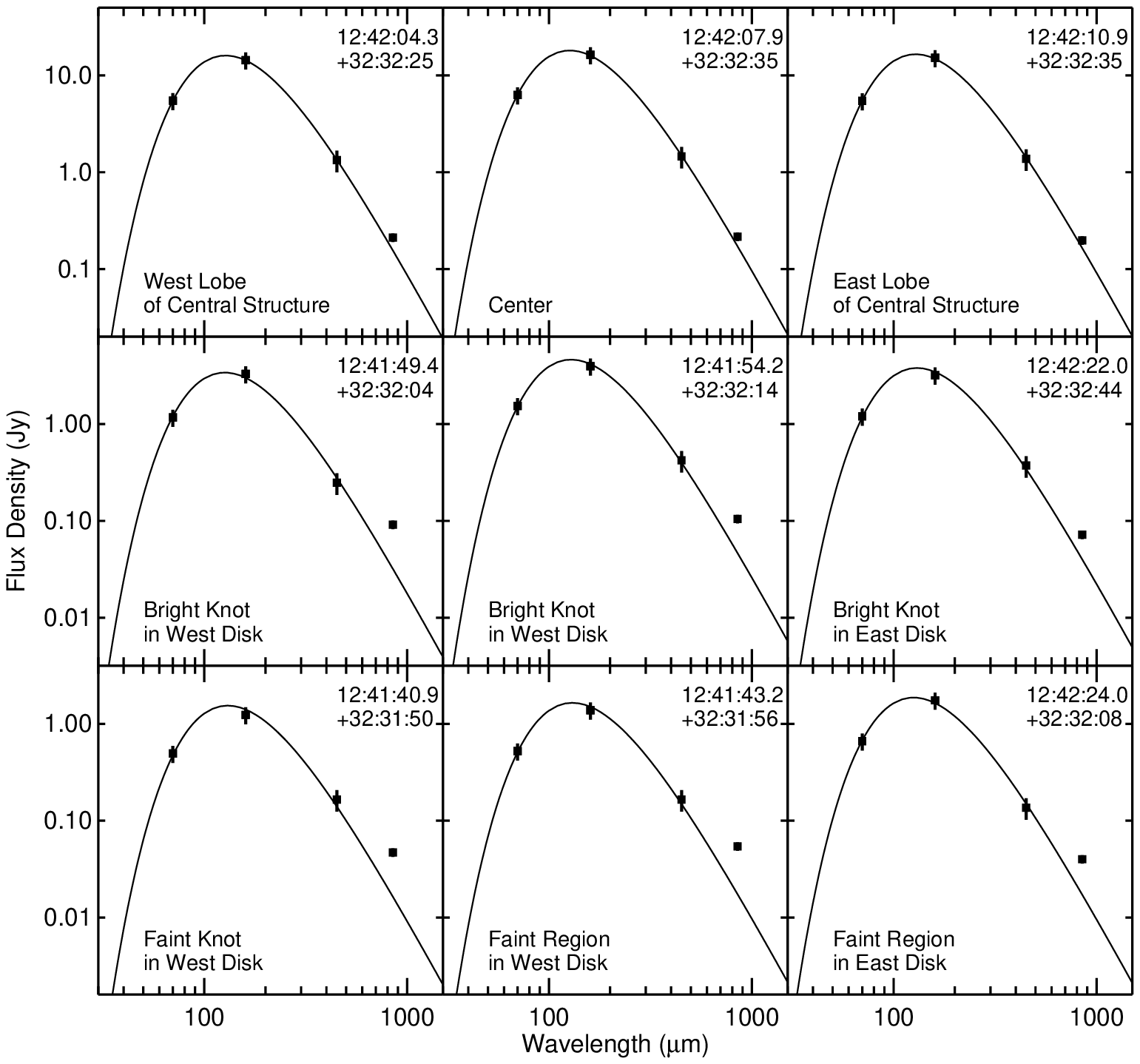}
\caption{70 - 850~$\mu$m SEDs for example $40^{\prime\prime}$ regions
listed in Table~\ref{t_fd40arcsec} with the best fitting blackbody
modified by a $\lambda^{-2}$ emissivity law overlaid on the data.  The
top row consists of data from regions in the central structure.  The
middle row consists of data from three bright knots in the outer disk.
The bottom row consists of data from three faint regions in the outer
disk.  The J2000 coordinates of the regions are listed in the top
right corners of each plot.  Except where uncertainty bars are shown,
the uncertainties are smaller than the symbols in this plot.}
\label{f_discretesed_bbfit}
\end{figure}

\clearpage

\begin{figure}
\plotone{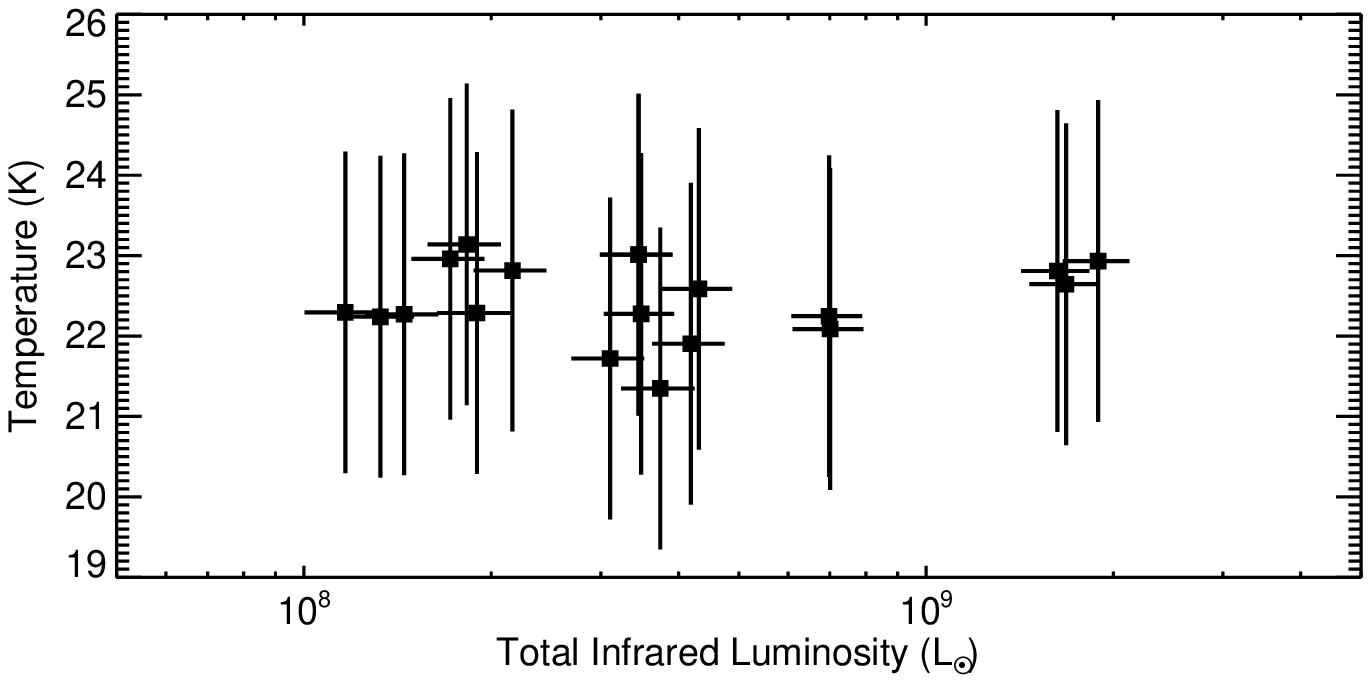}
\caption{A comparison of the temperatures of blackbodies modified with
$\lambda^{-2}$ emissivity laws fit to the 70 - 450~$\mu$m data to the
total infrared luminosities for the $40^{\prime\prime}$ regions listed
in Table~\ref{t_fd40arcsec}.}
\label{f_comp_tir_temp_bbfit}
\end{figure}

\clearpage

\begin{figure}
\plotone{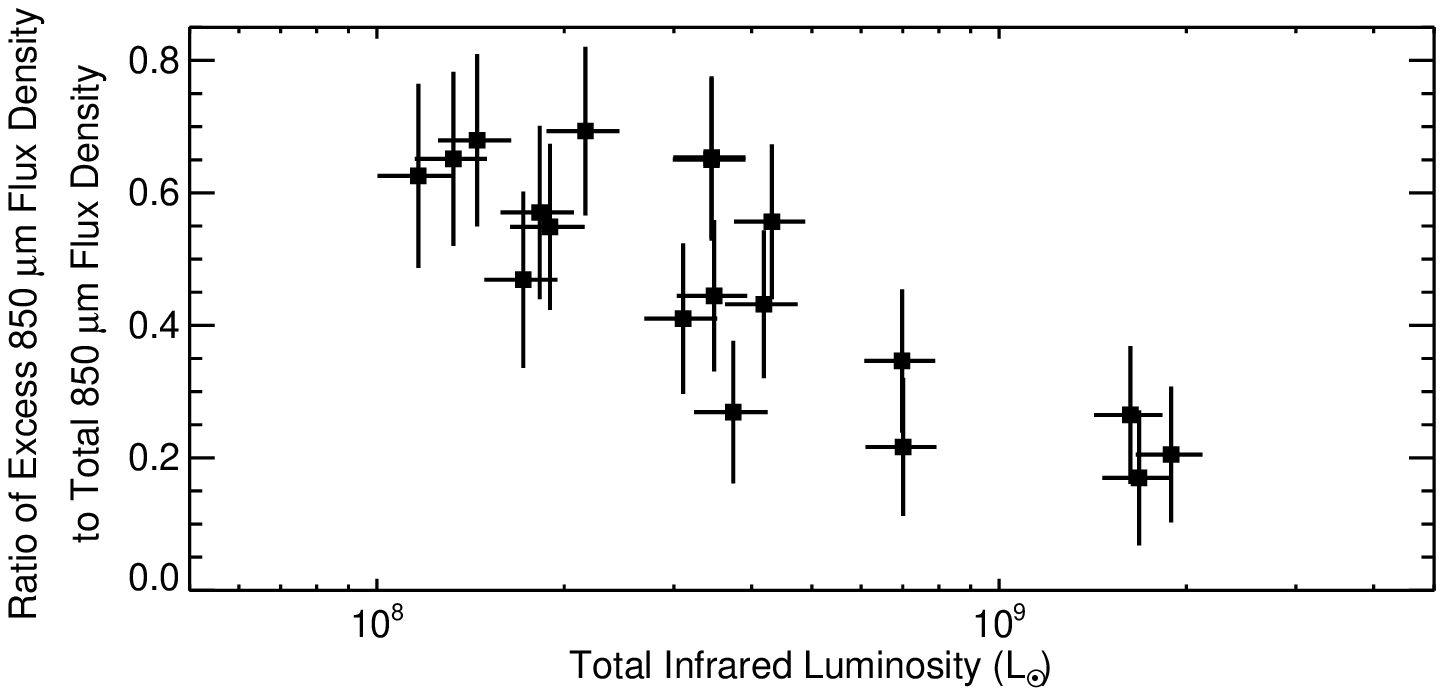}
\caption{A comparison of the ratio of the excess 850~$\mu$m flux
density to total 850~$\mu$m flux density to the total infrared
luminosities for the $40^{\prime\prime}$ regions listed in
Table~\ref{t_fd40arcsec}.  The excess emission in this plot is derived
from the difference between the total 850~$\mu$m emission and the
850~$\mu$m emission expected from a blackbody modified with a
$\lambda^{-2}$ emissivity law fit to the 70 - 450~$\mu$m data.}
\label{f_comp_tir_excess850_bbfit}
\end{figure}

\clearpage

\begin{figure}
\plotone{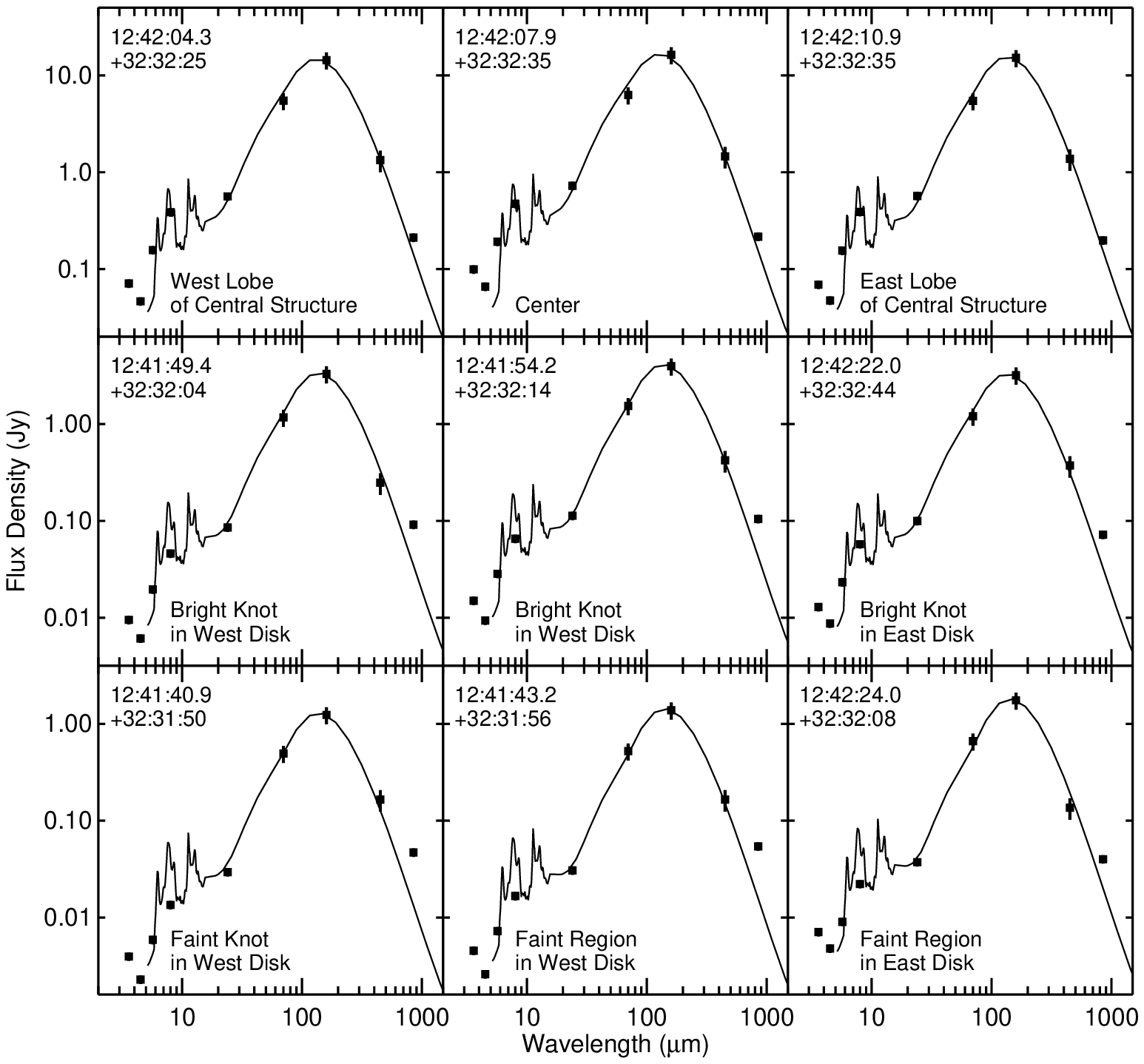}
\caption{3.6 - 850~$\mu$m SEDs for example $40^{\prime\prime}$ regions
listed in Table~\ref{t_fd40arcsec} with the semi-empirical models of
\citet{dhcsk01} that best fit the 5.7 - 450~$\mu$m overlaid on the
data.  These are the same regions shown in
Figure~\ref{f_discretesed_bbfit}.  The J2000 coordinates of the
regions are listed in the top left corners of each plot.  Except where
uncertainty bars are shown, the uncertainties are smaller than the
symbols in this plot.}
\label{f_discretesed_semimodelfit}
\end{figure}

\clearpage

\begin{figure}
\plotone{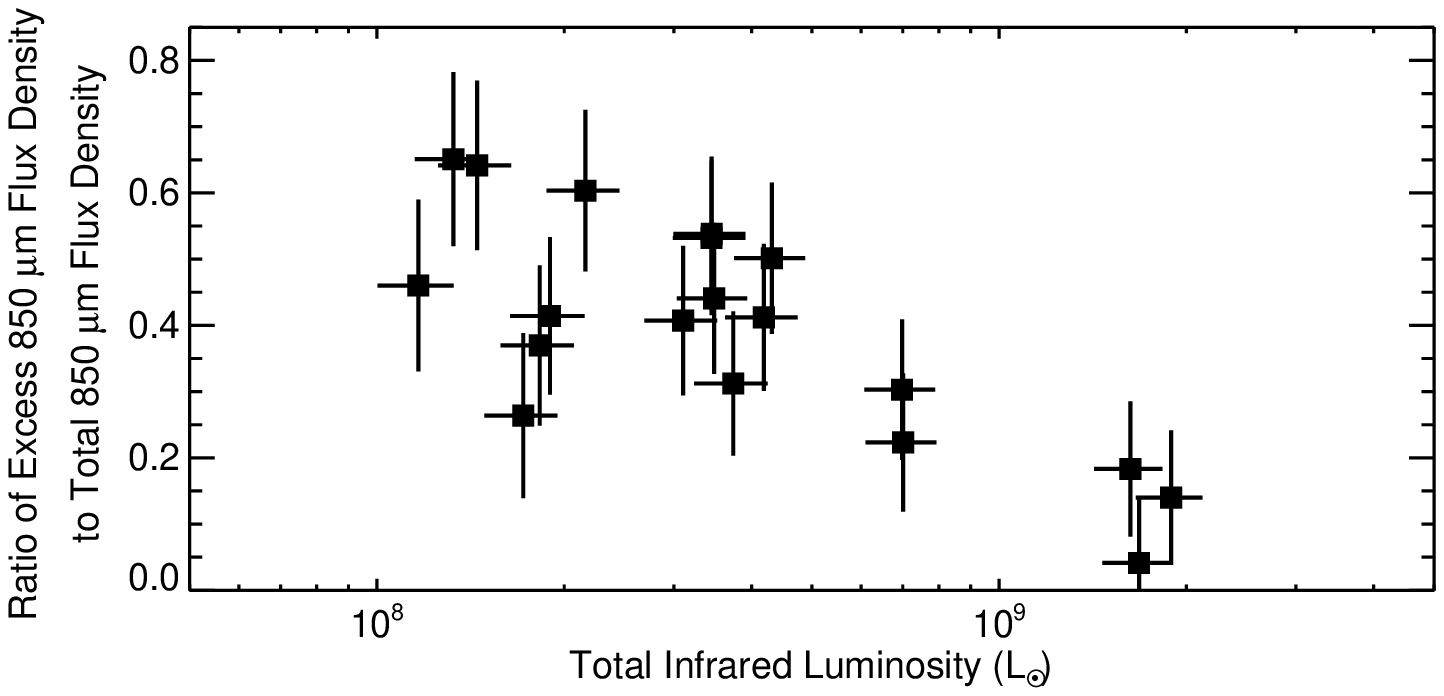}
\caption{A comparison of the ratio of the excess 850~$\mu$m flux
density to total 850~$\mu$m flux density to the total infrared
luminosities for the $40^{\prime\prime}$ regions listed in
Table~\ref{t_fd40arcsec}.  The excess emission in this plot is derived
from the difference between the total 850~$\mu$m emission and the
850~$\mu$m emission expected from the semi-empirical models of
\citet{dhcsk01} fit to the 5.7 - 450~$\mu$m data.}
\label{f_comp_tir_excess850_semimodelfit}
\end{figure}

\clearpage

\begin{figure}
\plotone{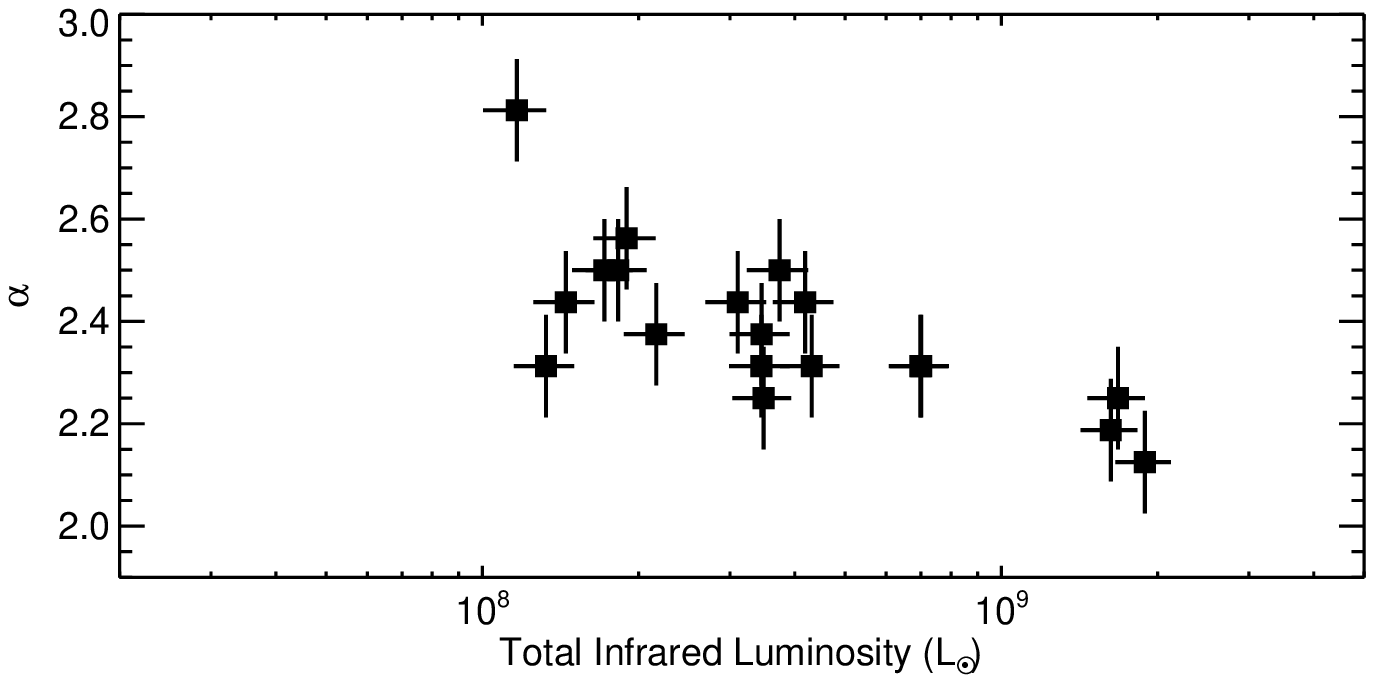}
\caption{A comparison of the $\alpha$ of the semi-empirical models fit
to the 5.7 - 450~$\mu$m data to the total infrared luminosities for
the $40^{\prime\prime}$ regions listed in Table~\ref{t_fd40arcsec}.
Lower $\alpha$ represent regions with proportionally more heating by
stronger radiation fields.}
\label{f_comp_tir_alpha_semimodelfit}
\end{figure}

\clearpage

\begin{figure}
\plotone{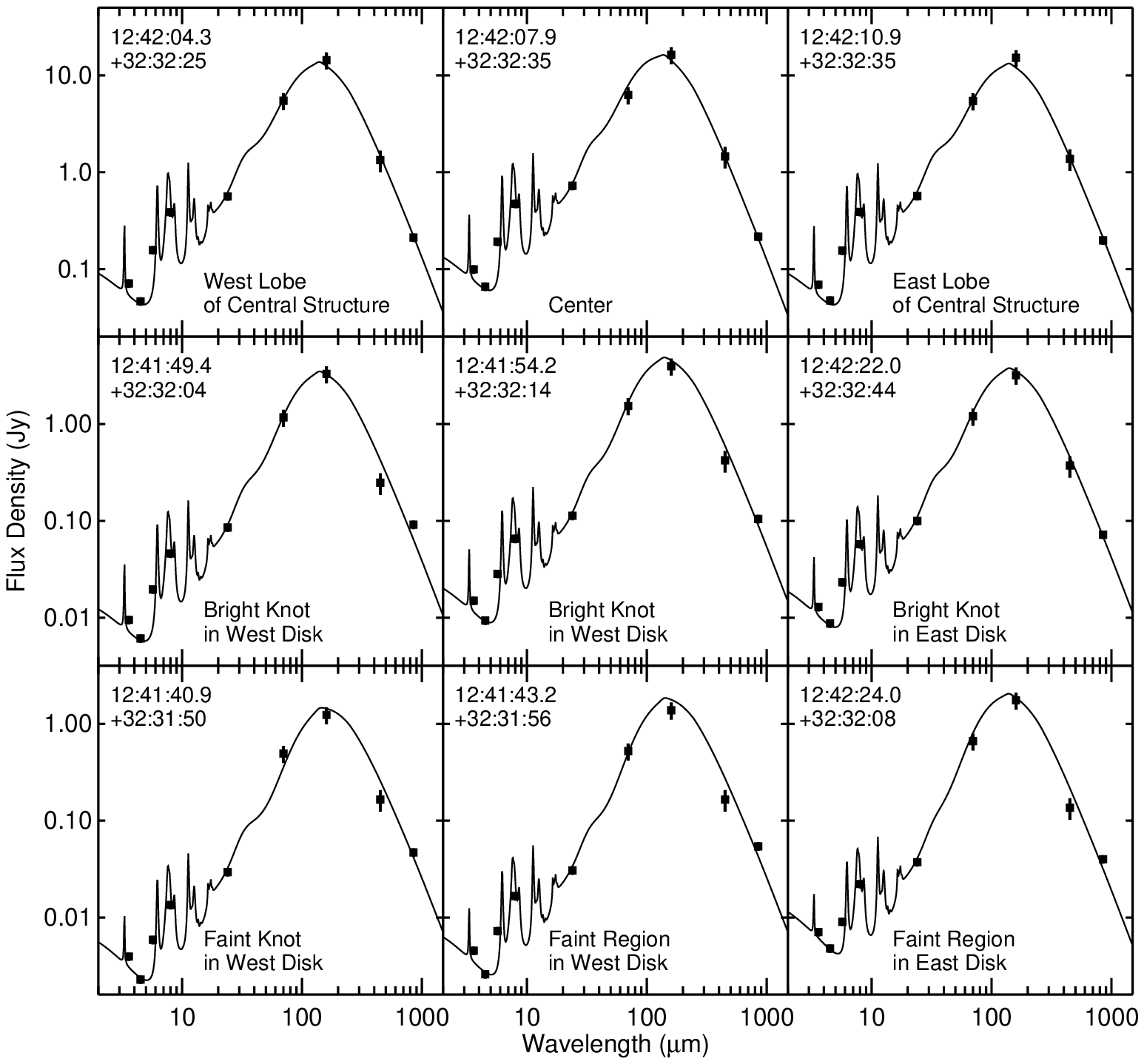}
\caption{The same as Figure~\ref{f_discretesed_semimodelfit} but with
the best fitting physical models overlaid on the data.}
\label{f_discretesed_drainemodelfit}
\end{figure}

\clearpage

\begin{figure}
\plotone{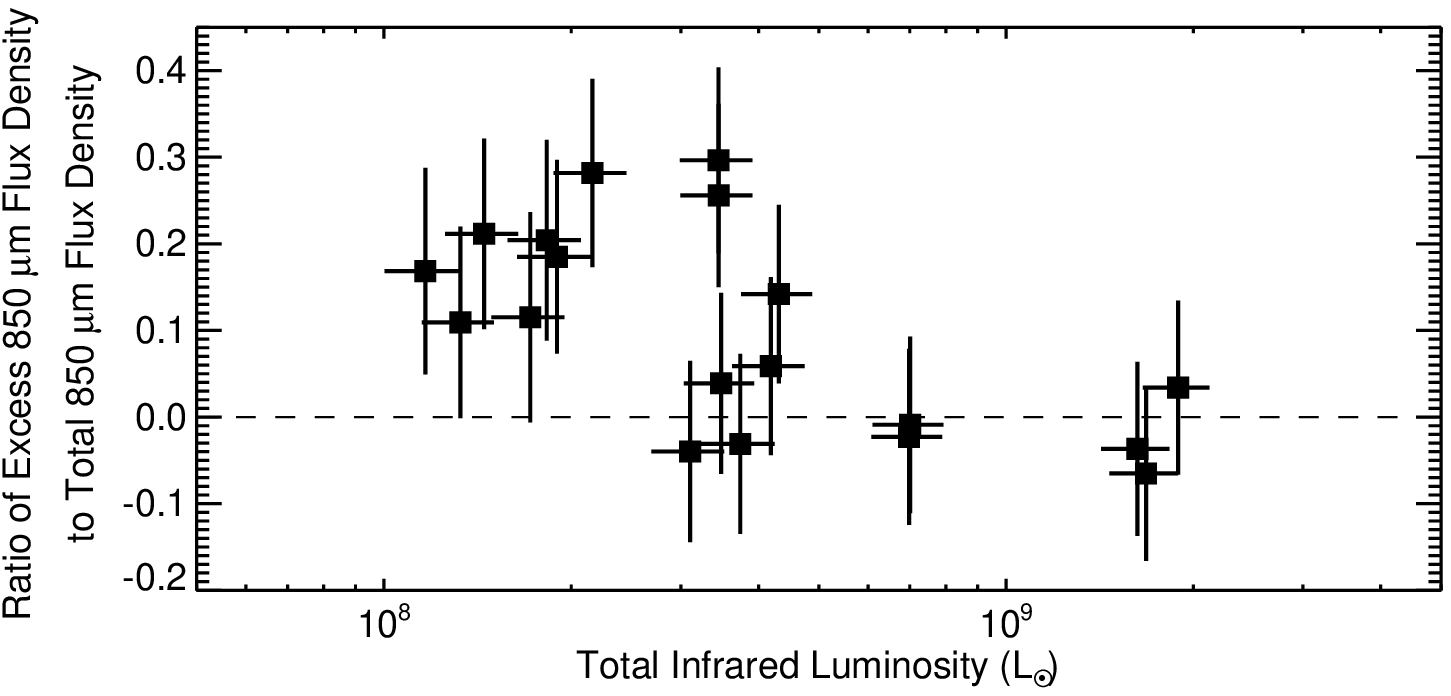}
\caption{A comparison of the ratio of the excess 850~$\mu$m flux
density to total 850~$\mu$m flux density to the total infrared
luminosities for the $40^{\prime\prime}$ regions listed in
Table~\ref{t_fd40arcsec}.  The excess emission in this plot is derived
from the difference between the total 850~$\mu$m emission and the
850~$\mu$m emission expected from the physical models fit to the
3.6-850~$\mu$m data.  The dashed line demarcates where the y-axis
equals 0.}
\label{f_comp_tir_excess850_drainemodelfit}
\end{figure}

\clearpage

\begin{figure}
\plotone{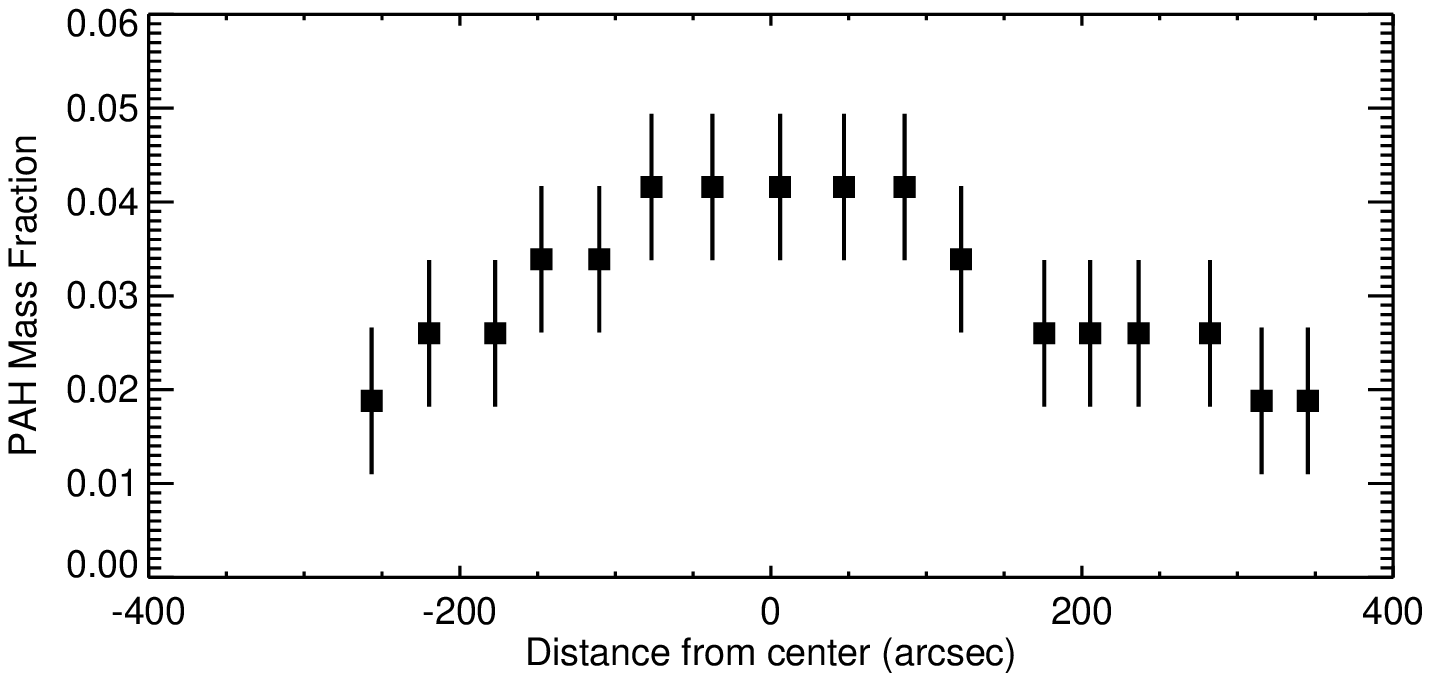}
\caption{The PAH mass fraction determined for the $40^{\prime\prime}$
regions listed in Table~\ref{t_fd40arcsec} plotted as a function of
distance from the center.  Data from the east side of the galaxy are
plotted to the left, and data from the west side are plotted to the
right.}
\label{f_comp_dist_pahdraine}
\end{figure}

\clearpage

\begin{figure}
\plotone{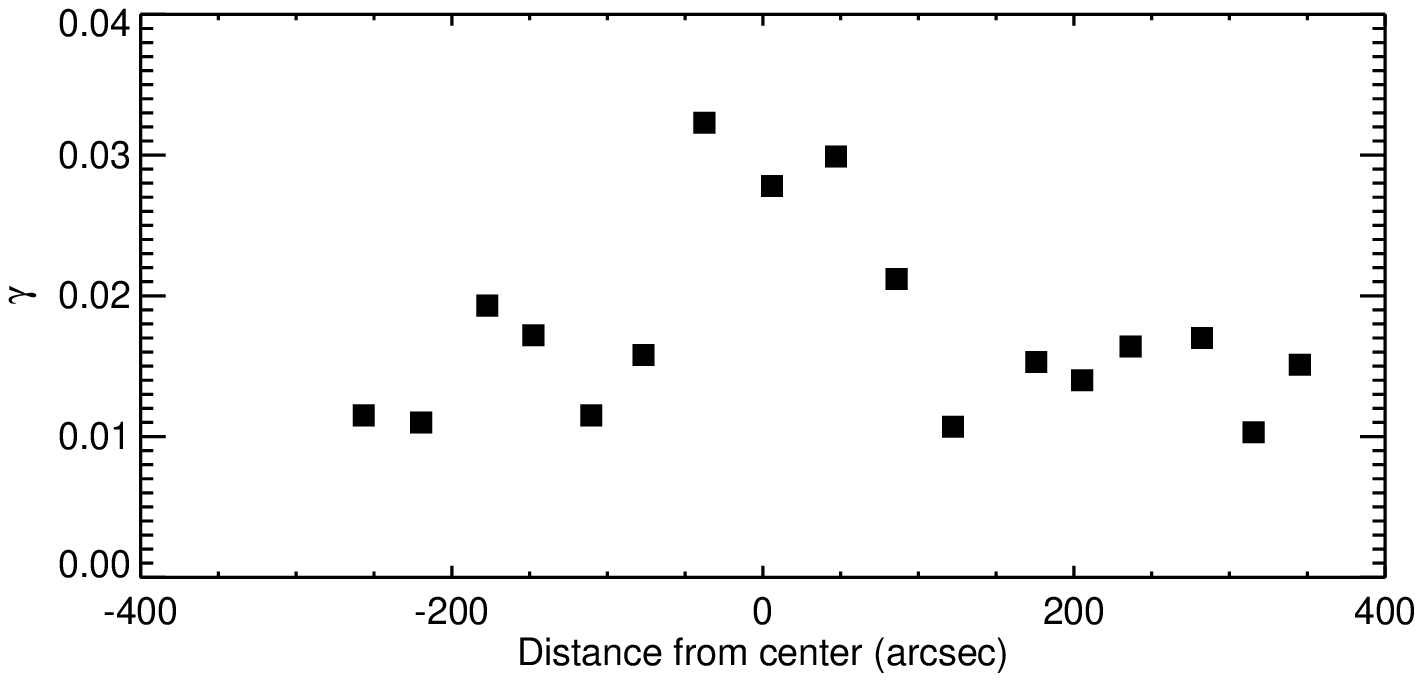}
\caption{The $\gamma$ term determined for the $40^{\prime\prime}$
regions listed in Table~\ref{t_fd40arcsec} plotted as a function of
distance from the center.  This $\gamma$ term represents the fraction
of the dust mass illuminated by the radiation fields described by the
power law (i.e.the dust heated in photodissociation regions).  Data
from the east side of the galaxy are plotted to the left, and data
from the west side are plotted to the right.}
\label{f_comp_dist_gammadraine}
\end{figure}

\clearpage

\begin{deluxetable}{ccccccccc}
\rotate
\tablecolumns{9}
\tablewidth{0pc}
\tabletypesize{\tiny}
\tablecaption{Flux Densities Measured within Discrete 
    $15^{\prime\prime}$ Regions of NGC~4631
    \label{t_fd15arcsec}}
\tablehead{
\multicolumn{2}{c}{Position (J2000)} &
\multicolumn{7}{c}{Flux Density (Jy)\tablenotemark{a}}\\
\colhead{R.A.} &     \colhead{Dec.} &
    \colhead{3.6~$\mu$m\tablenotemark{b}} &     
    \colhead{4.5~$\mu$m\tablenotemark{c}} &
    \colhead{5.7~$\mu$m\tablenotemark{d}} &
    \colhead{8.0~$\mu$m\tablenotemark{e}} &
    \colhead{24~$\mu$m\tablenotemark{f}} &
    \colhead{70~$\mu$m\tablenotemark{g}} &
    \colhead{850~$\mu$m\tablenotemark{h}} 
}
\startdata
12 41 37.6 &     +32 31 34 &
    $3.93 \times 10^{-4}$ &     $2.02 \times 10^{-4}$ &
    $6.71 \times 10^{-4}$ &     $1.582 \times 10^{-3}$ &
    $1.916 \times 10^{-3}$ &    $3.96 \times 10^{-2}$ &
    $1.07 \times 10^{-2}$\\
12 41 39.0 &     +32 31 43 &
    $6.86 \times 10^{-4}$ &     $3.90 \times 10^{-4}$ &
    $1.290 \times 10^{-3}$ &    $3.105 \times 10^{-3}$ &
    $4.569 \times 10^{-3}$ &    $8.65 \times 10^{-2}$ &
    $1.56 \times 10^{-2}$ \\
12 41 40.9 &     +32 31 50 &
    $9.16 \times 10^{-4}$ &     $5.92 \times 10^{-4}$ &
    $1.841 \times 10^{-3}$ &    $4.470 \times 10^{-3}$ &
    $9.053 \times 10^{-3}$ &    $1.513 \times 10^{-1}$ &
    $1.08 \times 10^{-2}$ \\
12 41 43.2 &     +32 31 56 &
    $8.03 \times 10^{-4}$ &     $4.62 \times 10^{-4}$ &
    $1.621 \times 10^{-3}$ &    $3.963 \times 10^{-3}$ &
    $5.086 \times 10^{-3}$ &    $9.33 \times 10^{-2}$ &
    $1.36 \times 10^{-2}$ \\
12 41 44.4 &     +32 31 56 &
    $9.92 \times 10^{-4}$ &     $5.90 \times 10^{-4}$ &
    $2.054 \times 10^{-3}$ &    $4.960 \times 10^{-3}$ &
    $7.107 \times 10^{-3}$ &    $1.199 \times 10^{-1}$ &
    $1.37 \times 10^{-2}$ \\
12 41 45.8 &     +32 32 00 &
    $1.501 \times 10^{-3}$ &    $9.40 \times 10^{-4}$ &
    $3.529 \times 10^{-3}$ &    $8.712 \times 10^{-3}$ &
    $1.3703 \times 10^{-2}$ &   $2.025 \times 10^{-1}$ &
    $2.22 \times 10^{-2}$ \\
12 41 47.5 &     +32 32 04 &
    $1.660 \times 10^{-3}$ &    $1.057 \times 10^{-3}$ &
    $4.569 \times 10^{-3}$ &    $1.1681 \times 10^{-2}$ &
    $1.4688 \times 10^{-2}$ &   $2.338 \times 10^{-1}$ &
    $2.16 \times 10^{-2}$ \\
12 41 49.4 &     +32 32 04 &
    $2.412 \times 10^{-3}$ &    $1.739 \times 10^{-3}$ &
    $6.470 \times 10^{-3}$ &    $1.6239 \times 10^{-2}$ &
    $3.0248 \times 10^{-2}$ &   $3.707 \times 10^{-1}$ &
    $3.01 \times 10^{-2}$ \\
12 41 51.0 &     +32 32 07 &
    $2.147 \times 10^{-3}$ &    $1.352 \times 10^{-3}$ &
    $5.369 \times 10^{-3}$ &    $1.3157 \times 10^{-2}$ &
    $1.4947 \times 10^{-2}$ &   $2.414 \times 10^{-1}$ &
    $2.01 \times 10^{-2}$ \\
12 41 51.7 &     +32 32 15 &
    $2.492 \times 10^{-3}$ &    $1.563 \times 10^{-3}$ &
    $6.549 \times 10^{-3}$ &    $1.6302 \times 10^{-2}$ &
    $1.8910 \times 10^{-2}$ &   $2.851 \times 10^{-1}$ &
    $2.80 \times 10^{-2}$ \\
12 41 52.7 &     +32 32 17 &
    $2.540 \times 10^{-3}$ &    $1.604 \times 10^{-3}$ &
    $6.815 \times 10^{-3}$ &    $1.7143 \times 10^{-2}$ &
    $2.0836 \times 10^{-2}$ &   $3.230 \times 10^{-1}$ &
    $2.89 \times 10^{-2}$ \\
12 41 54.2 &     +32 32 14 &
    $3.255 \times 10^{-3}$ &    $2.144 \times 10^{-3}$ &
    $8.177 \times 10^{-3}$ &    $2.0339 \times 10^{-2}$ &
    $3.4288 \times 10^{-2}$ &   $4.251 \times 10^{-1}$ &
    $2.80 \times 10^{-2}$ \\
12 41 55.8 &     +32 32 15 &
    $4.143 \times 10^{-3}$ &    $2.591 \times 10^{-3}$ &
    $8.506 \times 10^{-3}$ &    $2.0240 \times 10^{-2}$ &
    $2.6269 \times 10^{-2}$ &   $3.474 \times 10^{-1}$ &
    $2.57 \times 10^{-2}$ \\
12 41 56.6 &     +32 32 20 &
    $3.716 \times 10^{-3}$ &    $2.283 \times 10^{-3}$ &
    $7.458 \times 10^{-3}$ &    $1.7646 \times 10^{-2}$ &
    $1.9052 \times 10^{-2}$ &   $2.537 \times 10^{-1}$ &
    $2.33 \times 10^{-2}$ \\
12 41 58.1 &     +32 32 19 &
    $3.836 \times 10^{-3}$ &    $2.347 \times 10^{-3}$ &
    $7.053 \times 10^{-3}$ &    $1.6051 \times 10^{-2}$ &
    $1.4480 \times 10^{-2}$ &   $2.213 \times 10^{-1}$ &
    $2.07 \times 10^{-2}$ \\
12 41 59.2 &     +32 32 09 &
    $3.992 \times 10^{-3}$ &    $2.506 \times 10^{-3}$ &
    $7.578 \times 10^{-3}$ &    $1.7191 \times 10^{-2}$ &
    $1.5373 \times 10^{-2}$ &   $2.592 \times 10^{-1}$ &
    $1.60 \times 10^{-2}$ \\
12 42 01.3 &     +32 32 17 &
    $7.075 \times 10^{-3}$ &    $4.527 \times 10^{-3}$ &
    $1.7353 \times 10^{-2}$ &   $4.3708 \times 10^{-2}$ &
    $4.2201 \times 10^{-2}$ &   $5.105 \times 10^{-1}$ &
    $3.06 \times 10^{-2}$ \\
12 42 02.7 &     +32 32 14 &
    $9.800 \times 10^{-3}$ &    $6.404 \times 10^{-3}$ &
    $2.7375 \times 10^{-2}$ &   $7.0464 \times 10^{-2}$ &
    $6.9974 \times 10^{-2}$ &   $7.196 \times 10^{-1}$ &
    $4.28 \times 10^{-2}$ \\
12 42 04.3 &     +32 32 25 &
    $1.8805 \times 10^{-2}$ &   $1.2728 \times 10^{-2}$ &
    $5.8619 \times 10^{-2}$ &   $1.57674 \times 10^{-1}$ &
    $1.86839 \times 10^{-1}$ &  $1.7222 \times 10^{0}$ &
    $7.95 \times 10^{-2}$ \\
12 42 05.5 &     +32 32 29 &
    $1.9807 \times 10^{-2}$ &   $1.3120 \times 10^{-2}$ &
    $5.1955 \times 10^{-2}$ &   $1.38605 \times 10^{-1}$ &
    $1.43971 \times 10^{-1}$ &  $1.5470 \times 10^{0}$ &
    $5.67 \times 10^{-2}$ \\
12 42 06.6 &     +32 32 32 &
    $2.5688 \times 10^{-2}$ &   $1.6932 \times 10^{-2}$ &
    $5.9281 \times 10^{-2}$ &   $1.55280 \times 10^{-1}$ &
    $1.72257 \times 10^{-1}$ &  $1.6262 \times 10^{0}$ &
    $5.72 \times 10^{-2}$ \\
12 42 07.9 &     +32 32 35 &
    $2.8965 \times 10^{-2}$ &   $1.9612 \times 10^{-2}$ &
    $6.5750 \times 10^{-2}$ &   $1.72558 \times 10^{-1}$ &
    $2.43515 \times 10^{-1}$ &  $1.7079 \times 10^{0}$ &
    $6.29 \times 10^{-2}$ \\
12 42 09.3 &     +32 32 36 &
    $2.5118 \times 10^{-2}$ &   $1.7414 \times 10^{-2}$ &
    $6.4930 \times 10^{-2}$ &   $1.74174 \times 10^{-1}$ &
    $1.73071 \times 10^{-1}$ &  $1.4880 \times 10^{0}$ &
    $7.58 \times 10^{-2}$ \\
12 42 10.9 &     +32 32 35 &
    $2.0184 \times 10^{-2}$ &   $1.4550 \times 10^{-2}$ &
    $6.5122 \times 10^{-2}$ &   $1.73029 \times 10^{-1}$ &
    $2.20046 \times 10^{-1}$ &  $1.9241 \times 10^{0}$ &
    $8.65 \times 10^{-2}$ \\
12 42 12.5 &     +32 32 38 &
    $1.1142 \times 10^{-2}$ &   $7.770 \times 10^{-3}$ &
    $3.1484 \times 10^{-2}$ &   $8.3306 \times 10^{-2}$ &
    $7.6281 \times 10^{-2}$ &   $8.775 \times 10^{-1}$ &
    $4.65 \times 10^{-2}$ \\
12 42 13.7 &     +32 32 41 &
    $5.714 \times 10^{-3}$ &    $3.836 \times 10^{-3}$ &
    $1.4242 \times 10^{-2}$ &   $3.6708 \times 10^{-2}$ &
    $3.0358 \times 10^{-2}$ &   $4.084 \times 10^{-1}$ &
    $1.95 \times 10^{-2}$ \\
12 42 15.0 &     +32 32 42 &
    $4.336 \times 10^{-3}$ &    $2.838 \times 10^{-3}$ &
    $9.661 \times 10^{-3}$ &    $2.4563 \times 10^{-2}$ &
    $1.8495 \times 10^{-2}$ &   $2.479 \times 10^{-1}$ &
    $1.67 \times 10^{-2}$ \\
12 42 15.2 &     +32 32 21 &
    $3.336 \times 10^{-3}$ &    $2.182 \times 10^{-3}$ &
    $6.637 \times 10^{-3}$ &    $1.5839 \times 10^{-2}$ &
    $1.2145 \times 10^{-2}$ &   $1.866 \times 10^{-1}$ &
    $1.22 \times 10^{-2}$ \\
12 42 16.6 &     +32 32 48 &
    $3.400 \times 10^{-3}$ &    $2.157 \times 10^{-3}$ &
    $6.804 \times 10^{-3}$ &    $1.7377 \times 10^{-2}$ &
    $1.2480 \times 10^{-2}$ &   $1.823 \times 10^{-1}$ &
    $1.50 \times 10^{-2}$ \\
12 42 16.7 &     +32 32 24 &
    $2.847 \times 10^{-3}$ &    $1.872 \times 10^{-3}$ &
    $5.596 \times 10^{-3}$ &    $1.3527 \times 10^{-2}$ &
    $1.2081 \times 10^{-2}$ &   $1.899 \times 10^{-1}$ &
    $1.25 \times 10^{-2}$ \\
12 42 17.3 &     +32 32 37 &
    $2.911 \times 10^{-3}$ &    $1.874 \times 10^{-3}$ &
    $5.775 \times 10^{-3}$ &    $1.4400 \times 10^{-2}$ &
    $1.0929 \times 10^{-2}$ &   $1.694 \times 10^{-1}$ &
    $1.42 \times 10^{-2}$ \\
12 42 18.1 &     +32 32 46 &
    $3.199 \times 10^{-3}$ &    $2.030 \times 10^{-3}$ &
    $6.365 \times 10^{-3}$ &    $1.6041 \times 10^{-2}$ &
    $1.1849 \times 10^{-2}$ &   $1.837 \times 10^{-1}$ &
    $1.96 \times 10^{-2}$ \\
12 42 19.0 &     +32 32 10 &
    $1.495 \times 10^{-3}$ &    $9.80 \times 10^{-4}$ &
    $2.422 \times 10^{-3}$ &    $5.729 \times 10^{-3}$ &
    $5.789 \times 10^{-3}$ &    $9.78 \times 10^{-2}$ &
    $5.8 \times 10^{-3}$ \\
12 42 19.7 &     +32 32 37 &
    $2.483 \times 10^{-3}$ &    $1.639 \times 10^{-3}$ &
    $5.428 \times 10^{-3}$ &    $1.3756 \times 10^{-2}$ &
    $1.3018 \times 10^{-2}$ &   $1.744 \times 10^{-1}$ &
    $1.59 \times 10^{-2}$ \\
12 42 20.0 &     +32 32 00 &
    $1.313 \times 10^{-3}$ &    $8.64 \times 10^{-4}$ &
    $1.966 \times 10^{-3}$ &    $4.709 \times 10^{-3}$ &
    $5.600 \times 10^{-3}$ &    $1.070 \times 10^{-1}$ &
    $7.2 \times 10^{-3}$ \\
12 42 20.0 &     +32 32 50 &
    $2.392 \times 10^{-3}$ &    $1.575 \times 10^{-3}$ &
    $5.157 \times 10^{-3}$ &    $1.3298 \times 10^{-2}$ &
    $1.2519 \times 10^{-2}$ &   $1.667 \times 10^{-1}$ &
    $1.00 \times 10^{-2}$ \\
12 42 20.7 &     +32 31 46 &
    $8.41 \times 10^{-4}$ &     $5.66 \times 10^{-4}$ &
    $1.173 \times 10^{-3}$ &    $2.798 \times 10^{-3}$ &
    $3.791 \times 10^{-3}$ &    $7.13 \times 10^{-2}$ &
    $4.6 \times 10^{-3}$ \\
12 42 21.4 &     +32 33 05 &
    $1.645 \times 10^{-3}$ &    $1.157 \times 10^{-3}$ &
    $4.135 \times 10^{-3}$ &    $1.1073 \times 10^{-2}$ &
    $1.5873 \times 10^{-2}$ &   $1.659 \times 10^{-1}$ &
    $8.6 \times 10^{-3}$ \\
12 42 21.5 &     +32 32 23 &
    $1.882 \times 10^{-3}$ &    $1.246 \times 10^{-3}$ &
    $3.499 \times 10^{-3}$ &    $8.500 \times 10^{-3}$ &
    $9.754 \times 10^{-3}$ &    $1.467 \times 10^{-1}$ &
    $8.2 \times 10^{-3}$ \\
12 42 22.0 &     +32 32 44 &
    $2.926 \times 10^{-3}$ &    $2.067 \times 10^{-3}$ &
    $7.723 \times 10^{-3}$ &    $2.0298 \times 10^{-2}$ &
    $3.3467 \times 10^{-2}$ &   $3.708 \times 10^{-1}$ &
    $1.65 \times 10^{-2}$ \\
12 42 23.2 &     +32 31 44 &
    $5.83 \times 10^{-4}$ &     $4.07 \times 10^{-4}$ &
    $7.78 \times 10^{-4}$ &     $2.070 \times 10^{-3}$ &
    $2.687 \times 10^{-3}$ &    $5.82 \times 10^{-2}$ &
    $5.6 \times 10^{-3}$ \\
12 42 23.5 &     +32 32 44 &
    $2.036 \times 10^{-3}$ &    $1.417 \times 10^{-3}$ &
    $4.648 \times 10^{-3}$ &    $1.1665 \times 10^{-2}$ &
    $1.4880 \times 10^{-2}$ &   $2.060 \times 10^{-1}$ &
    $1.35 \times 10^{-2}$ \\
12 42 23.6 &     +32 32 25 &
    $1.491 \times 10^{-3}$ &    $9.89 \times 10^{-4}$ &
    $2.432 \times 10^{-3}$ &    $5.961 \times 10^{-3}$ &
    $7.083 \times 10^{-3}$ &    $1.189 \times 10^{-1}$ &
    $1.19 \times 10^{-2}$ \\
12 42 24.0 &     +32 32 08 &
    $1.133 \times 10^{-3}$ &    $7.57 \times 10^{-4}$ &
    $1.575 \times 10^{-3}$ &    $3.909 \times 10^{-3}$ &
    $5.996 \times 10^{-3}$ &    $1.187 \times 10^{-1}$ &
    $7.4 \times 10^{-3}$ \\
12 42 24.8 &     +32 32 14 &
    $1.211 \times 10^{-3}$ &    $8.21 \times 10^{-4}$ &
    $1.717 \times 10^{-3}$ &    $4.296 \times 10^{-3}$ &
    $6.017 \times 10^{-3}$ &    $1.155 \times 10^{-1}$ &
    $1.07 \times 10^{-2}$ \\
12 42 25.3 &     +32 32 50 &
    $1.377 \times 10^{-3}$ &    $9.21 \times 10^{-4}$ &
    $2.344 \times 10^{-3}$ &    $5.977 \times 10^{-3}$ &
    $5.666 \times 10^{-3}$ &    $9.11 \times 10^{-2}$ &
    $1.12 \times 10^{-2}$ \\
12 42 26.0 &     +32 32 31 &
    $1.098 \times 10^{-3}$ &    $7.39 \times 10^{-4}$ &
    $1.423 \times 10^{-3}$ &    $3.563 \times 10^{-3}$ &
    $3.900 \times 10^{-3}$ &    $7.69 \times 10^{-2}$ &
    $6.1 \times 10^{-3}$ \\
12 42 27.3 &     +32 32 21 &
    $8.46 \times 10^{-4}$ &     $5.64 \times 10^{-4}$ &
    $9.54 \times 10^{-4}$ &     $2.553 \times 10^{-3}$ &
    $2.687 \times 10^{-3}$ &    $5.94 \times 10^{-2}$ &
    $2.7 \times 10^{-3}$ \\
12 42 27.6 &     +32 32 41 &
    $9.95 \times 10^{-4}$ &     $6.62 \times 10^{-4}$ &
    $1.188 \times 10^{-3}$ &    $3.017 \times 10^{-3}$ &
    $3.343 \times 10^{-3}$ &    $5.72 \times 10^{-2}$ &
    $6.5 \times 10^{-3}$ \\
12 42 28.7 &     +32 32 40 &
    $9.94 \times 10^{-4}$ &     $6.65 \times 10^{-4}$ &
    $1.246 \times 10^{-3}$ &    $3.322 \times 10^{-3}$ &
    $3.894 \times 10^{-3}$ &    $7.37 \times 10^{-2}$ &
    $1.38 \times 10^{-2}$ \\
12 42 29.3 &     +32 32 49 &
    $8.25 \times 10^{-4}$ &     $5.59 \times 10^{-4}$ &
    $1.043 \times 10^{-3}$ &    $2.833 \times 10^{-3}$ &
    $3.544 \times 10^{-3}$ &    $6.66 \times 10^{-2}$ &
    $8.6 \times 10^{-3}$ \\
12 42 30.8 &     +32 32 31 &
    $6.98 \times 10^{-4}$ &     $4.94 \times 10^{-4}$ &
    $7.80 \times 10^{-4}$ &     $2.277 \times 10^{-3}$ &
    $4.414 \times 10^{-3}$ &    $8.35 \times 10^{-2}$ &
    $4.8 \times 10^{-3}$ \\
12 42 32.5 &     +32 32 37 &
    $4.48 \times 10^{-4}$ &     $3.25 \times 10^{-4}$ &
    $2.61 \times 10^{-4}$ &     $1.105 \times 10^{-3}$ &
    $1.485 \times 10^{-3}$ &    $3.34 \times 10^{-2}$ &
    \nodata \\
\enddata
\tablenotetext{a}
    {The 3.6 - 24~$\mu$m flux densities are measured in maps that have been 
    convolved with the kernels described in Section~\ref{s_obs_kernel} so as
    to match the higher-resolution PSFs to the PSF of the MIPS 70~$\mu$m 
    data.} 
\tablenotetext{b}
    {Calibration uncertainty: 10\%; background noise uncertainty: 
    $4 \times 10^{-6}$ Jy}
\tablenotetext{c}
    {Calibration uncertainty: 10\%; background noise uncertainty: 
    $6 \times 10^{-6}$ Jy}
\tablenotetext{d}
    {Calibration uncertainty: 10\%; background noise uncertainty:  
    $1.7 \times 10^{-5}$ Jy}
\tablenotetext{e}
    {Calibration uncertainty: 10\%; background noise uncertainty:  
    $1.4 \times 10^{-5}$ Jy;
    note that these data are not stellar continuum subtracted.}
\tablenotetext{f}
    {Calibration uncertainty: 10\%; background noise uncertainty:  
    $1.1 \times 10^{-5}$ Jy}
\tablenotetext{g}
    {Calibration uncertainty: 20\%; background noise uncertainty:  
    $7 \times 10^{-4}$ Jy}
\tablenotetext{h}
    {Calibration uncertainty: 10\%; background noise uncertainty:  
    $1.0 \times 10^{-3}$ Jy;
    note that CO, free-free, and synchrotron emission have not been subtracted
    from these data, but since the sum of these emission components only 
    contributes less than 10\% to the global flux density, the absence of
    a correction for these data should not strongly affect the results.}
\end{deluxetable}

\begin{deluxetable}{ccccccccccc}
\rotate
\tablecolumns{11}
\tablewidth{0pc}
\tabletypesize{\tiny}
\tablecaption{Flux Densities Measured within Discrete 
    $40^{\prime\prime}$ Regions in NGC~4631
    \label{t_fd40arcsec}}
\tablehead{
\multicolumn{2}{c}{Position (J2000)} &
\multicolumn{9}{c}{Flux Density (Jy)\tablenotemark{a}}\\
\colhead{R.A.} &     \colhead{Dec.} &
    \colhead{3.6~$\mu$m\tablenotemark{b}} &     
    \colhead{4.5~$\mu$m\tablenotemark{c}} &
    \colhead{5.7~$\mu$m\tablenotemark{d}} &
    \colhead{8.0~$\mu$m\tablenotemark{e}} &
    \colhead{24~$\mu$m\tablenotemark{f}} &
    \colhead{70~$\mu$m\tablenotemark{g}} &
    \colhead{160~$\mu$m\tablenotemark{h}} &
    \colhead{450~$\mu$m\tablenotemark{i}} &
    \colhead{850~$\mu$m\tablenotemark{j}}}
\startdata
12 41 37.6 &     +32 31 34 &
    $1.94 \times 10^{-3}$ &     $1.00 \times 10^{-3}$  &
    $2.48 \times 10^{-3}$ &     $5.49 \times 10^{-3}$ &
    $9.85 \times 10^{-3}$ &     $1.849 \times 10^{-1}$ &
    $4.72 \times 10^{-1}$ &
    \nodata &                   $3.7 \times 10^{-2}$ \\
12 41 40.9 &     +32 31 50 &
    $3.95 \times 10^{-3}$ &     $2.28 \times 10^{-3}$ &
    $5.87 \times 10^{-3}$ &     $1.349 \times 10^{-2}$ &
    $2.945 \times 10^{-2}$ &    $4.938 \times 10^{-1}$ &
    $1.236 \times 10^{0}$ &
    $1.66 \times 10^{-1}$ &     $4.7 \times 10^{-2}$ \\
12 41 43.2 &     +32 31 56 &
    $4.54 \times 10^{-3}$ &     $2.60 \times 10^{-3}$ &
    $7.25 \times 10^{-3}$ &     $1.667 \times 10^{-2}$ &
    $3.064 \times 10^{-2}$ &    $5.232 \times 10^{-1}$ &
    $1.383 \times 10^{0}$ &
    $1.65 \times 10^{-1}$ &     $5.4 \times 10^{-2}$ \\
12 41 45.8 &     +32 32 00 &
    $6.29 \times 10^{-3}$ &     $3.77 \times 10^{-3}$ &
    $1.177 \times 10^{-2}$ &    $2.739 \times 10^{-2}$ &
    $4.954 \times 10^{-2}$ &    $7.547 \times 10^{-1}$ &
    $2.070 \times 10^{0}$ &
    $1.77 \times 10^{-1}$ &     $7.1 \times 10^{-2}$ \\
12 41 49.4 &     +32 32 04 &
    $9.49 \times 10^{-3}$ &     $6.11 \times 10^{-3}$ &
    $1.959 \times 10^{-2}$ &    $4.593 \times 10^{-2}$ &
    $8.539 \times 10^{-2}$ &    $1.1708 \times 10^{0}$ &
    $3.294 \times 10^{0}$ &
    $2.48 \times 10^{-1}$ &     $9.1 \times 10^{-2}$ \\
12 41 51.8 &     +32 32 10 &
    $1.083 \times 10^{-2}$ &    $6.74 \times 10^{-3}$ &
    $2.211 \times 10^{-2}$ &    $5.146 \times 10^{-2}$ &
    $8.308 \times 10^{-2}$ &    $1.2284 \times 10^{0}$ &
    $3.242 \times 10^{0}$ &
    $2.72 \times 10^{-1}$ &     $9.5 \times 10^{-2}$ \\
12 41 54.2 &     +32 32 14 &
    $1.497 \times 10^{-2}$ &    $9.35 \times 10^{-3}$ &
    $2.841 \times 10^{-2}$ &    $6.537 \times 10^{-2}$ &
    $1.1305 \times 10^{-1}$ &   $1.5426 \times 10^{0}$ &
    $3.967 \times 10^{0}$ &
    $4.22 \times 10^{-1}$ &     $1.05 \times 10^{-1}$ \\
12 41 58.4 &     +32 32 12 &
    $2.153 \times 10^{-2}$ &    $1.328 \times 10^{-2}$ &
    $3.617 \times 10^{-2}$ &    $8.042 \times 10^{-2}$ &
    $9.937 \times 10^{-2}$ &    $1.4413 \times 10^{0}$ &
    $4.006 \times 10^{0}$ &
    $5.19 \times 10^{-1}$ &     $9.5 \times 10^{-2}$ \\
12 42 01.3 &     +32 32 17 &
    $3.402 \times 10^{-2}$ &     $2.141 \times 10^{-2}$ &
    $6.719 \times 10^{-2}$ &     $1.5801 \times 10^{-1}$ &
    $2.0379 \times 10^{-1}$ &   $2.3387 \times 10^{0}$ &
    $6.470 \times 10^{0}$ &
    $7.15 \times 10^{-1}$ &     $1.21 \times 10^{-1}$ \\
12 42 04.3 &     +32 32 25 &
    $7.091 \times 10^{-2}$ &     $4.637 \times 10^{-2}$ &
    $1.5671 \times 10^{-1}$ &     $3.8625 \times 10^{-1}$ &
    $5.6175 \times 10^{-1}$ &   $5.4659 \times 10^{0}$ &
    $1.4361 \times 10^{1}$ &
    $1.332 \times 10^{0}$ &     $2.11 \times 10^{-1}$ \\
12 42 07.9 &     +32 32 35 &
    $9.913 \times 10^{-2}$ &    $6.584 \times 10^{-2}$ &
    $1.9151 \times 10^{-1}$ &   $4.7155 \times 10^{-1}$ &
    $7.2293 \times 10^{-1}$ &   $6.2692 \times 10^{0}$ &
    $1.6281 \times 10^{1}$ &
    $1.457 \times 10^{0}$ &     $2.16 \times 10^{-1}$ \\
12 42 10.9 &     +32 32 35 &
    $6.881 \times 10^{-2}$ &    $4.729 \times 10^{-2}$ &
    $1.5475 \times 10^{-1}$ &   $3.8830 \times 10^{-1}$ &
    $5.6790 \times 10^{-1}$ &   $5.4380 \times 10^{0}$ &
    $1.5191 \times 10^{1}$ &
    $1.375 \times 10^{0}$ &     $1.98 \times 10^{-1}$ \\
12 42 14.0 &     +32 32 38 &
    $3.300 \times 10^{-2}$ &    $2.198 \times 10^{-2}$ &
    $6.523 \times 10^{-2}$ &    $1.6114 \times 10^{-1}$ &
    $1.9198 \times 10^{-1}$ &   $2.4415 \times 10^{0}$ &
    $6.474 \times 10^{0}$ &
    $8.26 \times 10^{-1}$ &     $1.10 \times 10^{-1}$ \\
12 42 16.7 &     +32 32 39 &
    $1.850 \times 10^{-2}$ &    $1.200 \times 10^{-2}$ &
    $3.205 \times 10^{-2}$ &    $7.759 \times 10^{-2}$ &
    $8.832 \times 10^{-2}$ &    $1.2228 \times 10^{0}$ &
    $3.664 \times 10^{0}$ &
    $5.55 \times 10^{-1}$ &     $7.6 \times 10^{-2}$ \\
12 42 19.7 &     +32 32 37 &
    $1.409 \times 10^{-2}$ &    $9.20 \times 10^{-3}$ &
    $2.351 \times 10^{-2}$ &    $5.692 \times 10^{-2}$ &
    $7.538 \times 10^{-2}$ &    $1.0356 \times 10^{0}$ &
    $3.002 \times 10^{0}$ &
    $3.98 \times 10^{-1}$ &     $7.1 \times 10^{-2}$ \\
12 42 20.2 &     +32 31 59 &
    $7.91 \times 10^{-3}$ &     $5.23 \times 10^{-3}$ &
    $1.071 \times 10^{-2}$ &     $2.518 \times 10^{-2}$ &
    $3.669 \times 10^{-2}$ &    $6.253 \times 10^{-1}$ &
    $1.632 \times 10^{0}$ &
    $1.42 \times 10^{-1}$ &     $3.2 \times 10^{-2}$ \\
12 42 22.0 &     +32 32 44 &
    $1.288 \times 10^{-2}$ &    $8.71 \times 10^{-3}$ &
    $2.323 \times 10^{-2}$ &    $5.746 \times 10^{-2}$ &
    $9.974 \times 10^{-2}$ &    $1.2033 \times 10^{0}$ &
    $3.196 \times 10^{0}$ &
    $3.73 \times 10^{-1}$ &     $7.2 \times 10^{-2}$ \\
12 42 24.0 &     +32 32 08 &
    $7.09 \times 10^{-3}$ &     $4.78 \times 10^{-3}$ &
    $9.03 \times 10^{-3}$ &     $2.219 \times 10^{-2}$ &
    $3.726 \times 10^{-2}$ &    $6.627 \times 10^{-1}$ &
    $1.750 \times 10^{0}$ &
    $1.36 \times 10^{-1}$ &     $4.0 \times 10^{-2}$ \\
12 42 25.3 &     +32 32 50 &
    $7.91 \times 10^{-3}$ &     $5.30 \times 10^{-3}$ &
    $1.106 \times 10^{-2}$ &    $2.746 \times 10^{-2}$ &
    $4.000 \times 10^{-2}$ &    $6.223 \times 10^{-1}$ &
    $1.894 \times 10^{0}$ &
    $1.75 \times 10^{-1}$ &     $4.7 \times 10^{-2}$ \\
12 42 28.3 &     +32 32 44 &
    $5.26 \times 10^{-3}$ &     $3.55 \times 10^{-3}$ &
    $5.52 \times 10^{-3}$ &     $1.428 \times 10^{-2}$ &
    $2.177 \times 10^{-2}$ &    $3.789 \times 10^{-1}$ &
    $1.189 \times 10^{0}$ &
    $1.01 \times 10^{-1}$ &     $3.5 \times 10^{-2}$ \\
12 42 30.8 &     +32 32 31 &
    $3.32 \times 10^{-3}$ &     $2.34 \times 10^{-3}$ &
    $2.64 \times 10^{-3}$ &     $7.98 \times 10^{-3}$ &
    $1.477 \times 10^{-2}$ &    $2.889 \times 10^{-1}$ &
    $9.00 \times 10^{-1}$ &
    $2.7 \times 10^{-2}$ &      \nodata \\
\enddata
\tablenotetext{a}
    {The 3.6 - 70 and 450 - 850~$\mu$m flux densities are measured in maps 
    that have been convolved with the kernels described in 
    Section~\ref{s_obs_kernel} so as to match the higher-resolution PSFs 
    to the PSF of the MIPS 160~$\mu$m data.} 
\tablenotetext{b}
    {Calibration uncertainty: 10\%; background noise uncertainty:  
    $3 \times 10^{-5}$ Jy; 
    the ``infinite'' aperture corrections from \citet{retal05} have been 
    applied to these data.}
\tablenotetext{c}
    {Calibration uncertainty: 10\%; background noise uncertainty:  
    $3 \times 10^{-5}$ Jy; 
    the ``infinite'' aperture corrections from \citet{retal05} have been 
    applied to these data.}
\tablenotetext{d}
    {Calibration uncertainty: 10\%; background noise uncertainty:  
    $6 \times 10^{-5}$ Jy; 
    the ``infinite'' aperture corrections from \citet{retal05} have been 
    applied to these data.}
\tablenotetext{e}
    {Calibration uncertainty: 10\%; background noise uncertainty:  
    $7 \times 10^{-5}$ Jy; 
    the ``infinite'' aperture corrections from \citet{retal05} have been 
    applied to these data; note that these data are not stellar continuum 
    subtracted.}
\tablenotetext{f}
    {Calibration uncertainty: 10\%; background noise uncertainty:  
    $3 \times 10^{-5}$ Jy.}
\tablenotetext{g}
    {Calibration uncertainty: 20\%; background noise uncertainty:  
    $5 \times 10^{-4}$ Jy.}
\tablenotetext{h}
    {Calibration uncertainty: 20\%; background noise uncertainty:  
    $3 \times 10^{-3}$ Jy.}
\tablenotetext{i}
    {Calibration uncertainty: 25\%; background noise uncertainty:  
    $1.9 \times 10^{-2}$ Jy.}
\tablenotetext{j}
    {Calibration uncertainty: 10\%; background noise uncertainty:  
    $2 \times 10^{-3}$ Jy;
    note that CO, free-free, and synchrotron emission have not been subtracted
    from these data, but since the sum of these emission components only 
    contributes less than 10\% to the global flux density, the absence of
    a correction for these data should not strongly affect the results.}
\end{deluxetable}

\begin{deluxetable}{cccc}
\tablecolumns{4}
\tablewidth{0pc}
\tablecaption{Global Spectral Energy Distribution of NGC~4631 
    \label{t_globalsed}}
\tablehead{ 
    \colhead{Wavelength ($\mu$m)} &
        \colhead{Flux Density (Jy)} &
        \colhead{Source} &
        \colhead{Reference}}
\startdata 
1.3 &    $1.56 \pm 0.03$ &                     2MASS &                 1\\
1.6 &    $1.91 \pm 0.04$ &                     2MASS &                 1\\
2.2 &    $1.73 \pm 0.04$ &                     2MASS &                 1\\
3.6 &    $1.30 \pm 0.13$\tablenotemark{a} &    {\it Spitzer}/IRAC &    2\\
4.5 &    $0.83 \pm 0.08$\tablenotemark{a} &    {\it Spitzer}/IRAC &    2\\
5.7 &    $2.5 \pm 0.3$\tablenotemark{a} &      {\it Spitzer}/IRAC &    2\\
8.0 &    $5.8 \pm 0.6$\tablenotemark{a} &      {\it Spitzer}/IRAC &    2\\
12 &     $5.5 \pm 0.8$ &                       IRAS &                  3\\  
24 &     $8.0 \pm 0.8$ &                       {\it Spitzer}/MIPS &    2\\
60 &     $83 \pm 12$ &                         IRAS &                  3\\    
70 &     $100 \pm 20$ &                        {\it Spitzer}/MIPS &    2\\
100 &    $210 \pm 30$ &                        IRAS &                  3\\
160 &    $270 \pm 50$ &                        {\it Spitzer}/MIPS &    2\\
450 &    $24 \pm 6$ &                          JCMT/SCUBA &      
    (this paper)\\
850 &    $4.5 \pm 0.5$\tablenotemark{b} &      JCMT/SCUBA &      
    (this paper)\\
1230 &   $2.0 \pm 0.2$\tablenotemark{b} &      IRAM &                  4
\enddata 
\tablenotetext{a}{The ``infinite'' aperture corrections from \citet{retal05}
    have been applied to these data.}
\tablenotetext{b}{CO, free-free, and synchrotron emission have been 
    subtracted from these global flux density measurements using data
    from \citet{dkw04}.}
\tablerefs{{1} \citet{jccsh03}, {2} \citet{detal05}, {3} \citet{rlsnkldh88}, 
    {4} \citet{dkw04}}
\end{deluxetable}

\begin{deluxetable}{cccccc}
\tabletypesize{\tiny}
\tablecolumns{6}
\tablewidth{0pc}
\tablecaption{Results of Fitting Blackbodies Modified with $\lambda^{-2}$
    Emissivity Laws to the 70 - 450~$\mu$m SEDs of $40^{\prime\prime}$ 
    Regions in NGC~4631\tablenotemark{a}
    \label{t_discretesed_bbfit}}
\tablehead{ 
    \multicolumn{2}{c}{Position (J2000)} &
        \colhead{Temperature} &
        \colhead{Excess 850~$\mu$m} &
        \colhead{Ratio Excess to Total} &
        \colhead{Total Infrared} \\
    \colhead{R.A.} &     \colhead{Dec.} &
        \colhead{(K)\tablenotemark{b}} &
        \colhead{Flux Density (Jy)} &
        \colhead{850~$\mu$m Flux Density} &
        \colhead{Luminosity (L$_\odot$)}}
\startdata 
12 41 40.9 &     +32 31 50 &
    22.2 &
    $0.031 \pm 0.005$ &     $0.65 \pm 0.13$ &
    $1.33 \pm 0.18. \times 10^8$
    \\
12 41 43.2 &     +32 31 56 &
    22.3 &
    $0.037 \pm 0.006$ &     $0.68 \pm 0.13$ &
    $1.45 \pm 0.19 \times 10^8$
    \\
12 41 45.8 &     +32 32 00 &
    22.8 &
    $0.049 \pm 0.007$ &     $0.69 \pm 0.13$ &
    $2.2 \pm 0.3. \times 10^8$
    \\
12 41 49.4 &     +32 32 04 &
    23.0 &
    $0.059 \pm 0.009$ &     $0.65 \pm 0.12$ &
    $3.4 \pm 5. \times 10^8$
    \\
12 41 51.8 &     +32 32 10 &
    23.0 &
    $0.062 \pm 0.010$ &     $0.65 \pm 0.12$ &
    $3.5 \pm 0.5 \times 10^8$
    \\
12 41 54.2 &     +32 32 14 &
    22.6 &
    $0.058 \pm 0.011$ &     $0.56 \pm 0.12$ &
    $4.3 \pm 0.6 \times 10^8$
    \\
12 41 58.4 &     +32 32 12 &
    21.9 &
    $0.041 \pm 0.010$ &     $0.43 \pm 0.11$ &
    $4.2 \pm 0.6 \times 10^8$
    \\
12 42 01.3 &     +32 32 17 &
    22.2 &
    $0.042 \pm 0.012$ &     $0.35 \pm 0.11$ &
    $7.0 \pm 0.9 \times 10^8$
    \\
12 42 04.3 &     +32 32 25 &
    22.8 &
    $0.056 \pm 0.021$ &     $0.26 \pm 0.10$ &
    $1.6 \pm 0.2 \times 10^9$
    \\
12 42 07.9 &     +32 32 35 &
    22.9 &
    $0.044 \pm 0.022$ &     $0.20 \pm 0.10$ &
    $1.9 \pm 0.2 \times 10^9$
    \\
12 42 10.9 &     +32 32 35 &
    22.6 &
    $0.034 \pm 0.020$ &     $0.17 \pm 0.10$ &
    $1.7 \pm 0.2 \times 10^9$
    \\
12 42 14.0 &     +32 32 38 &
    22.1 &
    $0.024 \pm 0.011$ &     $0.22 \pm 0.10$ &
    $7.0 \pm 0.9 \times 10^8$
    \\
12 42 16.7 &     +32 32 39 &
    21.3 &
    $0.020 \pm 0.008$ &     $0.27 \pm 0.11$ &
    $3.7 \pm 0.5 \times 10^8$
    \\
12 42 19.7 &     +32 32 37 &
    21.7 &
    $0.029 \pm 0.007$ &     $0.41 \pm 0.11$ &
    $3.1 \pm 0.4 \times 10^8$
    \\
12 42 20.2 &     +32 31 59 &
    23.0 &
    $0.015 \pm 0.004$ &     $0.47 \pm 0.13$ &
    $1.7 \pm 0.2 \times 10^8$
    \\
12 42 22.0 &     +32 32 44 &
    22.3 &
    $0.032 \pm 0.008$ &     $0.44 \pm 0.11$ &
    $3.5 \pm 0.5 \times 10^8$
    \\
12 42 24.0 &     +32 32 08 &
    23.2 &
    $0.023 \pm 0.005$ &     $0.57 \pm 0.13$ &
    $1.8 \pm 0.2 \times 10^8$
    \\
12 42 25.3 &     +32 32 50 &
    22.3 &
    $0.026 \pm 0.005$ &     $0.55 \pm 0.13$ &
    $1.9 \pm 0.3 \times 10^8$
    \\
12 42 28.3 &     +32 32 44 &
    22.3 &
    $0.022 \pm 0.004$ &     $0.63 \pm 0.14$ &
    $1.17 \pm 0.16 \times 10^8$
 \enddata 
\tablenotetext{a}{The flux densities for these regions are listed in 
    Table~\ref{t_fd40arcsec}.}
\tablenotetext{b}{Uncertainties in the temperatures are 2~K.}
\end{deluxetable}

\begin{deluxetable}{cccccc}
\tabletypesize{\scriptsize}
\tablecolumns{6}
\tablewidth{0pc}
\tablecaption{Results of Fitting Semi-Empirical Models to the 
    5.7 - 450~$\mu$m SEDs of $40^{\prime\prime}$ Regions in 
    NGC~4631\tablenotemark{a}
    \label{t_discretesed_semimodelfit}}
\tablehead{ 
    \multicolumn{2}{c}{Position (J2000)} &
        \colhead{$\alpha$\tablenotemark{b}} &
        \colhead{Excess 850~$\mu$m} &
        \colhead{Ratio Excess to Total} &
        \colhead{Total Infrared} \\
    \colhead{R.A.} &     \colhead{Dec.} &
        \colhead{} &
        \colhead{Flux Density (Jy)} &
        \colhead{850~$\mu$m Flux Density} &
        \colhead{Luminosity (L$_\odot$)}}
\startdata 
12 41 40.9 &     +32 31 50 &
    2.31 &
    $0.031 \pm 0.005$ &     $0.65 \pm 0.13$ &
    $1.33 \pm 0.18 \times 10^8$
    \\
12 41 43.2 &     +32 31 56 &
    2.44 &
    $0.035 \pm 0.006$ &     $0.64 \pm 0.13$ &
    $1.45 \pm 0.19 \times 10^8$
    \\
12 41 45.8 &     +32 32 00 &
    2.38 &
    $0.042 \pm 0.007$ &     $0.60 \pm 0.12$ &
    $2.2 \pm 0.3 \times 10^8$
    \\
12 41 49.4 &     +32 32 04 &
    2.31 &
    $0.049 \pm 0.009$ &     $0.53 \pm 0.12$ &
    $3.4 \pm 0.5 \times 10^8$
    \\
12 41 51.8 &     +32 32 10 &
    2.38 &
    $0.051 \pm 0.010$ &     $0.54 \pm 0.12$ &
    $4.3 \pm 0.6 \times 10^8$
    \\
12 41 54.2 &     +32 32 14 &
    2.31 &
    $0.053 \pm 0.011$ &     $0.50 \pm 0.11$ &
    $4.2 \pm 0.6 \times 10^8$
    \\
12 41 58.4 &     +32 32 12 &
    2.44 &
    $0.039 \pm 0.010$ &     $0.41 \pm 0.11$ &
    $4.2 \pm 0.6 \times 10^8$
    \\
12 42 01.3 &     +32 32 17 &
    2.31 &
    $0.037 \pm 0.012$ &     $0.30 \pm 0.11$ &
    $7.0 \pm 0.9 \times 10^8$
    \\
12 42 04.3 &     +32 32 25 &
    2.19 &
    $0.039 \pm 0.021$ &     $0.18 \pm 0.10$ &
    $1.6 \pm 0.2 \times 10^9$
    \\
12 42 07.9 &     +32 32 35 &
    2.13 &
    $0.030 \pm 0.022$ &     $0.14 \pm 0.10$ &
    $1.9 \pm 0.2 \times 10^9$
    \\
12 42 10.9 &     +32 32 35 &
    2.25 &
    $0.008 \pm 0.020$ &     $0.04 \pm 0.10$ &
    $1.7 \pm 0.2 \times 10^9$
    \\
12 42 14.0 &     +32 32 38 &
    2.31 &
    $0.024 \pm 0.011$ &     $0.22 \pm 0.10$ &
    $7.0 \pm 0.9 \times 10^8$
    \\
12 42 16.7 &     +32 32 39 &
    2.50 &
    $0.024 \pm 0.008$ &     $0.31 \pm 0.11$ &
    $3.7 \pm 0.5 \times 10^8$
    \\
12 42 19.7 &     +32 32 37 &
    2.44 &
    $0.029 \pm 0.007$ &     $0.41 \pm 0.11$ &
    $3.1 \pm 0.4 \times 10^8$
    \\
12 42 20.2 &     +32 31 59 &
    2.50 &
    $0.008 \pm 0.004$ &     $0.26 \pm 0.12$ &
    $1.7 \pm 0.2 \times 10^8$
    \\
12 42 22.0 &     +32 32 44 &
    2.25 &
    $0.032 \pm 0.008$ &     $0.44 \pm 0.11$ &
    $3.5 \pm 0.5 \times 10^8$
    \\
12 42 24.0 &     +32 32 08 &
    2.50 &
    $0.014 \pm 0.005$ &     $0.37 \pm 0.12$ &
    $1.8 \pm 0.2 \times 10^8$
    \\
12 42 25.3 &     +32 32 50 &
    2.56 &
    $0.019 \pm 0.005$ &     $0.41 \pm 0.12$ &
    $1.9 \pm 0.3 \times 10^8$
    \\
12 42 28.3 &     +32 32 44 &
    2.81 &
    $0.016 \pm 0.004$ &     $0.47 \pm 0.13$ &
    $1.17 \pm 0.16 \times 10^8$
\enddata 
\tablenotetext{a}{The flux densities for these regions are listed in 
    Table~\ref{t_fd40arcsec}.}
\tablenotetext{b}{This is the index of the power law that describes the
    range of radiation fields heating the dust.  Uncertainties in the 
    values are $\sim0.10$.}
\end{deluxetable}

\begin{deluxetable}{ccccccc}
\tabletypesize{\scriptsize}
\tablecolumns{7}
\tablewidth{0pc}
\tablecaption{Parameters of Physical Models fit to the 
    5.7 - 850~$\mu$m SEDs of $40^{\prime\prime}$ Regions in 
    NGC~4631\tablenotemark{a}
    \label{t_discretesed_drainemodelfit_param}}
\tablehead{ 
    \multicolumn{2}{c}{Position (J2000)} &
        \colhead{Total} &
        \colhead{PAH Mass} &
        \colhead{$\gamma$\tablenotemark{c}} &
        \colhead{Minimum Illuminating} \\
    \colhead{R.A.} &     \colhead{Dec.} &
        \colhead{Mass (M$_\odot$)} &
        \colhead{Fraction\tablenotemark{b}} &
	\colhead{} &
        \colhead{Radiation Field\tablenotemark{d}}}
\startdata 
12 41 40.9 &     +32 31 50 &
    $1.1\times10^6$ &   0.019 &     0.0151 &    0.7 \\
12 41 43.2 &     +32 31 56 &
    $9.8\times10^5$&    0.019 &     0.0103 &    1.0 \\
12 41 45.8 &     +32 32 00 &
    $1.2\times10^6$&    0.026 &     0.0170 &    1.0 \\
12 41 49.4 &     +32 32 04 &
    $1.3\times10^6$&    0.026 &     0.0164 &    1.5 \\
12 41 51.8 &     +32 32 10 &
    $1.5\times10^6$&    0.026 &     0.0140 &    1.5 \\
12 41 54.2 &     +32 32 14 &
    $1.9\times10^6$&    0.026 &     0.0153 &    1.5 \\
12 41 58.4 &     +32 32 12 &
    $1.9\times10^6$&    0.034 &     0.0107 &    1.5 \\
12 42 01.3 &     +32 32 17 &
    $2.7\times10^6$&    0.042 &     0.0212 &    1.5 \\
12 42 04.3 &     +32 32 25 &
    $4.5\times10^6$&    0.042 &     0.0299 &    2.0 \\
12 42 07.9 &     +32 32 35 &
    $3.8\times10^6$&    0.042 &     0.0278 &    3.0 \\
12 42 10.9 &     +32 32 35 &
    $4.3\times10^6$&    0.042 &     0.0323 &    2.0 \\
12 42 14.0 &     +32 32 38 &
    $2.2\times10^6$&    0.042 &     0.0158 &    2.0 \\
12 42 16.7 &     +32 32 39 &
    $1.7\times10^6$&    0.034 &     0.0115 &    1.5 \\
12 42 19.7 &     +32 32 37 &
    $1.7\times10^6$&    0.034 &     0.0172 &    1.0 \\
12 42 20.2 &     +32 31 59 &
    $5.5\times10^5$&    0.026 &     0.0110 &    2.0 \\
12 42 22.0 &     +32 32 44 &
    $1.4\times10^6$&    0.026 &     0.0193 &    1.5 \\
12 42 24.0 &     +32 32 08 &
    $6.2\times10^5$&    0.019 &     0.00921 &   2.0 \\
12 42 25.3 &     +32 32 50 &
    $8.1\times10^5$&    0.026 &     0.0110 &    1.5 \\
12 42 28.3 &     +32 32 44 &
    $6.6\times10^5$&    0.019 &     0.0115 &    1.0 
\enddata 
\tablenotetext{a}{The flux densities for these regions are listed in 
    Table~\ref{t_fd40arcsec}.}
\tablenotetext{b}{This is the fraction of the total dust mass that is in 
    PAH molecules that are smaller than $10^3$ carbon atoms. The value for 
    the Milky way is 0.0488 $\sim0.0078$.}
\tablenotetext{c}{This is the fraction of dust heated by a range of radiation
    fields described by a power law distribution.  The rest of the dust is
    assumed to be heated by a low level radiation field.}
\tablenotetext{d}{This is expressed in units of the local interstellar
    radiation field.}
\end{deluxetable}

\begin{deluxetable}{cccccccc}
\rotate
\tabletypesize{\scriptsize}
\tablecolumns{12}
\tablewidth{0pc}
\tablecaption{Results of Fitting Physical Models to the 
    5.7 - 850~$\mu$m SEDs of $40^{\prime\prime}$ Regions in 
    NGC~4631\tablenotemark{a}
    \label{t_discretesed_drainemodelfit}}
\tablehead{ 
    \multicolumn{2}{c}{Position (J2000)} &
        \colhead{Excess 450~$\mu$m} &
        \colhead{Ratio Excess to Total} &
        \colhead{Excess 850~$\mu$m} &
        \colhead{Ratio Excess to Total} &
        \colhead{Total Infrared} \\
    \colhead{R.A.} &     \colhead{Dec.} &
        \colhead{Flux Density (Jy)\tablenotemark{b}} &
        \colhead{450~$\mu$m Flux Density\tablenotemark{b}} &
        \colhead{Flux Density (Jy)\tablenotemark{b}} &
        \colhead{850~$\mu$m Flux Density\tablenotemark{b}} &
        \colhead{Luminosity (L$_\odot$)}}
\startdata 
12 41 40.9 &     +32 31 50 &
    $-0.09 \pm 0.05$ &      $-0.6 \pm 0.3$ &
    $0.005 \pm 0.005$ &     $0.11 \pm 0.11$ &
    $1.33 \pm 0.18 \times 10^8$
    \\
12 41 43.2 &     +32 31 56 &
    $-0.10 \pm 0.05$ &      $-0.6 \pm 0.3$ &
    $0.011 \pm 0.006$ &     $0.21 \pm 0.11$ &
    $1.45 \pm 0.19 \times 10^8$
    \\
12 41 45.8 &     +32 32 00 &
    -0.15$ \pm 0.05$ &      $-0.8 \pm 0.3$ &
    $0.020 \pm 0.007$ &     $0.28 \pm 0.10$ &
    $2.2 \pm 0.3 \times 10^8$
    \\
12 41 49.4 &     +32 32 04 &
    $-0.18 \pm 0.06$ &      $-0.7 \pm 0.3$ &
    $0.027 \pm 0.009$ &     $0.30 \pm 0.10$ &
    $3.4 \pm 0.5 \times 10^8$
    \\
12 41 51.8 &     +32 32 10 &
    $-0.20 \pm 0.07$ &      $-0.7 \pm 0.3$ &
    $0.024 \pm 0.010$ &     $0.26 \pm 0.10$ &
    $4.3 \pm 0.6 \times 10^8$
    \\
12 41 54.2 &     +32 32 14 &
    $-0.17 \pm 0.11$ &      $-0.4 \pm 0.3$ &
    $0.015 \pm 0.011$ &     $0.14 \pm 0.10$ &
    $4.2 \pm 0.6 \times 10^8$
    \\
12 41 58.4 &     +32 32 12 &
    $-0.07 \pm 0.13$ &      $-0.1 \pm 0.3$ &
    $0.006 \pm 0.010$ &     $0.06 \pm 0.10$ &
    $4.2 \pm 0.6 \times 10^8$
    \\
12 42 01.3 &     +32 32 17 &
    $-0.09 \pm 0.18$ &      $-0.1 \pm 0.3$ &
    $-0.003 \pm 0.012$ &    $-0.02 \pm 0.10$ &
    $7.0 \pm 0.9 \times 10^8$
    \\
12 42 04.3 &     +32 32 25 &
    $-0.1 \pm 0.4$ &        $-0.1 \pm 0.3$ &
    $-0.01 \pm 0.02$ &      $-0.04 \pm 0.10$ &
    $1.6 \pm 0.2 \times 10^9$
    \\
12 42 07.9 &     +32 32 35 &
    $0.0 \pm 0.4$ &         $0.0 \pm 0.3$ &
    $0.01 \pm 0.02$ &       $0.03 \pm 0.10$ &
    $1.9 \pm 0.2 \times 10^9$
    \\
12 42 10.9 &     +32 32 35 &
    $0.0 \pm 0.3$ &         $0.0 \pm 0.3$ &
    $-0.01 \pm 0.02$ &      $-0.07 \pm 0.10$ &
    $1.7 \pm 0.2 \times 10^9$
    \\
12 42 14.0 &     +32 32 38 &
    $0.1 \pm 0.2$ &         $0.1 \pm 0.3$ &
    $-0.001 \pm 0.011$ &    $-0.01 \pm 0.10$ &
    $7.0 \pm 0.9 \times 10^8$
    \\
12 42 16.7 &     +32 32 39 &
    $0.04 \pm 0.14$ &       $0.1 \pm 0.3$ &
    $-0.002 \pm 0.008$ &    $-0.03 \pm 0.10$ &
    $3.7 \pm 0.5 \times 10^8$
    \\
12 42 19.7 &     +32 32 37 &
    $-0.07 \pm 0.10$ &      $-0.2 \pm 0.3$ &
    $-0.003 \pm 0.007$ &    $-0.04 \pm 0.10$ &
    $3.1 \pm 0.4 \times 10^8$
    \\
12 42 20.2 &     +32 31 59 &
    $-0.05 \pm 0.04$ &      $-0.4 \pm 0.3$ &
    $0.004 \pm 0.004$ &     $0.12 \pm 0.12$ &
    $1.7 \pm 0.2 \times 10^8$
    \\
12 42 22.0 &     +32 32 44 &
    $-0.08 \pm 0.10$ &      $-0.2 \pm 0.3$ &
    $0.003 \pm 0.008$ &     $0.04 \pm 0.10$ &
    $3.5 \pm 0.5 \times 10^8$
    \\
12 42 24.0 &     +32 32 08 &
    $-0.08 \pm 0.04$ &      $-0.6 \pm 0.3$ &
    $0.008 \pm 0.005$ &     $0.20 \pm 0.11$ &
    $1.8 \pm 0.2 \times 10^8$
    \\
12 42 25.3 &     +32 32 50 &
    $-0.08 \pm 0.05$ &      $-0.4 \pm 0.3$ &
    $0.009\pm 0.005$ &      $0.19 \pm 0.11$ &
    $1.9 \pm 0.3 \times 10^8$
    \\
12 42 28.3 &     +32 32 44 &
    $-0.08 \pm 0.03$ &      $-0.8 \pm 0.4$ &
    $0.006 \pm 0.004$ &     $0.17 \pm 0.12$ &
    $1.17 \pm 0.16 \times 10^8$
\enddata 
\tablenotetext{a}{The flux densities for these regions are listed in 
    Table~\ref{t_fd40arcsec}.}
\tablenotetext{b}{Negative numbers correspond to regions where the model
    fits predict a higher flux density than the observations.} 
\end{deluxetable}

\end{document}